\documentclass[prd,aps,showpacs,superscriptaddress,nofootinbib,eqsecnum,amsfonts,amsmath,preprintnumbers,floatfix,floats]{revtex4}

\usepackage{epsfig}
\usepackage{bm}
\usepackage{graphics}
\usepackage{amssymb}
\usepackage{amsmath}
\usepackage{color} 
\definecolor{AliceBlue}{rgb}{0.94,0.97,1.00}
\definecolor{AntiqueWhite1}{rgb}{1.00,0.94,0.86}
\definecolor{AntiqueWhite2}{rgb}{0.93,0.87,0.80}
\definecolor{AntiqueWhite3}{rgb}{0.80,0.75,0.69}
\definecolor{AntiqueWhite4}{rgb}{0.55,0.51,0.47}
\definecolor{AntiqueWhite}{rgb}{0.98,0.92,0.84}
\definecolor{BlanchedAlmond}{rgb}{1.00,0.92,0.80}
\definecolor{BlueViolet}{rgb}{0.54,0.17,0.89}
\definecolor{CadetBlue1}{rgb}{0.60,0.96,1.00}
\definecolor{CadetBlue2}{rgb}{0.56,0.90,0.93}
\definecolor{CadetBlue3}{rgb}{0.48,0.77,0.80}
\definecolor{CadetBlue4}{rgb}{0.33,0.53,0.55}
\definecolor{CadetBlue}{rgb}{0.37,0.62,0.63}
\definecolor{CornflowerBlue}{rgb}{0.39,0.58,0.93}
\definecolor{DarkBlue}{rgb}{0.00,0.00,0.55}
\definecolor{DarkCyan}{rgb}{0.00,0.55,0.55}
\definecolor{DarkGoldenrod1}{rgb}{1.00,0.73,0.06}
\definecolor{DarkGoldenrod2}{rgb}{0.93,0.68,0.05}
\definecolor{DarkGoldenrod3}{rgb}{0.80,0.58,0.05}
\definecolor{DarkGoldenrod4}{rgb}{0.55,0.40,0.03}
\definecolor{DarkGoldenrod}{rgb}{0.72,0.53,0.04}
\definecolor{DarkGray}{rgb}{0.66,0.66,0.66}
\definecolor{DarkGreen}{rgb}{0.00,0.39,0.00}
\definecolor{DarkGrey}{rgb}{0.66,0.66,0.66}
\definecolor{DarkKhaki}{rgb}{0.74,0.72,0.42}
\definecolor{DarkMagenta}{rgb}{0.55,0.00,0.55}
\definecolor{DarkOliveGreen1}{rgb}{0.79,1.00,0.44}
\definecolor{DarkOliveGreen2}{rgb}{0.74,0.93,0.41}
\definecolor{DarkOliveGreen3}{rgb}{0.64,0.80,0.35}
\definecolor{DarkOliveGreen4}{rgb}{0.43,0.55,0.24}
\definecolor{DarkOliveGreen}{rgb}{0.33,0.42,0.18}
\definecolor{DarkOrange1}{rgb}{1.00,0.50,0.00}
\definecolor{DarkOrange2}{rgb}{0.93,0.46,0.00}
\definecolor{DarkOrange3}{rgb}{0.80,0.40,0.00}
\definecolor{DarkOrange4}{rgb}{0.55,0.27,0.00}
\definecolor{DarkOrange}{rgb}{1.00,0.55,0.00}
\definecolor{DarkOrchid1}{rgb}{0.75,0.24,1.00}
\definecolor{DarkOrchid2}{rgb}{0.70,0.23,0.93}
\definecolor{DarkOrchid3}{rgb}{0.60,0.20,0.80}
\definecolor{DarkOrchid4}{rgb}{0.41,0.13,0.55}
\definecolor{DarkOrchid}{rgb}{0.60,0.20,0.80}
\definecolor{DarkRed}{rgb}{0.55,0.00,0.00}
\definecolor{DarkSalmon}{rgb}{0.91,0.59,0.48}
\definecolor{DarkSeaGreen1}{rgb}{0.76,1.00,0.76}
\definecolor{DarkSeaGreen2}{rgb}{0.71,0.93,0.71}
\definecolor{DarkSeaGreen3}{rgb}{0.61,0.80,0.61}
\definecolor{DarkSeaGreen4}{rgb}{0.41,0.55,0.41}
\definecolor{DarkSeaGreen}{rgb}{0.56,0.74,0.56}
\definecolor{DarkSlateBlue}{rgb}{0.28,0.24,0.55}
\definecolor{DarkSlateGray1}{rgb}{0.59,1.00,1.00}
\definecolor{DarkSlateGray2}{rgb}{0.55,0.93,0.93}
\definecolor{DarkSlateGray3}{rgb}{0.47,0.80,0.80}
\definecolor{DarkSlateGray4}{rgb}{0.32,0.55,0.55}
\definecolor{DarkSlateGray}{rgb}{0.18,0.31,0.31}
\definecolor{DarkSlateGrey}{rgb}{0.18,0.31,0.31}
\definecolor{DarkTurquoise}{rgb}{0.00,0.81,0.82}
\definecolor{DarkViolet}{rgb}{0.58,0.00,0.83}
\definecolor{DeepPink1}{rgb}{1.00,0.08,0.58}
\definecolor{DeepPink2}{rgb}{0.93,0.07,0.54}
\definecolor{DeepPink3}{rgb}{0.80,0.06,0.46}
\definecolor{DeepPink4}{rgb}{0.55,0.04,0.31}
\definecolor{DeepPink}{rgb}{1.00,0.08,0.58}
\definecolor{DeepSkyBlue1}{rgb}{0.00,0.75,1.00}
\definecolor{DeepSkyBlue2}{rgb}{0.00,0.70,0.93}
\definecolor{DeepSkyBlue3}{rgb}{0.00,0.60,0.80}
\definecolor{DeepSkyBlue4}{rgb}{0.00,0.41,0.55}
\definecolor{DeepSkyBlue}{rgb}{0.00,0.75,1.00}
\definecolor{DimGray}{rgb}{0.41,0.41,0.41}
\definecolor{DimGrey}{rgb}{0.41,0.41,0.41}
\definecolor{DodgerBlue1}{rgb}{0.12,0.56,1.00}
\definecolor{DodgerBlue2}{rgb}{0.11,0.53,0.93}
\definecolor{DodgerBlue3}{rgb}{0.09,0.45,0.80}
\definecolor{DodgerBlue4}{rgb}{0.06,0.31,0.55}
\definecolor{DodgerBlue}{rgb}{0.12,0.56,1.00}
\definecolor{FloralWhite}{rgb}{1.00,0.98,0.94}
\definecolor{ForestGreen}{rgb}{0.13,0.55,0.13}
\definecolor{GhostWhite}{rgb}{0.97,0.97,1.00}
\definecolor{GreenYellow}{rgb}{0.68,1.00,0.18}
\definecolor{HotPink1}{rgb}{1.00,0.43,0.71}
\definecolor{HotPink2}{rgb}{0.93,0.42,0.65}
\definecolor{HotPink3}{rgb}{0.80,0.38,0.56}
\definecolor{HotPink4}{rgb}{0.55,0.23,0.38}
\definecolor{HotPink}{rgb}{1.00,0.41,0.71}
\definecolor{IndianRed1}{rgb}{1.00,0.42,0.42}
\definecolor{IndianRed2}{rgb}{0.93,0.39,0.39}
\definecolor{IndianRed3}{rgb}{0.80,0.33,0.33}
\definecolor{IndianRed4}{rgb}{0.55,0.23,0.23}
\definecolor{IndianRed}{rgb}{0.80,0.36,0.36}
\definecolor{LavenderBlush1}{rgb}{1.00,0.94,0.96}
\definecolor{LavenderBlush2}{rgb}{0.93,0.88,0.90}
\definecolor{LavenderBlush3}{rgb}{0.80,0.76,0.77}
\definecolor{LavenderBlush4}{rgb}{0.55,0.51,0.53}
\definecolor{LavenderBlush}{rgb}{1.00,0.94,0.96}
\definecolor{LawnGreen}{rgb}{0.49,0.99,0.00}
\definecolor{LemonChiffon1}{rgb}{1.00,0.98,0.80}
\definecolor{LemonChiffon2}{rgb}{0.93,0.91,0.75}
\definecolor{LemonChiffon3}{rgb}{0.80,0.79,0.65}
\definecolor{LemonChiffon4}{rgb}{0.55,0.54,0.44}
\definecolor{LemonChiffon}{rgb}{1.00,0.98,0.80}
\definecolor{LightBlue1}{rgb}{0.75,0.94,1.00}
\definecolor{LightBlue2}{rgb}{0.70,0.87,0.93}
\definecolor{LightBlue3}{rgb}{0.60,0.75,0.80}
\definecolor{LightBlue4}{rgb}{0.41,0.51,0.55}
\definecolor{LightBlue}{rgb}{0.68,0.85,0.90}
\definecolor{LightCoral}{rgb}{0.94,0.50,0.50}
\definecolor{LightCyan1}{rgb}{0.88,1.00,1.00}
\definecolor{LightCyan2}{rgb}{0.82,0.93,0.93}
\definecolor{LightCyan3}{rgb}{0.71,0.80,0.80}
\definecolor{LightCyan4}{rgb}{0.48,0.55,0.55}
\definecolor{LightCyan}{rgb}{0.88,1.00,1.00}
\definecolor{LightGoldenrod1}{rgb}{1.00,0.93,0.55}
\definecolor{LightGoldenrod2}{rgb}{0.93,0.86,0.51}
\definecolor{LightGoldenrod3}{rgb}{0.80,0.75,0.44}
\definecolor{LightGoldenrod4}{rgb}{0.55,0.51,0.30}
\definecolor{LightGoldenrodYellow}{rgb}{0.98,0.98,0.82}
\definecolor{LightGoldenrod}{rgb}{0.93,0.87,0.51}
\definecolor{LightGray}{rgb}{0.83,0.83,0.83}
\definecolor{LightGreen}{rgb}{0.56,0.93,0.56}
\definecolor{LightGrey}{rgb}{0.83,0.83,0.83}
\definecolor{LightPink1}{rgb}{1.00,0.68,0.73}
\definecolor{LightPink2}{rgb}{0.93,0.64,0.68}
\definecolor{LightPink3}{rgb}{0.80,0.55,0.58}
\definecolor{LightPink4}{rgb}{0.55,0.37,0.40}
\definecolor{LightPink}{rgb}{1.00,0.71,0.76}
\definecolor{LightSalmon1}{rgb}{1.00,0.63,0.48}
\definecolor{LightSalmon2}{rgb}{0.93,0.58,0.45}
\definecolor{LightSalmon3}{rgb}{0.80,0.51,0.38}
\definecolor{LightSalmon4}{rgb}{0.55,0.34,0.26}
\definecolor{LightSalmon}{rgb}{1.00,0.63,0.48}
\definecolor{LightSeaGreen}{rgb}{0.13,0.70,0.67}
\definecolor{LightSkyBlue1}{rgb}{0.69,0.89,1.00}
\definecolor{LightSkyBlue2}{rgb}{0.64,0.83,0.93}
\definecolor{LightSkyBlue3}{rgb}{0.55,0.71,0.80}
\definecolor{LightSkyBlue4}{rgb}{0.38,0.48,0.55}
\definecolor{LightSkyBlue}{rgb}{0.53,0.81,0.98}
\definecolor{LightSlateBlue}{rgb}{0.52,0.44,1.00}
\definecolor{LightSlateGray}{rgb}{0.47,0.53,0.60}
\definecolor{LightSlateGrey}{rgb}{0.47,0.53,0.60}
\definecolor{LightSteelBlue1}{rgb}{0.79,0.88,1.00}
\definecolor{LightSteelBlue2}{rgb}{0.74,0.82,0.93}
\definecolor{LightSteelBlue3}{rgb}{0.64,0.71,0.80}
\definecolor{LightSteelBlue4}{rgb}{0.43,0.48,0.55}
\definecolor{LightSteelBlue}{rgb}{0.69,0.77,0.87}
\definecolor{LightYellow1}{rgb}{1.00,1.00,0.88}
\definecolor{LightYellow2}{rgb}{0.93,0.93,0.82}
\definecolor{LightYellow3}{rgb}{0.80,0.80,0.71}
\definecolor{LightYellow4}{rgb}{0.55,0.55,0.48}
\definecolor{LightYellow}{rgb}{1.00,1.00,0.88}
\definecolor{LimeGreen}{rgb}{0.20,0.80,0.20}
\definecolor{MediumAquamarine}{rgb}{0.40,0.80,0.67}
\definecolor{MediumBlue}{rgb}{0.00,0.00,0.80}
\definecolor{MediumOrchid1}{rgb}{0.88,0.40,1.00}
\definecolor{MediumOrchid2}{rgb}{0.82,0.37,0.93}
\definecolor{MediumOrchid3}{rgb}{0.71,0.32,0.80}
\definecolor{MediumOrchid4}{rgb}{0.48,0.22,0.55}
\definecolor{MediumOrchid}{rgb}{0.73,0.33,0.83}
\definecolor{MediumPurple1}{rgb}{0.67,0.51,1.00}
\definecolor{MediumPurple2}{rgb}{0.62,0.47,0.93}
\definecolor{MediumPurple3}{rgb}{0.54,0.41,0.80}
\definecolor{MediumPurple4}{rgb}{0.36,0.28,0.55}
\definecolor{MediumPurple}{rgb}{0.58,0.44,0.86}
\definecolor{MediumSeaGreen}{rgb}{0.24,0.70,0.44}
\definecolor{MediumSlateBlue}{rgb}{0.48,0.41,0.93}
\definecolor{MediumSpringGreen}{rgb}{0.00,0.98,0.60}
\definecolor{MediumTurquoise}{rgb}{0.28,0.82,0.80}
\definecolor{MediumVioletRed}{rgb}{0.78,0.08,0.52}
\definecolor{MidnightBlue}{rgb}{0.10,0.10,0.44}
\definecolor{MintCream}{rgb}{0.96,1.00,0.98}
\definecolor{MistyRose1}{rgb}{1.00,0.89,0.88}
\definecolor{MistyRose2}{rgb}{0.93,0.84,0.82}
\definecolor{MistyRose3}{rgb}{0.80,0.72,0.71}
\definecolor{MistyRose4}{rgb}{0.55,0.49,0.48}
\definecolor{MistyRose}{rgb}{1.00,0.89,0.88}
\definecolor{NavajoWhite1}{rgb}{1.00,0.87,0.68}
\definecolor{NavajoWhite2}{rgb}{0.93,0.81,0.63}
\definecolor{NavajoWhite3}{rgb}{0.80,0.70,0.55}
\definecolor{NavajoWhite4}{rgb}{0.55,0.47,0.37}
\definecolor{NavajoWhite}{rgb}{1.00,0.87,0.68}
\definecolor{NavyBlue}{rgb}{0.00,0.00,0.50}
\definecolor{OldLace}{rgb}{0.99,0.96,0.90}
\definecolor{OliveDrab1}{rgb}{0.75,1.00,0.24}
\definecolor{OliveDrab2}{rgb}{0.70,0.93,0.23}
\definecolor{OliveDrab3}{rgb}{0.60,0.80,0.20}
\definecolor{OliveDrab4}{rgb}{0.41,0.55,0.13}
\definecolor{OliveDrab}{rgb}{0.42,0.56,0.14}
\definecolor{OrangeRed1}{rgb}{1.00,0.27,0.00}
\definecolor{OrangeRed2}{rgb}{0.93,0.25,0.00}
\definecolor{OrangeRed3}{rgb}{0.80,0.22,0.00}
\definecolor{OrangeRed4}{rgb}{0.55,0.15,0.00}
\definecolor{OrangeRed}{rgb}{1.00,0.27,0.00}
\definecolor{PaleGoldenrod}{rgb}{0.93,0.91,0.67}
\definecolor{PaleGreen1}{rgb}{0.60,1.00,0.60}
\definecolor{PaleGreen2}{rgb}{0.56,0.93,0.56}
\definecolor{PaleGreen3}{rgb}{0.49,0.80,0.49}
\definecolor{PaleGreen4}{rgb}{0.33,0.55,0.33}
\definecolor{PaleGreen}{rgb}{0.60,0.98,0.60}
\definecolor{PaleTurquoise1}{rgb}{0.73,1.00,1.00}
\definecolor{PaleTurquoise2}{rgb}{0.68,0.93,0.93}
\definecolor{PaleTurquoise3}{rgb}{0.59,0.80,0.80}
\definecolor{PaleTurquoise4}{rgb}{0.40,0.55,0.55}
\definecolor{PaleTurquoise}{rgb}{0.69,0.93,0.93}
\definecolor{PaleVioletRed1}{rgb}{1.00,0.51,0.67}
\definecolor{PaleVioletRed2}{rgb}{0.93,0.47,0.62}
\definecolor{PaleVioletRed3}{rgb}{0.80,0.41,0.54}
\definecolor{PaleVioletRed4}{rgb}{0.55,0.28,0.36}
\definecolor{PaleVioletRed}{rgb}{0.86,0.44,0.58}
\definecolor{PapayaWhip}{rgb}{1.00,0.94,0.84}
\definecolor{PeachPuff1}{rgb}{1.00,0.85,0.73}
\definecolor{PeachPuff2}{rgb}{0.93,0.80,0.68}
\definecolor{PeachPuff3}{rgb}{0.80,0.69,0.58}
\definecolor{PeachPuff4}{rgb}{0.55,0.47,0.40}
\definecolor{PeachPuff}{rgb}{1.00,0.85,0.73}
\definecolor{PowderBlue}{rgb}{0.69,0.88,0.90}
\definecolor{RosyBrown1}{rgb}{1.00,0.76,0.76}
\definecolor{RosyBrown2}{rgb}{0.93,0.71,0.71}
\definecolor{RosyBrown3}{rgb}{0.80,0.61,0.61}
\definecolor{RosyBrown4}{rgb}{0.55,0.41,0.41}
\definecolor{RosyBrown}{rgb}{0.74,0.56,0.56}
\definecolor{RoyalBlue1}{rgb}{0.28,0.46,1.00}
\definecolor{RoyalBlue2}{rgb}{0.26,0.43,0.93}
\definecolor{RoyalBlue3}{rgb}{0.23,0.37,0.80}
\definecolor{RoyalBlue4}{rgb}{0.15,0.25,0.55}
\definecolor{RoyalBlue}{rgb}{0.25,0.41,0.88}
\definecolor{SaddleBrown}{rgb}{0.55,0.27,0.07}
\definecolor{SandyBrown}{rgb}{0.96,0.64,0.38}
\definecolor{SeaGreen1}{rgb}{0.33,1.00,0.62}
\definecolor{SeaGreen2}{rgb}{0.31,0.93,0.58}
\definecolor{SeaGreen3}{rgb}{0.26,0.80,0.50}
\definecolor{SeaGreen4}{rgb}{0.18,0.55,0.34}
\definecolor{SeaGreen}{rgb}{0.18,0.55,0.34}
\definecolor{SkyBlue1}{rgb}{0.53,0.81,1.00}
\definecolor{SkyBlue2}{rgb}{0.49,0.75,0.93}
\definecolor{SkyBlue3}{rgb}{0.42,0.65,0.80}
\definecolor{SkyBlue4}{rgb}{0.29,0.44,0.55}
\definecolor{SkyBlue}{rgb}{0.53,0.81,0.92}
\definecolor{SlateBlue1}{rgb}{0.51,0.44,1.00}
\definecolor{SlateBlue2}{rgb}{0.48,0.40,0.93}
\definecolor{SlateBlue3}{rgb}{0.41,0.35,0.80}
\definecolor{SlateBlue4}{rgb}{0.28,0.24,0.55}
\definecolor{SlateBlue}{rgb}{0.42,0.35,0.80}
\definecolor{SlateGray1}{rgb}{0.78,0.89,1.00}
\definecolor{SlateGray2}{rgb}{0.73,0.83,0.93}
\definecolor{SlateGray3}{rgb}{0.62,0.71,0.80}
\definecolor{SlateGray4}{rgb}{0.42,0.48,0.55}
\definecolor{SlateGray}{rgb}{0.44,0.50,0.56}
\definecolor{SlateGrey}{rgb}{0.44,0.50,0.56}
\definecolor{SpringGreen1}{rgb}{0.00,1.00,0.50}
\definecolor{SpringGreen2}{rgb}{0.00,0.93,0.46}
\definecolor{SpringGreen3}{rgb}{0.00,0.80,0.40}
\definecolor{SpringGreen4}{rgb}{0.00,0.55,0.27}
\definecolor{SpringGreen}{rgb}{0.00,1.00,0.50}
\definecolor{SteelBlue1}{rgb}{0.39,0.72,1.00}
\definecolor{SteelBlue2}{rgb}{0.36,0.67,0.93}
\definecolor{SteelBlue3}{rgb}{0.31,0.58,0.80}
\definecolor{SteelBlue4}{rgb}{0.21,0.39,0.55}
\definecolor{SteelBlue}{rgb}{0.27,0.51,0.71}
\definecolor{VioletRed1}{rgb}{1.00,0.24,0.59}
\definecolor{VioletRed2}{rgb}{0.93,0.23,0.55}
\definecolor{VioletRed3}{rgb}{0.80,0.20,0.47}
\definecolor{VioletRed4}{rgb}{0.55,0.13,0.32}
\definecolor{VioletRed}{rgb}{0.82,0.13,0.56}
\definecolor{WhiteSmoke}{rgb}{0.96,0.96,0.96}
\definecolor{YellowGreen}{rgb}{0.60,0.80,0.20}
\definecolor{aliceblue}{rgb}{0.94,0.97,1.00}
\definecolor{antiquewhite}{rgb}{0.98,0.92,0.84}
\definecolor{aquamarine1}{rgb}{0.50,1.00,0.83}
\definecolor{aquamarine2}{rgb}{0.46,0.93,0.78}
\definecolor{aquamarine3}{rgb}{0.40,0.80,0.67}
\definecolor{aquamarine4}{rgb}{0.27,0.55,0.45}
\definecolor{aquamarine}{rgb}{0.50,1.00,0.83}
\definecolor{azure1}{rgb}{0.94,1.00,1.00}
\definecolor{azure2}{rgb}{0.88,0.93,0.93}
\definecolor{azure3}{rgb}{0.76,0.80,0.80}
\definecolor{azure4}{rgb}{0.51,0.55,0.55}
\definecolor{azure}{rgb}{0.94,1.00,1.00}
\definecolor{beige}{rgb}{0.96,0.96,0.86}
\definecolor{bisque1}{rgb}{1.00,0.89,0.77}
\definecolor{bisque2}{rgb}{0.93,0.84,0.72}
\definecolor{bisque3}{rgb}{0.80,0.72,0.62}
\definecolor{bisque4}{rgb}{0.55,0.49,0.42}
\definecolor{bisque}{rgb}{1.00,0.89,0.77}
\definecolor{black}{rgb}{0.00,0.00,0.00}
\definecolor{blanchedalmond}{rgb}{1.00,0.92,0.80}
\definecolor{blue1}{rgb}{0.00,0.00,1.00}
\definecolor{blue2}{rgb}{0.00,0.00,0.93}
\definecolor{blue3}{rgb}{0.00,0.00,0.80}
\definecolor{blue4}{rgb}{0.00,0.00,0.55}
\definecolor{blueviolet}{rgb}{0.54,0.17,0.89}
\definecolor{blue}{rgb}{0.00,0.00,1.00}
\definecolor{brown1}{rgb}{1.00,0.25,0.25}
\definecolor{brown2}{rgb}{0.93,0.23,0.23}
\definecolor{brown3}{rgb}{0.80,0.20,0.20}
\definecolor{brown4}{rgb}{0.55,0.14,0.14}
\definecolor{brown}{rgb}{0.65,0.16,0.16}
\definecolor{burlywood1}{rgb}{1.00,0.83,0.61}
\definecolor{burlywood2}{rgb}{0.93,0.77,0.57}
\definecolor{burlywood3}{rgb}{0.80,0.67,0.49}
\definecolor{burlywood4}{rgb}{0.55,0.45,0.33}
\definecolor{burlywood}{rgb}{0.87,0.72,0.53}
\definecolor{cadetblue}{rgb}{0.37,0.62,0.63}
\definecolor{chartreuse1}{rgb}{0.50,1.00,0.00}
\definecolor{chartreuse2}{rgb}{0.46,0.93,0.00}
\definecolor{chartreuse3}{rgb}{0.40,0.80,0.00}
\definecolor{chartreuse4}{rgb}{0.27,0.55,0.00}
\definecolor{chartreuse}{rgb}{0.50,1.00,0.00}
\definecolor{chocolate1}{rgb}{1.00,0.50,0.14}
\definecolor{chocolate2}{rgb}{0.93,0.46,0.13}
\definecolor{chocolate3}{rgb}{0.80,0.40,0.11}
\definecolor{chocolate4}{rgb}{0.55,0.27,0.07}
\definecolor{chocolate}{rgb}{0.82,0.41,0.12}
\definecolor{coral1}{rgb}{1.00,0.45,0.34}
\definecolor{coral2}{rgb}{0.93,0.42,0.31}
\definecolor{coral3}{rgb}{0.80,0.36,0.27}
\definecolor{coral4}{rgb}{0.55,0.24,0.18}
\definecolor{coral}{rgb}{1.00,0.50,0.31}
\definecolor{cornflowerblue}{rgb}{0.39,0.58,0.93}
\definecolor{cornsilk1}{rgb}{1.00,0.97,0.86}
\definecolor{cornsilk2}{rgb}{0.93,0.91,0.80}
\definecolor{cornsilk3}{rgb}{0.80,0.78,0.69}
\definecolor{cornsilk4}{rgb}{0.55,0.53,0.47}
\definecolor{cornsilk}{rgb}{1.00,0.97,0.86}
\definecolor{cyan1}{rgb}{0.00,1.00,1.00}
\definecolor{cyan2}{rgb}{0.00,0.93,0.93}
\definecolor{cyan3}{rgb}{0.00,0.80,0.80}
\definecolor{cyan4}{rgb}{0.00,0.55,0.55}
\definecolor{cyan}{rgb}{0.00,1.00,1.00}
\definecolor{darkblue}{rgb}{0.00,0.00,0.55}
\definecolor{darkcyan}{rgb}{0.00,0.55,0.55}
\definecolor{darkgoldenrod}{rgb}{0.72,0.53,0.04}
\definecolor{darkgray}{rgb}{0.66,0.66,0.66}
\definecolor{darkgreen}{rgb}{0.00,0.39,0.00}
\definecolor{darkgrey}{rgb}{0.66,0.66,0.66}
\definecolor{darkkhaki}{rgb}{0.74,0.72,0.42}
\definecolor{darkmagenta}{rgb}{0.55,0.00,0.55}
\definecolor{darkolive}{rgb}{0.33,0.42,0.18}
\definecolor{darkorange}{rgb}{1.00,0.55,0.00}
\definecolor{darkorchid}{rgb}{0.60,0.20,0.80}
\definecolor{darkred}{rgb}{0.55,0.00,0.00}
\definecolor{darksalmon}{rgb}{0.91,0.59,0.48}
\definecolor{darksea}{rgb}{0.56,0.74,0.56}
\definecolor{darkslate}{rgb}{0.18,0.31,0.31}
\definecolor{darkslate}{rgb}{0.18,0.31,0.31}
\definecolor{darkslate}{rgb}{0.28,0.24,0.55}
\definecolor{darkturquoise}{rgb}{0.00,0.81,0.82}
\definecolor{darkviolet}{rgb}{0.58,0.00,0.83}
\definecolor{deeppink}{rgb}{1.00,0.08,0.58}
\definecolor{deepsky}{rgb}{0.00,0.75,1.00}
\definecolor{dimgray}{rgb}{0.41,0.41,0.41}
\definecolor{dimgrey}{rgb}{0.41,0.41,0.41}
\definecolor{dodgerblue}{rgb}{0.12,0.56,1.00}
\definecolor{firebrick1}{rgb}{1.00,0.19,0.19}
\definecolor{firebrick2}{rgb}{0.93,0.17,0.17}
\definecolor{firebrick3}{rgb}{0.80,0.15,0.15}
\definecolor{firebrick4}{rgb}{0.55,0.10,0.10}
\definecolor{firebrick}{rgb}{0.70,0.13,0.13}
\definecolor{floralwhite}{rgb}{1.00,0.98,0.94}
\definecolor{forestgreen}{rgb}{0.13,0.55,0.13}
\definecolor{gainsboro}{rgb}{0.86,0.86,0.86}
\definecolor{ghostwhite}{rgb}{0.97,0.97,1.00}
\definecolor{gold1}{rgb}{1.00,0.84,0.00}
\definecolor{gold2}{rgb}{0.93,0.79,0.00}
\definecolor{gold3}{rgb}{0.80,0.68,0.00}
\definecolor{gold4}{rgb}{0.55,0.46,0.00}
\definecolor{goldenrod1}{rgb}{1.00,0.76,0.15}
\definecolor{goldenrod2}{rgb}{0.93,0.71,0.13}
\definecolor{goldenrod3}{rgb}{0.80,0.61,0.11}
\definecolor{goldenrod4}{rgb}{0.55,0.41,0.08}
\definecolor{goldenrod}{rgb}{0.85,0.65,0.13}
\definecolor{gold}{rgb}{1.00,0.84,0.00}
\definecolor{gray0}{rgb}{0.00,0.00,0.00}
\definecolor{gray100}{rgb}{1.00,1.00,1.00}
\definecolor{gray10}{rgb}{0.10,0.10,0.10}
\definecolor{gray11}{rgb}{0.11,0.11,0.11}
\definecolor{gray12}{rgb}{0.12,0.12,0.12}
\definecolor{gray13}{rgb}{0.13,0.13,0.13}
\definecolor{gray14}{rgb}{0.14,0.14,0.14}
\definecolor{gray15}{rgb}{0.15,0.15,0.15}
\definecolor{gray16}{rgb}{0.16,0.16,0.16}
\definecolor{gray17}{rgb}{0.17,0.17,0.17}
\definecolor{gray18}{rgb}{0.18,0.18,0.18}
\definecolor{gray19}{rgb}{0.19,0.19,0.19}
\definecolor{gray1}{rgb}{0.01,0.01,0.01}
\definecolor{gray20}{rgb}{0.20,0.20,0.20}
\definecolor{gray21}{rgb}{0.21,0.21,0.21}
\definecolor{gray22}{rgb}{0.22,0.22,0.22}
\definecolor{gray23}{rgb}{0.23,0.23,0.23}
\definecolor{gray24}{rgb}{0.24,0.24,0.24}
\definecolor{gray25}{rgb}{0.25,0.25,0.25}
\definecolor{gray26}{rgb}{0.26,0.26,0.26}
\definecolor{gray27}{rgb}{0.27,0.27,0.27}
\definecolor{gray28}{rgb}{0.28,0.28,0.28}
\definecolor{gray29}{rgb}{0.29,0.29,0.29}
\definecolor{gray2}{rgb}{0.02,0.02,0.02}
\definecolor{gray30}{rgb}{0.30,0.30,0.30}
\definecolor{gray31}{rgb}{0.31,0.31,0.31}
\definecolor{gray32}{rgb}{0.32,0.32,0.32}
\definecolor{gray33}{rgb}{0.33,0.33,0.33}
\definecolor{gray34}{rgb}{0.34,0.34,0.34}
\definecolor{gray35}{rgb}{0.35,0.35,0.35}
\definecolor{gray36}{rgb}{0.36,0.36,0.36}
\definecolor{gray37}{rgb}{0.37,0.37,0.37}
\definecolor{gray38}{rgb}{0.38,0.38,0.38}
\definecolor{gray39}{rgb}{0.39,0.39,0.39}
\definecolor{gray3}{rgb}{0.03,0.03,0.03}
\definecolor{gray40}{rgb}{0.40,0.40,0.40}
\definecolor{gray41}{rgb}{0.41,0.41,0.41}
\definecolor{gray42}{rgb}{0.42,0.42,0.42}
\definecolor{gray43}{rgb}{0.43,0.43,0.43}
\definecolor{gray44}{rgb}{0.44,0.44,0.44}
\definecolor{gray45}{rgb}{0.45,0.45,0.45}
\definecolor{gray46}{rgb}{0.46,0.46,0.46}
\definecolor{gray47}{rgb}{0.47,0.47,0.47}
\definecolor{gray48}{rgb}{0.48,0.48,0.48}
\definecolor{gray49}{rgb}{0.49,0.49,0.49}
\definecolor{gray4}{rgb}{0.04,0.04,0.04}
\definecolor{gray50}{rgb}{0.50,0.50,0.50}
\definecolor{gray51}{rgb}{0.51,0.51,0.51}
\definecolor{gray52}{rgb}{0.52,0.52,0.52}
\definecolor{gray53}{rgb}{0.53,0.53,0.53}
\definecolor{gray54}{rgb}{0.54,0.54,0.54}
\definecolor{gray55}{rgb}{0.55,0.55,0.55}
\definecolor{gray56}{rgb}{0.56,0.56,0.56}
\definecolor{gray57}{rgb}{0.57,0.57,0.57}
\definecolor{gray58}{rgb}{0.58,0.58,0.58}
\definecolor{gray59}{rgb}{0.59,0.59,0.59}
\definecolor{gray5}{rgb}{0.05,0.05,0.05}
\definecolor{gray60}{rgb}{0.60,0.60,0.60}
\definecolor{gray61}{rgb}{0.61,0.61,0.61}
\definecolor{gray62}{rgb}{0.62,0.62,0.62}
\definecolor{gray63}{rgb}{0.63,0.63,0.63}
\definecolor{gray64}{rgb}{0.64,0.64,0.64}
\definecolor{gray65}{rgb}{0.65,0.65,0.65}
\definecolor{gray66}{rgb}{0.66,0.66,0.66}
\definecolor{gray67}{rgb}{0.67,0.67,0.67}
\definecolor{gray68}{rgb}{0.68,0.68,0.68}
\definecolor{gray69}{rgb}{0.69,0.69,0.69}
\definecolor{gray6}{rgb}{0.06,0.06,0.06}
\definecolor{gray70}{rgb}{0.70,0.70,0.70}
\definecolor{gray71}{rgb}{0.71,0.71,0.71}
\definecolor{gray72}{rgb}{0.72,0.72,0.72}
\definecolor{gray73}{rgb}{0.73,0.73,0.73}
\definecolor{gray74}{rgb}{0.74,0.74,0.74}
\definecolor{gray75}{rgb}{0.75,0.75,0.75}
\definecolor{gray76}{rgb}{0.76,0.76,0.76}
\definecolor{gray77}{rgb}{0.77,0.77,0.77}
\definecolor{gray78}{rgb}{0.78,0.78,0.78}
\definecolor{gray79}{rgb}{0.79,0.79,0.79}
\definecolor{gray7}{rgb}{0.07,0.07,0.07}
\definecolor{gray80}{rgb}{0.80,0.80,0.80}
\definecolor{gray81}{rgb}{0.81,0.81,0.81}
\definecolor{gray82}{rgb}{0.82,0.82,0.82}
\definecolor{gray83}{rgb}{0.83,0.83,0.83}
\definecolor{gray84}{rgb}{0.84,0.84,0.84}
\definecolor{gray85}{rgb}{0.85,0.85,0.85}
\definecolor{gray86}{rgb}{0.86,0.86,0.86}
\definecolor{gray87}{rgb}{0.87,0.87,0.87}
\definecolor{gray88}{rgb}{0.88,0.88,0.88}
\definecolor{gray89}{rgb}{0.89,0.89,0.89}
\definecolor{gray8}{rgb}{0.08,0.08,0.08}
\definecolor{gray90}{rgb}{0.90,0.90,0.90}
\definecolor{gray91}{rgb}{0.91,0.91,0.91}
\definecolor{gray92}{rgb}{0.92,0.92,0.92}
\definecolor{gray93}{rgb}{0.93,0.93,0.93}
\definecolor{gray94}{rgb}{0.94,0.94,0.94}
\definecolor{gray95}{rgb}{0.95,0.95,0.95}
\definecolor{gray96}{rgb}{0.96,0.96,0.96}
\definecolor{gray97}{rgb}{0.97,0.97,0.97}
\definecolor{gray98}{rgb}{0.98,0.98,0.98}
\definecolor{gray99}{rgb}{0.99,0.99,0.99}
\definecolor{gray9}{rgb}{0.09,0.09,0.09}
\definecolor{gray}{rgb}{0.75,0.75,0.75}
\definecolor{green1}{rgb}{0.00,1.00,0.00}
\definecolor{green2}{rgb}{0.00,0.93,0.00}
\definecolor{green3}{rgb}{0.00,0.80,0.00}
\definecolor{green4}{rgb}{0.00,0.55,0.00}
\definecolor{greenyellow}{rgb}{0.68,1.00,0.18}
\definecolor{green}{rgb}{0.00,1.00,0.00}
\definecolor{grey0}{rgb}{0.00,0.00,0.00}
\definecolor{grey100}{rgb}{1.00,1.00,1.00}
\definecolor{grey10}{rgb}{0.10,0.10,0.10}
\definecolor{grey11}{rgb}{0.11,0.11,0.11}
\definecolor{grey12}{rgb}{0.12,0.12,0.12}
\definecolor{grey13}{rgb}{0.13,0.13,0.13}
\definecolor{grey14}{rgb}{0.14,0.14,0.14}
\definecolor{grey15}{rgb}{0.15,0.15,0.15}
\definecolor{grey16}{rgb}{0.16,0.16,0.16}
\definecolor{grey17}{rgb}{0.17,0.17,0.17}
\definecolor{grey18}{rgb}{0.18,0.18,0.18}
\definecolor{grey19}{rgb}{0.19,0.19,0.19}
\definecolor{grey1}{rgb}{0.01,0.01,0.01}
\definecolor{grey20}{rgb}{0.20,0.20,0.20}
\definecolor{grey21}{rgb}{0.21,0.21,0.21}
\definecolor{grey22}{rgb}{0.22,0.22,0.22}
\definecolor{grey23}{rgb}{0.23,0.23,0.23}
\definecolor{grey24}{rgb}{0.24,0.24,0.24}
\definecolor{grey25}{rgb}{0.25,0.25,0.25}
\definecolor{grey26}{rgb}{0.26,0.26,0.26}
\definecolor{grey27}{rgb}{0.27,0.27,0.27}
\definecolor{grey28}{rgb}{0.28,0.28,0.28}
\definecolor{grey29}{rgb}{0.29,0.29,0.29}
\definecolor{grey2}{rgb}{0.02,0.02,0.02}
\definecolor{grey30}{rgb}{0.30,0.30,0.30}
\definecolor{grey31}{rgb}{0.31,0.31,0.31}
\definecolor{grey32}{rgb}{0.32,0.32,0.32}
\definecolor{grey33}{rgb}{0.33,0.33,0.33}
\definecolor{grey34}{rgb}{0.34,0.34,0.34}
\definecolor{grey35}{rgb}{0.35,0.35,0.35}
\definecolor{grey36}{rgb}{0.36,0.36,0.36}
\definecolor{grey37}{rgb}{0.37,0.37,0.37}
\definecolor{grey38}{rgb}{0.38,0.38,0.38}
\definecolor{grey39}{rgb}{0.39,0.39,0.39}
\definecolor{grey3}{rgb}{0.03,0.03,0.03}
\definecolor{grey40}{rgb}{0.40,0.40,0.40}
\definecolor{grey41}{rgb}{0.41,0.41,0.41}
\definecolor{grey42}{rgb}{0.42,0.42,0.42}
\definecolor{grey43}{rgb}{0.43,0.43,0.43}
\definecolor{grey44}{rgb}{0.44,0.44,0.44}
\definecolor{grey45}{rgb}{0.45,0.45,0.45}
\definecolor{grey46}{rgb}{0.46,0.46,0.46}
\definecolor{grey47}{rgb}{0.47,0.47,0.47}
\definecolor{grey48}{rgb}{0.48,0.48,0.48}
\definecolor{grey49}{rgb}{0.49,0.49,0.49}
\definecolor{grey4}{rgb}{0.04,0.04,0.04}
\definecolor{grey50}{rgb}{0.50,0.50,0.50}
\definecolor{grey51}{rgb}{0.51,0.51,0.51}
\definecolor{grey52}{rgb}{0.52,0.52,0.52}
\definecolor{grey53}{rgb}{0.53,0.53,0.53}
\definecolor{grey54}{rgb}{0.54,0.54,0.54}
\definecolor{grey55}{rgb}{0.55,0.55,0.55}
\definecolor{grey56}{rgb}{0.56,0.56,0.56}
\definecolor{grey57}{rgb}{0.57,0.57,0.57}
\definecolor{grey58}{rgb}{0.58,0.58,0.58}
\definecolor{grey59}{rgb}{0.59,0.59,0.59}
\definecolor{grey5}{rgb}{0.05,0.05,0.05}
\definecolor{grey60}{rgb}{0.60,0.60,0.60}
\definecolor{grey61}{rgb}{0.61,0.61,0.61}
\definecolor{grey62}{rgb}{0.62,0.62,0.62}
\definecolor{grey63}{rgb}{0.63,0.63,0.63}
\definecolor{grey64}{rgb}{0.64,0.64,0.64}
\definecolor{grey65}{rgb}{0.65,0.65,0.65}
\definecolor{grey66}{rgb}{0.66,0.66,0.66}
\definecolor{grey67}{rgb}{0.67,0.67,0.67}
\definecolor{grey68}{rgb}{0.68,0.68,0.68}
\definecolor{grey69}{rgb}{0.69,0.69,0.69}
\definecolor{grey6}{rgb}{0.06,0.06,0.06}
\definecolor{grey70}{rgb}{0.70,0.70,0.70}
\definecolor{grey71}{rgb}{0.71,0.71,0.71}
\definecolor{grey72}{rgb}{0.72,0.72,0.72}
\definecolor{grey73}{rgb}{0.73,0.73,0.73}
\definecolor{grey74}{rgb}{0.74,0.74,0.74}
\definecolor{grey75}{rgb}{0.75,0.75,0.75}
\definecolor{grey76}{rgb}{0.76,0.76,0.76}
\definecolor{grey77}{rgb}{0.77,0.77,0.77}
\definecolor{grey78}{rgb}{0.78,0.78,0.78}
\definecolor{grey79}{rgb}{0.79,0.79,0.79}
\definecolor{grey7}{rgb}{0.07,0.07,0.07}
\definecolor{grey80}{rgb}{0.80,0.80,0.80}
\definecolor{grey81}{rgb}{0.81,0.81,0.81}
\definecolor{grey82}{rgb}{0.82,0.82,0.82}
\definecolor{grey83}{rgb}{0.83,0.83,0.83}
\definecolor{grey84}{rgb}{0.84,0.84,0.84}
\definecolor{grey85}{rgb}{0.85,0.85,0.85}
\definecolor{grey86}{rgb}{0.86,0.86,0.86}
\definecolor{grey87}{rgb}{0.87,0.87,0.87}
\definecolor{grey88}{rgb}{0.88,0.88,0.88}
\definecolor{grey89}{rgb}{0.89,0.89,0.89}
\definecolor{grey8}{rgb}{0.08,0.08,0.08}
\definecolor{grey90}{rgb}{0.90,0.90,0.90}
\definecolor{grey91}{rgb}{0.91,0.91,0.91}
\definecolor{grey92}{rgb}{0.92,0.92,0.92}
\definecolor{grey93}{rgb}{0.93,0.93,0.93}
\definecolor{grey94}{rgb}{0.94,0.94,0.94}
\definecolor{grey95}{rgb}{0.95,0.95,0.95}
\definecolor{grey96}{rgb}{0.96,0.96,0.96}
\definecolor{grey97}{rgb}{0.97,0.97,0.97}
\definecolor{grey98}{rgb}{0.98,0.98,0.98}
\definecolor{grey99}{rgb}{0.99,0.99,0.99}
\definecolor{grey9}{rgb}{0.09,0.09,0.09}
\definecolor{grey}{rgb}{0.75,0.75,0.75}
\definecolor{honeydew1}{rgb}{0.94,1.00,0.94}
\definecolor{honeydew2}{rgb}{0.88,0.93,0.88}
\definecolor{honeydew3}{rgb}{0.76,0.80,0.76}
\definecolor{honeydew4}{rgb}{0.51,0.55,0.51}
\definecolor{honeydew}{rgb}{0.94,1.00,0.94}
\definecolor{hotpink}{rgb}{1.00,0.41,0.71}
\definecolor{indianred}{rgb}{0.80,0.36,0.36}
\definecolor{ivory1}{rgb}{1.00,1.00,0.94}
\definecolor{ivory2}{rgb}{0.93,0.93,0.88}
\definecolor{ivory3}{rgb}{0.80,0.80,0.76}
\definecolor{ivory4}{rgb}{0.55,0.55,0.51}
\definecolor{ivory}{rgb}{1.00,1.00,0.94}
\definecolor{khaki1}{rgb}{1.00,0.96,0.56}
\definecolor{khaki2}{rgb}{0.93,0.90,0.52}
\definecolor{khaki3}{rgb}{0.80,0.78,0.45}
\definecolor{khaki4}{rgb}{0.55,0.53,0.31}
\definecolor{khaki}{rgb}{0.94,0.90,0.55}
\definecolor{lavenderblush}{rgb}{1.00,0.94,0.96}
\definecolor{lavender}{rgb}{0.90,0.90,0.98}
\definecolor{lawngreen}{rgb}{0.49,0.99,0.00}
\definecolor{lemonchiffon}{rgb}{1.00,0.98,0.80}
\definecolor{lightblue}{rgb}{0.68,0.85,0.90}
\definecolor{lightcoral}{rgb}{0.94,0.50,0.50}
\definecolor{lightcyan}{rgb}{0.88,1.00,1.00}
\definecolor{lightgoldenrod}{rgb}{0.93,0.87,0.51}
\definecolor{lightgoldenrod}{rgb}{0.98,0.98,0.82}
\definecolor{lightgray}{rgb}{0.83,0.83,0.83}
\definecolor{lightgreen}{rgb}{0.56,0.93,0.56}
\definecolor{lightgrey}{rgb}{0.83,0.83,0.83}
\definecolor{lightpink}{rgb}{1.00,0.71,0.76}
\definecolor{lightsalmon}{rgb}{1.00,0.63,0.48}
\definecolor{lightsea}{rgb}{0.13,0.70,0.67}
\definecolor{lightsky}{rgb}{0.53,0.81,0.98}
\definecolor{lightslate}{rgb}{0.47,0.53,0.60}
\definecolor{lightslate}{rgb}{0.47,0.53,0.60}
\definecolor{lightslate}{rgb}{0.52,0.44,1.00}
\definecolor{lightsteel}{rgb}{0.69,0.77,0.87}
\definecolor{lightyellow}{rgb}{1.00,1.00,0.88}
\definecolor{limegreen}{rgb}{0.20,0.80,0.20}
\definecolor{linen}{rgb}{0.98,0.94,0.90}
\definecolor{magenta1}{rgb}{1.00,0.00,1.00}
\definecolor{magenta2}{rgb}{0.93,0.00,0.93}
\definecolor{magenta3}{rgb}{0.80,0.00,0.80}
\definecolor{magenta4}{rgb}{0.55,0.00,0.55}
\definecolor{magenta}{rgb}{1.00,0.00,1.00}
\definecolor{maroon1}{rgb}{1.00,0.20,0.70}
\definecolor{maroon2}{rgb}{0.93,0.19,0.65}
\definecolor{maroon3}{rgb}{0.80,0.16,0.56}
\definecolor{maroon4}{rgb}{0.55,0.11,0.38}
\definecolor{maroon}{rgb}{0.69,0.19,0.38}
\definecolor{mediumaquamarine}{rgb}{0.40,0.80,0.67}
\definecolor{mediumblue}{rgb}{0.00,0.00,0.80}
\definecolor{mediumorchid}{rgb}{0.73,0.33,0.83}
\definecolor{mediumpurple}{rgb}{0.58,0.44,0.86}
\definecolor{mediumsea}{rgb}{0.24,0.70,0.44}
\definecolor{mediumslate}{rgb}{0.48,0.41,0.93}
\definecolor{mediumspring}{rgb}{0.00,0.98,0.60}
\definecolor{mediumturquoise}{rgb}{0.28,0.82,0.80}
\definecolor{mediumviolet}{rgb}{0.78,0.08,0.52}
\definecolor{midnightblue}{rgb}{0.10,0.10,0.44}
\definecolor{mintcream}{rgb}{0.96,1.00,0.98}
\definecolor{mistyrose}{rgb}{1.00,0.89,0.88}
\definecolor{moccasin}{rgb}{1.00,0.89,0.71}
\definecolor{navajowhite}{rgb}{1.00,0.87,0.68}
\definecolor{navyblue}{rgb}{0.00,0.00,0.50}
\definecolor{navy}{rgb}{0.00,0.00,0.50}
\definecolor{oldlace}{rgb}{0.99,0.96,0.90}
\definecolor{olivedrab}{rgb}{0.42,0.56,0.14}
\definecolor{orange1}{rgb}{1.00,0.65,0.00}
\definecolor{orange2}{rgb}{0.93,0.60,0.00}
\definecolor{orange3}{rgb}{0.80,0.52,0.00}
\definecolor{orange4}{rgb}{0.55,0.35,0.00}
\definecolor{orangered}{rgb}{1.00,0.27,0.00}
\definecolor{orange}{rgb}{1.00,0.65,0.00}
\definecolor{orchid1}{rgb}{1.00,0.51,0.98}
\definecolor{orchid2}{rgb}{0.93,0.48,0.91}
\definecolor{orchid3}{rgb}{0.80,0.41,0.79}
\definecolor{orchid4}{rgb}{0.55,0.28,0.54}
\definecolor{orchid}{rgb}{0.85,0.44,0.84}
\definecolor{palegoldenrod}{rgb}{0.93,0.91,0.67}
\definecolor{palegreen}{rgb}{0.60,0.98,0.60}
\definecolor{paleturquoise}{rgb}{0.69,0.93,0.93}
\definecolor{paleviolet}{rgb}{0.86,0.44,0.58}
\definecolor{papayawhip}{rgb}{1.00,0.94,0.84}
\definecolor{peachpuff}{rgb}{1.00,0.85,0.73}
\definecolor{peru}{rgb}{0.80,0.52,0.25}
\definecolor{pink1}{rgb}{1.00,0.71,0.77}
\definecolor{pink2}{rgb}{0.93,0.66,0.72}
\definecolor{pink3}{rgb}{0.80,0.57,0.62}
\definecolor{pink4}{rgb}{0.55,0.39,0.42}
\definecolor{pink}{rgb}{1.00,0.75,0.80}
\definecolor{plum1}{rgb}{1.00,0.73,1.00}
\definecolor{plum2}{rgb}{0.93,0.68,0.93}
\definecolor{plum3}{rgb}{0.80,0.59,0.80}
\definecolor{plum4}{rgb}{0.55,0.40,0.55}
\definecolor{plum}{rgb}{0.87,0.63,0.87}
\definecolor{powderblue}{rgb}{0.69,0.88,0.90}
\definecolor{purple1}{rgb}{0.61,0.19,1.00}
\definecolor{purple2}{rgb}{0.57,0.17,0.93}
\definecolor{purple3}{rgb}{0.49,0.15,0.80}
\definecolor{purple4}{rgb}{0.33,0.10,0.55}
\definecolor{purple}{rgb}{0.63,0.13,0.94}
\definecolor{red1}{rgb}{1.00,0.00,0.00}
\definecolor{red2}{rgb}{0.93,0.00,0.00}
\definecolor{red3}{rgb}{0.80,0.00,0.00}
\definecolor{red4}{rgb}{0.55,0.00,0.00}
\definecolor{red}{rgb}{1.00,0.00,0.00}
\definecolor{rosybrown}{rgb}{0.74,0.56,0.56}
\definecolor{royalblue}{rgb}{0.25,0.41,0.88}
\definecolor{saddlebrown}{rgb}{0.55,0.27,0.07}
\definecolor{salmon1}{rgb}{1.00,0.55,0.41}
\definecolor{salmon2}{rgb}{0.93,0.51,0.38}
\definecolor{salmon3}{rgb}{0.80,0.44,0.33}
\definecolor{salmon4}{rgb}{0.55,0.30,0.22}
\definecolor{salmon}{rgb}{0.98,0.50,0.45}
\definecolor{sandybrown}{rgb}{0.96,0.64,0.38}
\definecolor{seagreen}{rgb}{0.18,0.55,0.34}
\definecolor{seashell1}{rgb}{1.00,0.96,0.93}
\definecolor{seashell2}{rgb}{0.93,0.90,0.87}
\definecolor{seashell3}{rgb}{0.80,0.77,0.75}
\definecolor{seashell4}{rgb}{0.55,0.53,0.51}
\definecolor{seashell}{rgb}{1.00,0.96,0.93}
\definecolor{sienna1}{rgb}{1.00,0.51,0.28}
\definecolor{sienna2}{rgb}{0.93,0.47,0.26}
\definecolor{sienna3}{rgb}{0.80,0.41,0.22}
\definecolor{sienna4}{rgb}{0.55,0.28,0.15}
\definecolor{sienna}{rgb}{0.63,0.32,0.18}
\definecolor{skyblue}{rgb}{0.53,0.81,0.92}
\definecolor{slateblue}{rgb}{0.42,0.35,0.80}
\definecolor{slategray}{rgb}{0.44,0.50,0.56}
\definecolor{slategrey}{rgb}{0.44,0.50,0.56}
\definecolor{snow1}{rgb}{1.00,0.98,0.98}
\definecolor{snow2}{rgb}{0.93,0.91,0.91}
\definecolor{snow3}{rgb}{0.80,0.79,0.79}
\definecolor{snow4}{rgb}{0.55,0.54,0.54}
\definecolor{snow}{rgb}{1.00,0.98,0.98}
\definecolor{springgreen}{rgb}{0.00,1.00,0.50}
\definecolor{steelblue}{rgb}{0.27,0.51,0.71}
\definecolor{tan1}{rgb}{1.00,0.65,0.31}
\definecolor{tan2}{rgb}{0.93,0.60,0.29}
\definecolor{tan3}{rgb}{0.80,0.52,0.25}
\definecolor{tan4}{rgb}{0.55,0.35,0.17}
\definecolor{tan}{rgb}{0.82,0.71,0.55}
\definecolor{thistle1}{rgb}{1.00,0.88,1.00}
\definecolor{thistle2}{rgb}{0.93,0.82,0.93}
\definecolor{thistle3}{rgb}{0.80,0.71,0.80}
\definecolor{thistle4}{rgb}{0.55,0.48,0.55}
\definecolor{thistle}{rgb}{0.85,0.75,0.85}
\definecolor{tomato1}{rgb}{1.00,0.39,0.28}
\definecolor{tomato2}{rgb}{0.93,0.36,0.26}
\definecolor{tomato3}{rgb}{0.80,0.31,0.22}
\definecolor{tomato4}{rgb}{0.55,0.21,0.15}
\definecolor{tomato}{rgb}{1.00,0.39,0.28}
\definecolor{turquoise1}{rgb}{0.00,0.96,1.00}
\definecolor{turquoise2}{rgb}{0.00,0.90,0.93}
\definecolor{turquoise3}{rgb}{0.00,0.77,0.80}
\definecolor{turquoise4}{rgb}{0.00,0.53,0.55}
\definecolor{turquoise}{rgb}{0.25,0.88,0.82}
\definecolor{violetred}{rgb}{0.82,0.13,0.56}
\definecolor{violet}{rgb}{0.93,0.51,0.93}
\definecolor{wheat1}{rgb}{1.00,0.91,0.73}
\definecolor{wheat2}{rgb}{0.93,0.85,0.68}
\definecolor{wheat3}{rgb}{0.80,0.73,0.59}
\definecolor{wheat4}{rgb}{0.55,0.49,0.40}
\definecolor{wheat}{rgb}{0.96,0.87,0.70}
\definecolor{whitesmoke}{rgb}{0.96,0.96,0.96}
\definecolor{white}{rgb}{1.00,1.00,1.00}
\definecolor{yellow1}{rgb}{1.00,1.00,0.00}
\definecolor{yellow2}{rgb}{0.93,0.93,0.00}
\definecolor{yellow3}{rgb}{0.80,0.80,0.00}
\definecolor{yellow4}{rgb}{0.55,0.55,0.00}
\definecolor{yellowgreen}{rgb}{0.60,0.80,0.20}
\definecolor{yellow}{rgb}{1.00,1.00,0.00}

\def\bea{\begin{eqnarray}}
\def\eea{\end{eqnarray}}
\def\vt{\vartheta}

\begin{document}

\newcommand{\rhat}{\hat{r}}
\newcommand{\iotahat}{\hat{\iota}}
\newcommand{\phihat}{\hat{\phi}}
\newcommand{\h}{\mathfrak{h}}
\newcommand{\be}{\begin{equation}}
\newcommand{\ee}{\end{equation}}
\newcommand{\ber}{\begin{eqnarray}}
\newcommand{\eer}{\end{eqnarray}}
\newcommand{\fmerg}{f_{\rm merg}}
\newcommand{\fcut}{f_{\rm cut}}
\newcommand{\fring}{f_{\rm ring}}
\newcommand{\cA}{\mathcal{A}}
\newcommand{\ie}{i.e.}
\newcommand{\df}{{\mathrm{d}f}}
\newcommand{\rmi}{\mathrm{i}}
\newcommand{\rmd}{\mathrm{d}}
\newcommand{\rme}{\mathrm{e}}
\newcommand{\dt}{{\mathrm{d}t}}
\newcommand{\pj}{\partial_j}
\newcommand{\pk}{\partial_k}
\newcommand{\psifl}{\Psi(f; {\bm \lambda})}
\newcommand{\hp}{h_+(t)}
\newcommand{\hc}{h_\times(t)}
\newcommand{\Fp}{F_+}
\newcommand{\Fc}{F_\times}
\newcommand{\Ylm}{Y_{\ell m}^{-2}}
\def\no{\nonumber \\ & \quad}
\def\noQ{\nonumber \\}
\newcommand{\mc}{M_c}
\newcommand{\vek}[1]{\boldsymbol{#1}}
\newcommand{\vdag}{(v)^\dagger}
\newcommand{\bvtheta}{{\bm \vartheta}}
\newcommand{\btheta}{{\bm \theta}}
\newcommand{\brho}{{\bm \rho}}
\newcommand{\pa}{\partial_a}
\newcommand{\pb}{\partial_b}
\newcommand{\Psieff}{\Psi_{\rm eff}}
\newcommand{\Aeff}{A_{\rm eff}}
\newcommand{\deff}{d_{\rm eff}}
\newcommand{\corr}{\mathcal{C}}
\newcommand{\bvthat}{\hat{\mbox{\boldmath $\vt$}}}
\newcommand{\bvt}{\mbox{\boldmath $\vt$}}

\newcommand{\comment}[1]{{\textsf{#1}}}
\newcommand{\ajith}[1]{\textcolor{magenta}{\textit{Ajith: #1}}}
\newcommand{\sukanta}[1]{\textcolor{blue}{\textit{Sukanta: #1}}}

\newcommand{\AEIHann}{Max-Planck-Institut f\"ur Gravitationsphysik 
(Albert-Einstein-Institut) and Leibniz Universit\"at Hannover, 
Callinstr.~38, 30167~Hannover, Germany}
\newcommand{\WSU}{Department of Physics \& Astronomy, Washington State University,
1245 Webster, Pullman, WA 99164-2814, U.S.A.}
\newcommand{\IUCAA}{Inter-University Centre for Astronomy and Astrophysics, Post Bag 4, Ganeshkhind, Pune 411 007, India \\
}
\newcommand{\LIGOCaltech}{LIGO Laboratory, California Institute of Technology, 
Pasadena, CA 91125, U.S.A.}
\newcommand{\TAPIR}{Theoretical Astrophysics, California Institute of Technology, 
Pasadena, CA 91125, U.S.A.}


\title{Toward finding gravitational-wave signals from progenitors of short hard gamma-ray bursts and orphaned afterglows}

\preprint{LIGO-P1200169}

\author{Shaon Ghosh}
\email{shaonghosh@mail.wsu.edu}
\affiliation{\WSU}

\author{Sukanta Bose}
\email{sukanta@wsu.edu}
\affiliation{\WSU}
\affiliation{\IUCAA}


\begin{abstract}


With multiple observatories and missions being planned for detecting orphaned afterglows associated with gamma-ray bursts (GRBs) we emphasize the importance of developing data analysis strategies for searching their possible counterpart signals in the data of gravitational wave (GW) detectors in the advanced detector era. 
This is especially attractive since short hard gamma-ray bursts (SGRBs) may have compact binary coalescences involving neutron stars (CBCNSs) as their progenitors, which emit gravitational waves.
Joint electromagnetic (EM) and GW observations of these objects will enrich our understanding of their beaming, energetics, galactic environment, and shed light on a host of other outstanding questions related to them. Here we recognize some of the astrophysical factors that determine what fraction of CBCNS sources can generate orphaned afterglows.
Pipelines already exist that target the sky-position and time of occurrence of SGRBs, known from EM observations, to search for their counterparts in GW detector data. 
Modifying them to analyse extended periods of time in the GW data in the past of the afterglow detection, while targeting a single sky-position, can search for GWs from the common progenitor. 
We assess the improvement in GW detectability to be had from utilizing the sky-position information. We also propose a method for improving the detection efficiency of targeted searches of GW signals from the putative CBCNS sources of afterglows and short gamma ray bursts in the presence of errors in detector calibration or CBCNS waveform models used in the search. 
The improvement arises from searching in a wider patch of the sky even when the sky-position is known accurately from EM observations and utilizes the covariance of the errors in waveform parameters with those in the sky position. 

\end{abstract}
\maketitle

\section{Introduction}
\label{sec:intro}


Short duration gamma-ray bursts (SGRBs) are less of an enigma now than when the extraterrestrial nature of the first gamma-ray bursts (GRBs) was established in the seventies \cite{Klebesadel}. Since then we have learned a lot about their characteristics and how they differ from their long duration counterparts, the long GRBs or LGRBs. This knowledge includes 
the nature of their host galaxies, including their redshift, star formation rate, age, metallicity, distances of separation from the host galaxy, and distinguishing features in their light-curves compared to those of LGRBs \cite{Nysewander:2008ks,Nakar:2007yr,Gehrels:2009qy,Metzger:2011bv}.
Some critical aspects are still unknown or need unequivocal observational confirmation. While there is some evidence that the progenitor of a SGRB might be the compact binary coalescence (CBC) of a black hole (BH) and a neutron star (NS) or two neutron stars (see, e.g., the review in Ref. \cite{Nakar:2007yr}), the observation of gravitational waves (GWs) from them can provide direct confirmation of that hypothesis \cite{Nissanke:2009kt,Metzger:2011bv,Harry:2010fr,Kelley:2012tc,Dietz:2012qp,Nissanke:2012dj,Bartos:2012vd}. Heretofore, such CBCs, which include at least one neutron star, will be termed as ``CBCNS'' sources. Once GW observations are occurring regularly in the advanced detector era (ADE), GW associations with SGRBs can provide additional astrophysical information about these objects.

It is well established that gamma-ray bursts are transient events of gamma-ray flashes occurring at cosmological distances. 
These events are different from soft gamma repeaters specifically in being extremely intense, of short duration, and in being non-repeating. On the basis of spectral hardness, these events broadly divide into two categories, those of a short duration spanning less than 2 sec and with a harder gamma-ray spectrum and those of a duration longer than 2 sec and with a softer gamma-ray spectrum. The first one is called the short duration gamma-ray burst and the second is called the long duration gamma-ray burst. Because of the dichotomy in the time scales of these events and the fact that they systematically fall in different regions in spectral hardness, it can be conjectured that different physics is involved in their occurrence. The most likely progenitor of the long duration gamma-ray burst is a core collapse supernova of a massive star \cite{MacFadyen:1998vz} and that of a short duration one is the 
merger
of the two objects in a compact binary system involving at least one neutron star as a component \cite{Eichler:1989ve}. The physical time scales of a collapsar and the coalescence of the remnants of neutron star and black holes are commensurate with the time scales of the long duration and short duration gamma-ray bursts, respectively. 
A gravitational wave (GW) discovery coincident with a short gamma-ray burst observation 
will provide the strongest evidence to date of the merger model. The time delay between the gravitational wave signal and the GRB will 
provide clues to
the burst mechanism and additional information in the form of GW polarization will help us determine the source geometry. Coincident GW-GRB discovery will also enable us to measure the source distance independently of the cosmological distance ladder and, therefore, provide a test for it. 

Gravitational wave observations can complement electromagnetic (EM) studies to better understand the SGRB sources and even resolve some anomalies. In fact it is known that some SGRB light-curves last for quite a bit more than 2 sec and some LGRB light-curves resemble those of SGRBs. This has led to a recent proposal of categorizing GRBs based not on their light-curves but on their progenitor type: In this classification scheme, Type I GRBs are associated with CBCNS and Type II with a collapsar \cite{Zhang:2006dh}.
It is to be seen if this scheme will stand the scrutiny of GW observations.
Additionally, multi-baseline network of GW detectors can resolve the inclination of those sources accurately enough to constrain the beaming angle of the GRBs \cite{Nissanke:2009kt}. 
This is a nice alternative to the method based on electromagnetic afterglows proposed by Rhoads in 1997 \cite{Rhoads:1997ps}. An accurate determination of the beaming angle will resolve how energetic SGRBs really are and, thereby, unravel if SGRBs have a narrow or wide range of energy output and what the possible reasons for it might be. 
Rhoads' method depends upon the presence of an afterglow that presents clear evidence for breaks in the power-law in the emission spectrum of the putative GRB jet \cite{Rhoads:1997ps}. 
It can work even when the gamma-ray emission from the GRB itself goes unobserved, such as when the afterglow is an ``orphan''. On the other hand, an afterglow may not reveal the sought spectral breaks because it may not extend over a wide enough frequency range. (See Refs. \cite{Dalal:2001ym} for additional problems with this method.)
Such afterglows, e.g., in X-ray, optical, or radio, may still provide spectroscopic redshift \cite{Yu:2011bq,Zafar:2011jg,Kruhler:2010jw}, 
and distance to the source.
Joint EM and GW observations can determine both the distance and the inclination angle of the CBCNS more accurately than either one of them. 

It is obvious that this type of joint observation will also aid in improving theoretical models of SGRB afterglows. If indeed it is established that they are associated with CBCNS, then a host of outstanding questions can be answered. For example, how much later can afterglows in different bands of the EM spectrum occur after the CBCNS merger? How isotropic are the afterglows? How varied can the afterglow energetics be? How strongly beamed SGRBs are can be resolved by joint GW-EM observations, which in turn can shed light on afterglow energetics, as noted above. Is the source model rich enough to explain that observed variety? What strong gravity physics can be probed with joint GW-EM observations? On the other hand, if no GW signals are observed from multiple SGRBs within a distance of about 500Mpc, then it would debunk the hypothesis that CBCNS are the progenitors of SGRBs.

While most studies in the past have focused on targeted searches of GW counterparts of SGRBs here we broaden the category of EM triggers by including orphaned afterglows of SGRBs. By a targeted GW search we mean a search for GWs from a part of the sky where an EM or a neutrino signal was detected from a source that may also emit GWs. The EM or neutrino signal is then termed as an external trigger whose sky-position, time, or other characteristics, deemed relevant, are used to define the GW search. In this paper, we highlight the astrophysical factors that are required to estimate how many GW signals will be detected from SGRBs and orphaned afterglows in the ADE.
The Laser Interferometer Gravitational-wave Observatory (LIGO) and the Virgo detector,
have demonstrated successfully that large scale interferometers can be used to detect GWs that change their arm-lengths at sub-nuclear scales. With the advent of the next generation gravitational wave interferometers in the next few to several years, with roughly ten times increased strain sensitivity, we increase 
the event rate thousandfold. These interferometers include the advanced LIGO (aLIGO) \cite{AdLigoUrl} and the advanced Virgo (AdV) detectors, the Japanese detector KAGRA \cite{Kuroda2010} and a LIGO detector in India.
Collaboration with the gamma-ray astronomy community has helped us understand and use GRBs to trigger searches in the data of GW detectors \cite{Abbott:2008zzb}. 
Prior information about the sky position and time of the EM event improves GW detection confidence and
significantly reduces the data analysis computational cost of the search. It also helps us reduce the threshold for detection of possible candidates and false alarm probability, as we discuss here. The sensitivity of such a targeted GW search
depends strongly on the accuracy of the GRB sky position.
A substantial fraction of GRBs detected by some observatories, e.g., Fermi \cite{Abdo:2009zza}, can have error radii in the sky that are several degrees wide.
The interplanetary network (IPN) \cite{IPN} is a group of satellites with onboard gamma-ray detectors that are used to locate gamma-ray bursts in the sky through triangulation. Given the number of satellites detecting a particular GRB and also the uncertainty in the relative location of the satellites and clock synchronization, the sky positions of a good fraction of GRBs detected by them can have errors of several degrees or worse. In this paper we study the effect of the error in the SGRB sky position on the detectability of GW from its progenitor. In such cases, one needs to search over a wider patch in the sky for GW signals. Note that when there is ambiguity about the short or long nature of a GRB from studying its light-curve, it makes sense to allow for the possibility of it being a SGRB and, therefore, search for a GW counterpart. 

On the other hand, when the sky-position of a GRB is accurately known, searching for a GW counterpart only at that location and close to the GRB time would seem to be the best strategy to detect GWs. This is true unless there exists a significant mismatch between the modeled GW waveform and the CBCNS signal or if the detector calibration is erroneous. It turns out that each of these sources of error can cause a non-negligible drop in the signal-to-noise ratio (SNR). In such a case of systematic error, searching in a larger patch of the sky than just a single sky position can reduce the extent of that drop and improve the detection confidence. More importantly, when the SNR is close to the detection threshold it can make all the difference between a detection and a non-detection.


The layout of the paper is as follows. In Sec. \ref{sec:orphanagprospects} we show how the detectability of a GW signal from an orphaned afterglow improves owing to the localization of the source in the sky. Unless noted otherwise, the working assumption in the rest of the paper is that CBCNS sources are the progenitors of SGRBs. In that vein, we show how the possible beaming of the SGRB influences the fraction of CBCNS sources that will be detected as orphaned afterglows of SGRBs. 
The framework used for this calculation is broadened to enquire what fraction of all CBCs in a given volume will be detectable as GW events. For that general case of CBC sources we reproduce the well known result that GW strain of a detected source, when averaged over its sky position and orientation, is less than the maximum strain from an optimally oriented and located source of the same kind, by a factor of 2.26.
In Sec. \ref{sec:poorlyLocSource}, we study the effect of poor sky localization of SGRBs on the detectability of GWs from them. 
These results emphasize the importance of performing GW searches over a sky-grid involving multiple sky-positions covering the error regions of the SGRBs.
In Sec. \ref{sec:poorsky}, 
we study CBC triggers from previously analyzed LIGO data to check if any of them was concurrent with a GRB. 
In Sec. \ref{sec:noskyerror}, we study other causes of systematic errors, e.g., detector calibration errors and the mismatch between a GW signal and the waveform model used to search for it. Here we find that searching in a wider patch of the sky for the GW counterpart of a SGRB or an orphaned afterglow, even when the sky-position of the latter is known accurately through EM observations, can improve the detectability of the signal owing to the possible covariance of the errors in the waveform or calibration with the error in the source sky-position. We end with a discussion in Sec. \ref{sec:discussion} of some unresolved issues related to GW-EM targeted searches that
should be explored in the near future.

A note on conventions and terminology used in this paper is in order. Unless otherwise specified, a detector in this paper means a GW detector and should be distinguished from non-GW detectors, such as gamma-ray detectors. Similarly, a GW search refers to a targeted GW search unless it is explicitly stated that the search in question is of an un-targeted type where the sky position or timing information of an EM or neutrino observation is not used to search in GW detector data. Finally, the LIGO detectors, with 4km arm-lengths, in Hanford and Livingston (US) are labelled H1 and L1, respectively, while the Virgo detector in Cascina (Italy) is denoted as V1. These three detectors form three single baseline networks, H1L1, L1V1, H1V1, and one multi-baseline network H1L1V1. A second LIGO detector in Hanford, which participated in the first five LIGO science runs, had 2km long arms and is labelled H2.


\section{SGRBs, orphaned afterglows, and their prospects as GW candidates}
\label{sec:orphanagprospects}

In this section, we show how the rate of CBC detections would improve if GW counterparts of orphaned afterglows are sought in data of ADE detectors. Currently, SGRBs are the only targeted searches for GWs from CBCs. As we conclude below, the improvement in rates can be modest enough that orphaned afterglows should be added to the list of targeted GW searches. 


The first afterglow of a SGRB was observed in 2005 \cite{Berger:2005rv}. Afterglow emissions of short GRBs are similar to that of long GRBS, but are less luminous. 
While it is not confirmed yet what the progenitor of a SGRB is, it is widely believed that it might be the merger of a NS with another NS or a stellar mass BH. Alternative sources, e.g., long-lived magnetars \cite{Bernardini:2011ki}, have been proposed too. It is not clear if in the alternative model the magnetar is in itself the product of a CBCNS progenitor. GW observations will provide direct confirmation of the SGRB model or help rule it out. If the SGRB model is correct, then the GRB is powered by an accretion disk that is formed after the tidal disruption of a NS by its other compact companion. Matter falling into the central spinning object from the accretion disk can form bipolar jets via the Blandford-Znajek mechanism. Numerical relativistic simulations suggest this as a likely scenario, especially, if the central object is a highly spinning black hole and the NS equation of state (EOS) is relatively stiff, which allow for large enough accretion disks \cite{Foucart:2012vn}. Strong magnetic fields are also believed to play an important role in powering GRBs. The gamma-ray burst is generated by the shock-accelerated electrons in the relativistic jet. This is the so-called fireball model (see, e.g., the review \cite{Piran:1999kx}). A GRB afterglow is produced when the jet interacts with medium surrounding the burst. In the process it can produce radiation over a wide range of frequencies, from X-Rays to radio, as the jet slows while ploughing through that medium.

A kilonova could be another EM manifestation of a CBCNS progenitor. 
The ejecta from NS-NS merger is neutron rich. This results in the formation of heavier elements due to r-process neutron captures. These heavy elements undergo nuclear fission and beta decays on time scales of the order of a day.
These are also interesting candidates for external trigger study. The data analysis pipeline that is constructed for detecting GW counterparts to orphaned afterglows in archived data can also be applied to kilonovae because in both cases the sky position will be known accurately enough to launch a search in GW archived data for the time of arrival of the signal.




There are existing missions Swift and Fermi that are expected to overlap with aLIGO observations from 2015-2018 and, perhaps, even beyond.
There are a handful of planned observatories that will target GRB afterglows and orphaned afterglows in the next several years with expected concurrent observations with aLIGO and, possibly, AdV detectors. The Australian Square Kilometer Array Pathfinder (ASKAP) is currently under construction at the Murchison Radio-astronomy Observatory in Western Australia and is expected to start early operations in 2013 \cite{Murphy:2012fw}. Its initial five-years operation period will overlap with aLIGO science runs.
Among proposed observatories, Lobster is a NASA mission that will have a wide-field X-ray imager, which will be more sensitive than Swift's BAT but will have a smaller 0.5sr field of view, and a narrow-field followup IR telescope and slewing apparatus. The French-Chinese Space-based multi-band astronomical Variable Objects Monitor (SVOM) and the broad spectral band Indian Astronomy Satellite ASTROSAT \cite{astrosat} are a couple of other  missions that will have capabilities of detecting GRB afterglows and will likely overlap with aLIGO observations. Furthermore, the South Korean-led Ultra-fast Flash Observatory
Pathfinder (UFFO-P) mission intends to catch the rise of GRBs \cite{Grossan:2012it}.


\begin{figure}[tbh]
\centering
\includegraphics[width=8.0cm]{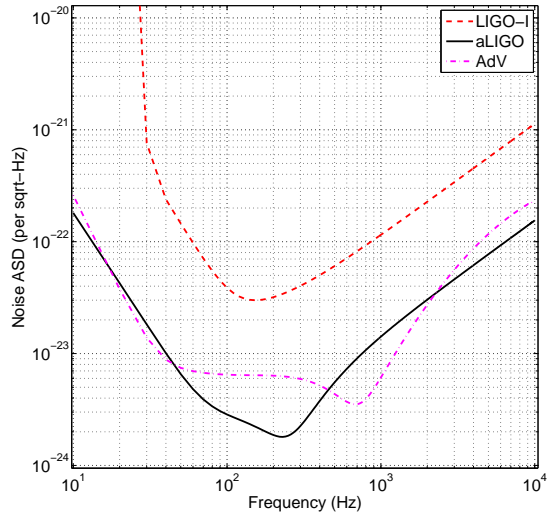}
\caption{The design noise amplitude spectral densities (ASDs) of LIGO-I, Advanced LIGO (aLIGO), and Advanced Virgo (AdV) detectors \cite{LAL}.}
\label{fig:sensCurves}
\end{figure}



\subsection{Comparing the detectabilities of GW counterparts of SGRBs and orphaned afterglows}
\label{ObsOrphAG}


We would like to estimate how much the detection probability improves when the sky-position of a GW source is accurately known but its time of occurrence has a window of uncertainty, $T_{\rm OAG}$. Following Ref. \cite{Dietz:2012qp}, let the desired FAP of a GW search be $10^{-4}$. Reference \cite{Dietz:2012qp} estimated that when the time of occurrence
is known accurately, the GW SNR threshold at that FAP is $\rho_{\rm th}^{\rm GRB}=9.0$, dropping from $\rho_{\rm th}^{\rm LM}=11.3$ at the same FAP for an all-sky, all-time (i.e., a blind) GW search of low-mass (LM) CBC signals in a period of $T_{\rm LM} = 3$ months, from sources with total mass of $\leq 25 M_\odot$ and each component mass between $1 M_\odot$ and $25 M_\odot$.
For an X-ray afterglow the putative CBCNS coalescence can occur an hour to a day in the past. 
For $T_{\rm OAG}=10^5$ sec, the FAP at a SNR threshold $\rho_{\rm th}$ obeys
\be
\label{FAPOAG}
\frac{{\rm FAP}_{\rm OAG}(\rho_{\rm th})}{{\rm FAP}_{\rm LM}(\rho_{\rm th})} = \frac{T_{\rm OAG}}{T_{\rm LM}} \approx \frac{10^5}{10^7} = 10^{-2} \,.
\ee
We would like to find $\rho_{\rm th} = \rho_{\rm th}^{\rm OAG}$ such that ${\rm FAP}_{\rm OAG} = 10^{-4}$. Equivalently, we ask what is $\rho_{\rm th}^{\rm OAG}$ for ${\rm FAR}_{\rm OAG} = 10^{-2}{\rm yr}^{-1}$? The answer, based on real data from the sixth science run (S6) of the LIGO H1 and L1 detectors, can be obtained from the data plotted in Fig. 3 of Ref. \cite{Colaboration:2011np}, namely, $\rho_{\rm th}^{\rm OAG} = 10.5$. This is analogous to the result obtained in Ref. \cite{Dietz:2012qp}, where it was shown that at a FAP of $10^{-4}$, the false-alarm rate (FAR) of a year-long low-mass CBC search at threshold $\rho_{\rm th}^{\rm LM}=11.3$ is ${\rm FAR}_{\rm LM} \left(\rho_{\rm th}^{\rm LM}=11.3 \right)= 10^{-4}{\rm yr}^{-1}$.
The threshold $\rho_{\rm th}^{\rm OAG}$ is about 17\% higher than the threshold for targeted GW searches of SGRBs with known sky position and time of occurrence but 8\% lower than than that of all-sky, all-time searches. 

There is an additional improvement that arises when one considers the fact that accurate information about the sky-position will further lower the FAR of a targeted GW search. 
With that information in hand, one need not search in the sky or the time-delays of the signals from the same source in different GW detectors. This reduces the FAR by a factor of a few to hundred depending on how big or small the duration of the noise artifacts are compared to the time-delays across the detector baselines. For instance, consider the single baseline H1L1, 
which has a light-travel time of $\pm 10$ msec. Let the FAP in each detector be $P_i$, where $i=1,2$ denotes the two detectors. Also, for every trigger in H1 let $N_2$ be the number of independent experiments of L1 coincidences one can perform in twice the light-travel time across the baseline. Then the joint FAP of H1L1 is 
\be\label{eq:skyfar}
P = P_1~[ 1 - ( 1 -P_2 )^{N_2} ] \,.
\ee
The filter response to noise artifacts in the data that masquerade as signals is similar to that on real signals, as expected. Studies of software injections of simulated signals in GW detector data show that the timescale of such a response is about 2 msec, which is roughly the inverse of the frequency where GW signals from binary neutron stars (BNSs) contribute maximally to the signal-to-noise ratio. Thus, $N_2$ is 10 for a 20 msec window. This means that if $P_1 \approx P_2$, then each one of them is about $3.2\times10^{-3}$ when $P=10^{-4}$. However, when the SGRB's sky position and time of occurrence are known and, therefore, the time-delays of its signals in the network detectors are known, the joint FAP is $10^{-5}$, which is, of course, 10 times smaller. With three baselines, a further reduction in FAP is possible. This will lower the thresholds discussed earlier somewhat. 
In this paper, we will assume that the FAP discussed in this section remains valid up to a total mass of about 43$M_{\odot}$, which is the maximum total mass used in targeted GW searches in LIGO-Virgo data. 
It is worth emphasizing that $P_{1,2}$ and, therefore, $P$ depend on the SNR. Results of Monte Carlo simulations discussed in Sec. \ref{sec:poorlyLocSource} bear out this property. 

However, it is worth noting that the improvement in detectability of a targeted search as estimated above is conservative. This is because it was deduced for a search that selected candidates after imposing thresholds on SNRs in individual detectors before requiring the threshold-crossing triggers to be coincident in mass and time. A more optimal external trigger search would instead analyze the data from multiple detectors coherently without imposing those thresholds \cite{Bose:2011km, Harry10a}, and can further improve the detection probability.

\subsection{The fraction of compact binary coalescences detectable as GW events}
\label{fracCBC}

Here we calculate the fraction of all compact binary coalescences occurring in the universe that are detectable as GW events in a single detector. For this calculation, we assume the CBCs to be distributed uniformly in the volume accessible to ADE detectors. Let $r_H$ be the horizon distance or the maximum distance to which an optimally oriented and optimally located source can be detected, for given component masses and spins. Let these parameters constitute the components of a vector $\bvtheta_{\rm in}$. Then the horizon distance $r_H(\bvtheta_{\rm in})$ depends on the values of these parameters.
Directions directly overhead or underneath of the plane containing the arms of an interferometric detector are optimal source locations. On the other hand, the optimal orientation of a CBC source is one where its inclination angle $\iota$, i.e., the angle between its orbital angular momentum vector and the negative of the line of sight vector, is zero.
If $\mathcal{P}(\theta, \phi)$ is the antenna power pattern of a single interferometer,
\bea
\mathcal{P}(\theta, \phi) = F_+(\theta, \phi, \psi)^2 + F_{\times}(\theta, \phi, \psi)^2 = \frac{1}{4}(1 + \cos^2\theta)^2\cos^22\phi + \cos^2\theta \sin^22\phi \,, \label{antpattern}
\eea
then $r_H\sqrt{\mathcal{P}(\theta, \phi)}$ is the greatest distance to which a CBC source is detectable in the $(\theta, \phi)$ direction, where $\mathcal{P}(\theta, \phi) \leq 1$.
For the rest of this section, we will take the binary component to be non-spinning
unless mentioned otherwise. At the greatest detectable distance the binary will have a face-on orientation, i.e., the orbital inclination angle $\iota$ will be zero or $\pi$ radians. 
As one decreases the distance, binaries with a wider variety of inclinations become detectable. So much so that for favorable directions the closest binaries will be detectable with any value of $\iota$, with the maximum possible value being $\pi/2$, namely, the edge-on orientation. By accounting for all allowed inclinations, 
at every $r \leq r_H\sqrt{\mathcal{P}(\theta, \phi)}$,
we find the fraction of CBCs that will be detectable as GW events.

The gravitational wave power received at a detector on earth depends on the sky position of the source $(\theta, \phi)$ and the orbital inclination angle 
$\iota$ as follows:
\bea
P_{\rm{rad}}(\iota, \theta, \phi) = \mathcal{P}(\theta, \phi)P_{\rm{rad}}(\iota = 0, \theta=0)\frac{(1 + 6\cos^2\iota + \cos^4\iota)}{8} \, ,
\eea
where $P_{\rm{rad}}(\iota = 0, \theta=0)$ is the GW power of a source that is optimally oriented and located in the sky.
The GW strain signal strength at a detector is measured in terms of the SNR \cite{Bose:2011km}. The SNR is inversely proportional to the source distance $r$. By the definition of the horizon distance $r_H$ an optimally oriented ($\iota = 0$) and located ($\theta = 0$ or $\theta = \pi$) source at that distance will be found with a SNR at the threshold of detection, $\rho_{\rm{th}}$.
Of course, this distance will vary for sources of different mass combinations. Thus we can write the SNR with above dependencies on distance and power
\bea\label{SNR}
\rho(r, \iota, \theta, \phi) = \rho_{\rm{th}}\frac{\sqrt{\mathcal{P}(\theta, \phi)}r_H}{r}  \sqrt{\frac{1}{8}(1 + 6\cos^2\iota + \cos^4\iota)} \, .
\eea
Whether a particular source of component masses $(m_1,m_2)$, at a distance 
$r \leq r_H$
will be detectable or not is determined by the inclination angle of the binary. 

\begin{figure}[tbh]
\centering
\includegraphics[width=12.7cm]{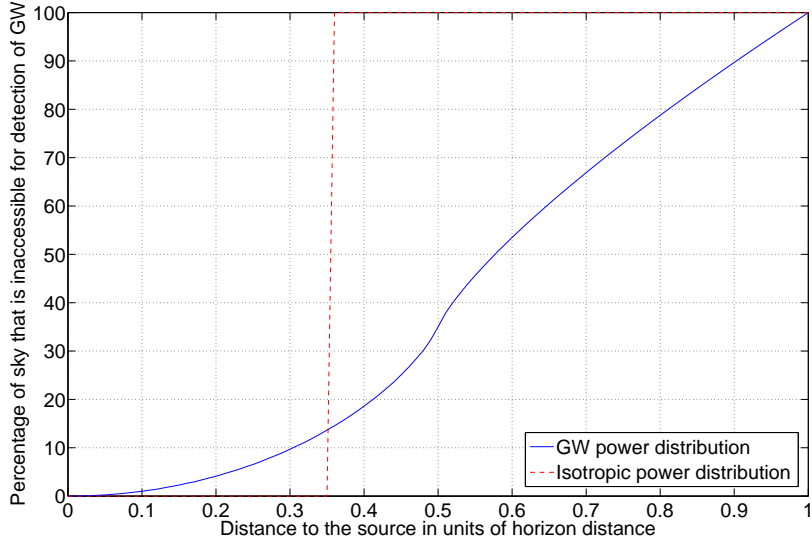}
\caption{The blue (solid) curve shows the percentage of sky inaccessible to interferometric detectors for the detection of GW signals from CBC sources (with all possible orientations), plotted as a function of the source distance. For comparison, if all sources were oriented face-on (i.e., with $\iota = 0$ or $\pi$) and if the detector had $\mathcal{P}(\theta, \phi) = 1$ everywhere on the sky, then all of those sources would be detectable in that hypothetical detector at all distances obeying $r \leq r_H$. On the other hand, if all sources were oriented edge-on (i.e., with $\iota = \pi/2$), then Eq. (\ref{SNR}) shows that, hypothetically,  for $\mathcal{P}(\theta, \phi) = 1$ everywhere on the sky, all of those sources would be inaccessible for $r > r_H/2\sqrt 2$ but accessible when closer. The red (dashed) step function represents those sources in that imaginary detector.
}
\label{fig:inaccessible}
\end{figure}

For every source at a particular distance and sky position there is a limiting inclination angle, $\iota_{\rm {max}}$, beyond which the signal from the source falls below the detection threshold. Thus, the probability that a source is detectable or not is the probability that the source has an inclination angle at most equal to $\iota_{\rm{max}}$. 
The probability distribution of the inclination angle $p(\iota) = \sin\iota$. 
Thus, the probability of detecting a source at a distance $r$ and sky position $(\theta, \phi)$ is given by,
\bea \label{probability_i_max}
P(0 \leq \iota \leq \iota_{\rm{max}}(r, \theta, \phi)) = \frac{ 2\int_0^{ \iota_{\rm{max}}(r, \theta, \phi) }{p(\iota)}d\iota }{ \int_0^{\pi} p(\iota)d\iota} = 1 - \cos \iota_{\rm{max}}(r, \theta, \phi) \,.
\eea
Since the SNR at this limiting inclination angle must be at the threshold of detection, one can use Eq. (\ref{SNR}) to find it as a function of $r, \theta$ and $\phi$:
\bea
\cos^2\iota_{\rm{max}} = -3 \pm \sqrt{8 + \frac{8r^2} {\mathcal{P}(\theta, \phi) r_H^2} }\,,
\eea
which 
has the following real solution:
\bea
  \cos \iota_{\rm{max}}(r, \theta, \phi)=\begin{cases}
    \left[\sqrt{8 + \frac{8r^2} {\mathcal{P}(\theta, \phi) r_H^2} } -3\right]^{1/2}, & \text{if $\frac{ 1 }{2\sqrt{2}} \leq \frac{r}{r_H \sqrt{\mathcal{P}(\theta, \phi)}} \leq 1$.}\\
    0, & \text{if $\frac{r}{r_H \sqrt{\mathcal{P}(\theta, \phi)}} < \frac{ 1 }{2\sqrt{2}}$.}
  \end{cases}
\eea
Above we used the fact that since $\cos^2 \iota_{\rm{max}}(r, \theta, \phi) $ must be in the interval $[0,1]$, 
the source distance must obey
\bea\frac{r}{r_H \sqrt{\mathcal{P}(\theta, \phi)}} \, \leq \, 1 \,.\label{bounds}\eea
The detection volume is defined by the bounding surface that satisfies the above inequality. The bound in Eq. (\ref{bounds}) is used in Fig. \ref{fig:inaccessible}, to show the percentage of the sky or $4\pi$ steradians that is inaccessible for detection as a function of the distance to the source.

\begin{figure}[tbh]
\centering
\includegraphics[width=12.7cm]{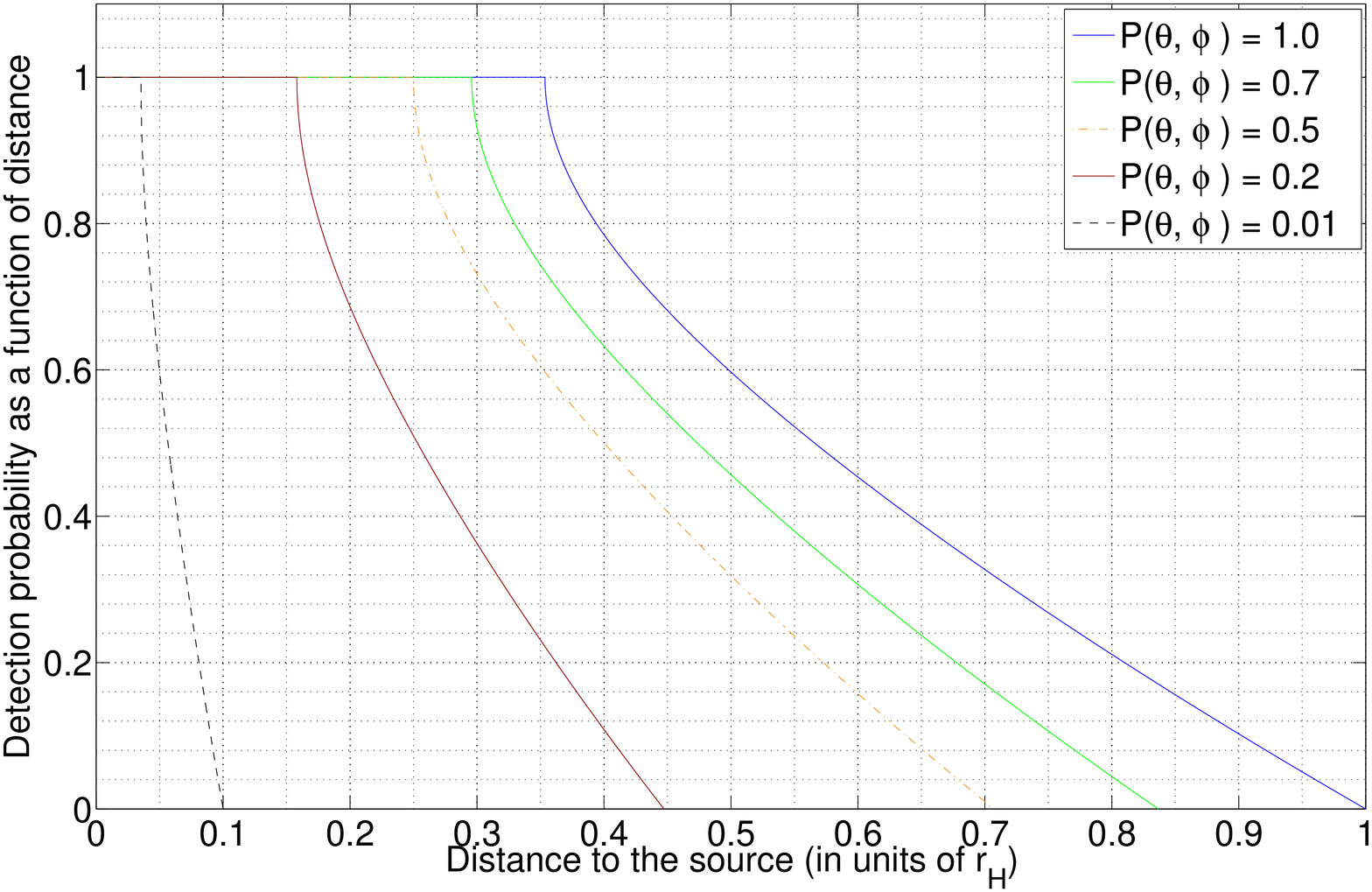}
\caption{
Probability of detecting a CBC source at a distance $r$ (given in units of the horizon distance $r_H$) assuming that the sources are distributed uniformly in the volume of interest. This plot is obtained from Eq. \ref{eq:probCBC}. The four different plots are 
for four different values of $\mathcal{P}(\theta, \phi)$. The blue curve shows the probability of a system that is overhead. For a given $\mathcal{P}(\theta, \phi)$, the largest distance at which the probability reaches unity is when a CBC is detectable for all possible values of its inclination angle. For an overhead source (i.e., with $\theta = 0$) this distance is $r = r_H/2\sqrt{2}$. For other sky positions, it is $r = r_H \sqrt{\mathcal{P}(\theta, \phi)}/2\sqrt{2}$.}
\label{fig:probIotaMax}
\end{figure}

The probability of detecting a CBC source is 
\bea
  P(0 \leq \iota \leq \iota_{\rm{max}}(r, \theta, \phi)) =\begin{cases}
    1 -  \left[\sqrt{8 + \frac{8r^2} {\mathcal{P}(\theta, \phi) r_H^2} } -3\right]^{1/2}, & \text{if $\frac{ 1 }{2\sqrt{2}} \leq \frac{r}{r_H \sqrt{\mathcal{P}(\theta, \phi)}} \leq 1$,}\\
    1, & \text{if $\frac{r}{r_H \sqrt{\mathcal{P}(\theta, \phi)}} < \frac{ 1 }{2\sqrt{2}}$,}
  \end{cases}
\label{eq:probCBC}
\eea
where Eq. (\ref{probability_i_max}) was used along with the constraint in Eq. (\ref{bounds}).
At $r = r_H$ the only value of $\mathcal{P}(\theta, \phi)$ possible is $1.0$ (see Eq. (\ref{bounds})). Thus, the probability of detecting a source at the horizon distance is vanishingly small. 
As one reduces the source distance the probability increases monotonically till it rearches unity at a distance of $r_H \sqrt{\mathcal{P}(\theta, \phi)} / (2\sqrt{2})$. This is shown in Fig. \ref{fig:probIotaMax} for five different lines of sight, namely, overhead (i.e, $\theta = 0$), corresponding to $\mathcal{P}(\theta, \phi) = 1.0$, and four other directions, corresponding to $\mathcal{P}(\theta, \phi) = 0.7, \, 0.5, \, 0.2$ and $0.01$, respectively.

\begin{figure}[tbh]
\centering
\includegraphics[width=7.8cm]{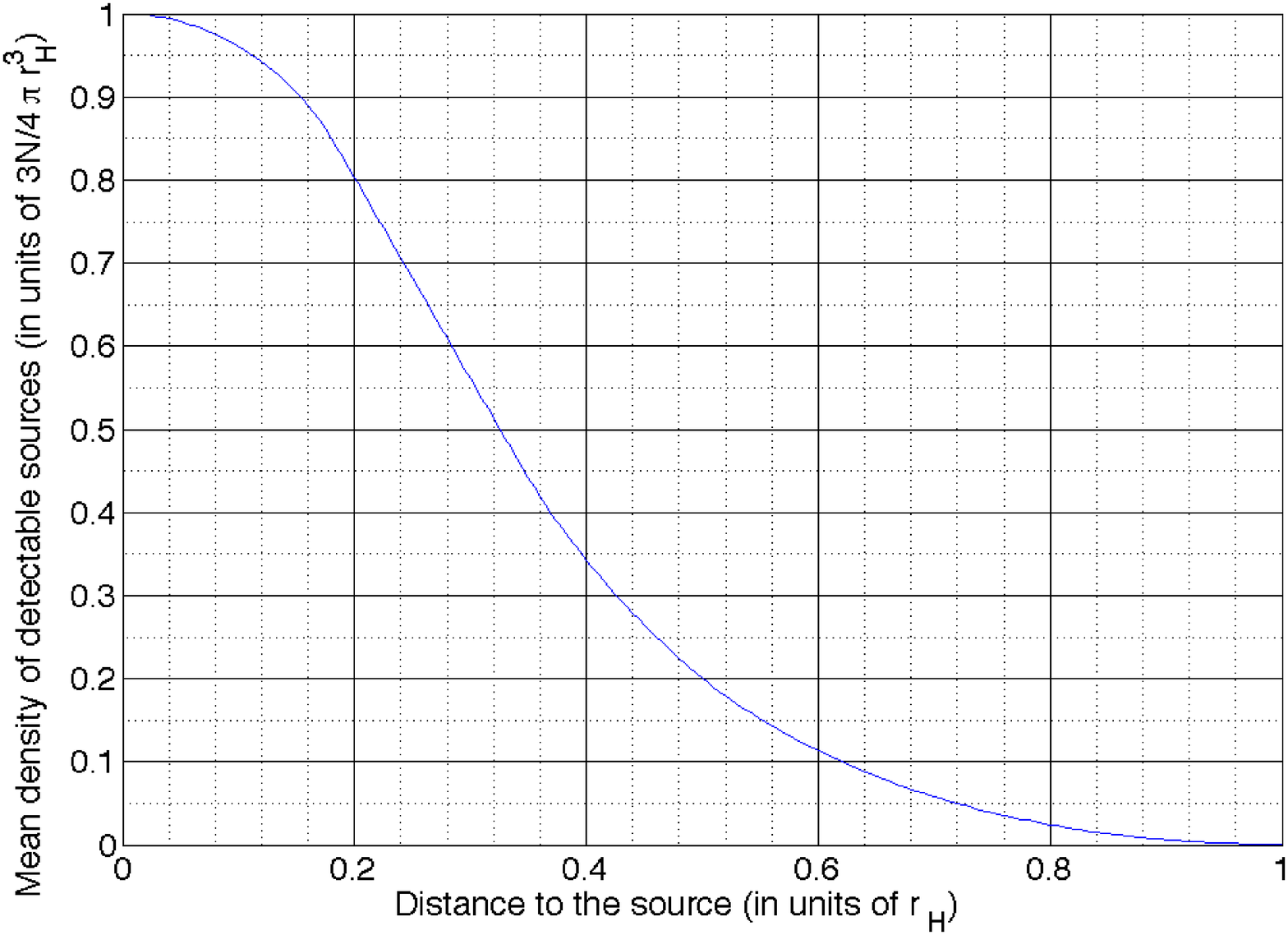}
\includegraphics[width=7.8cm]{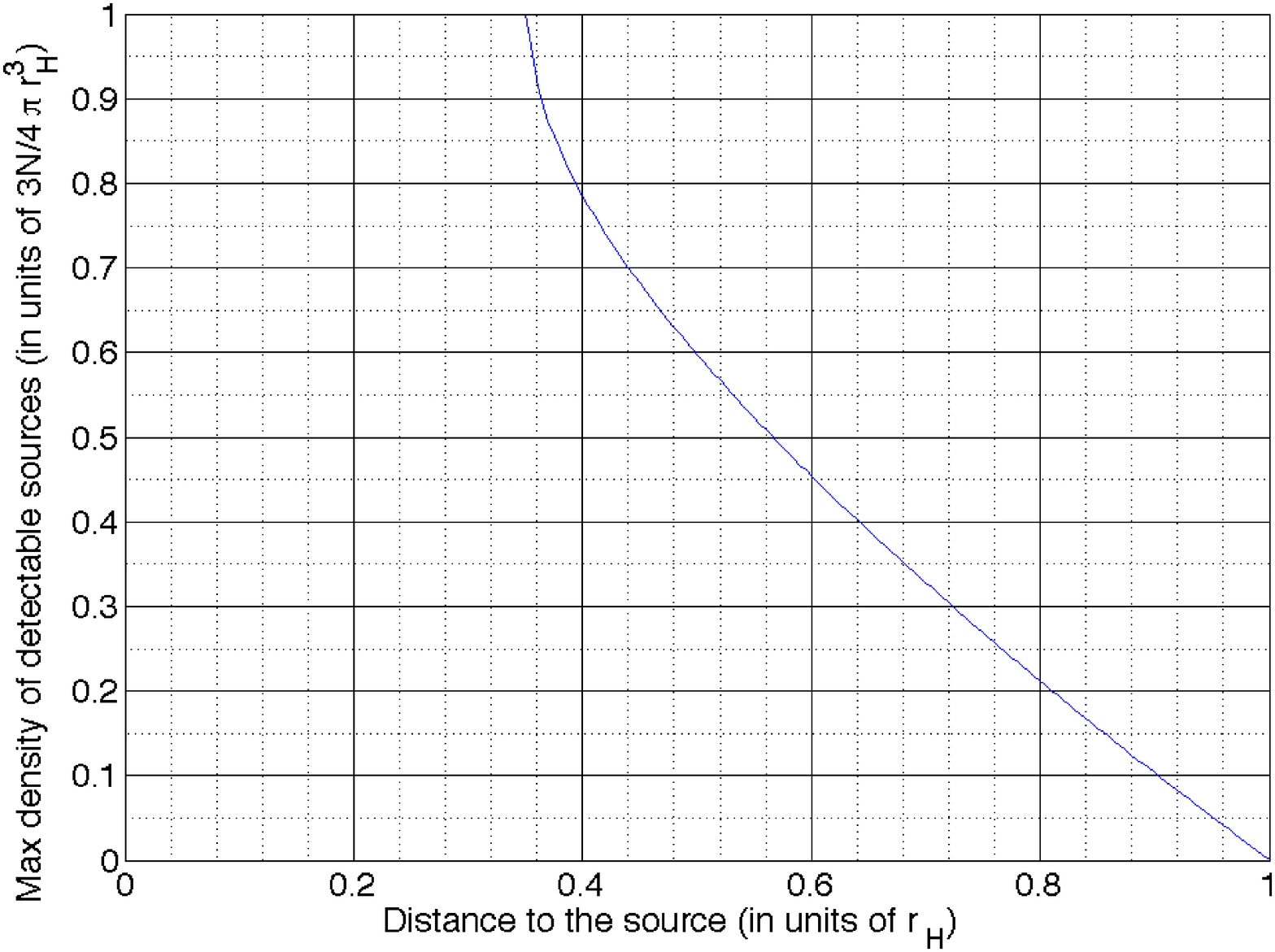}
\includegraphics[width=7.8cm]{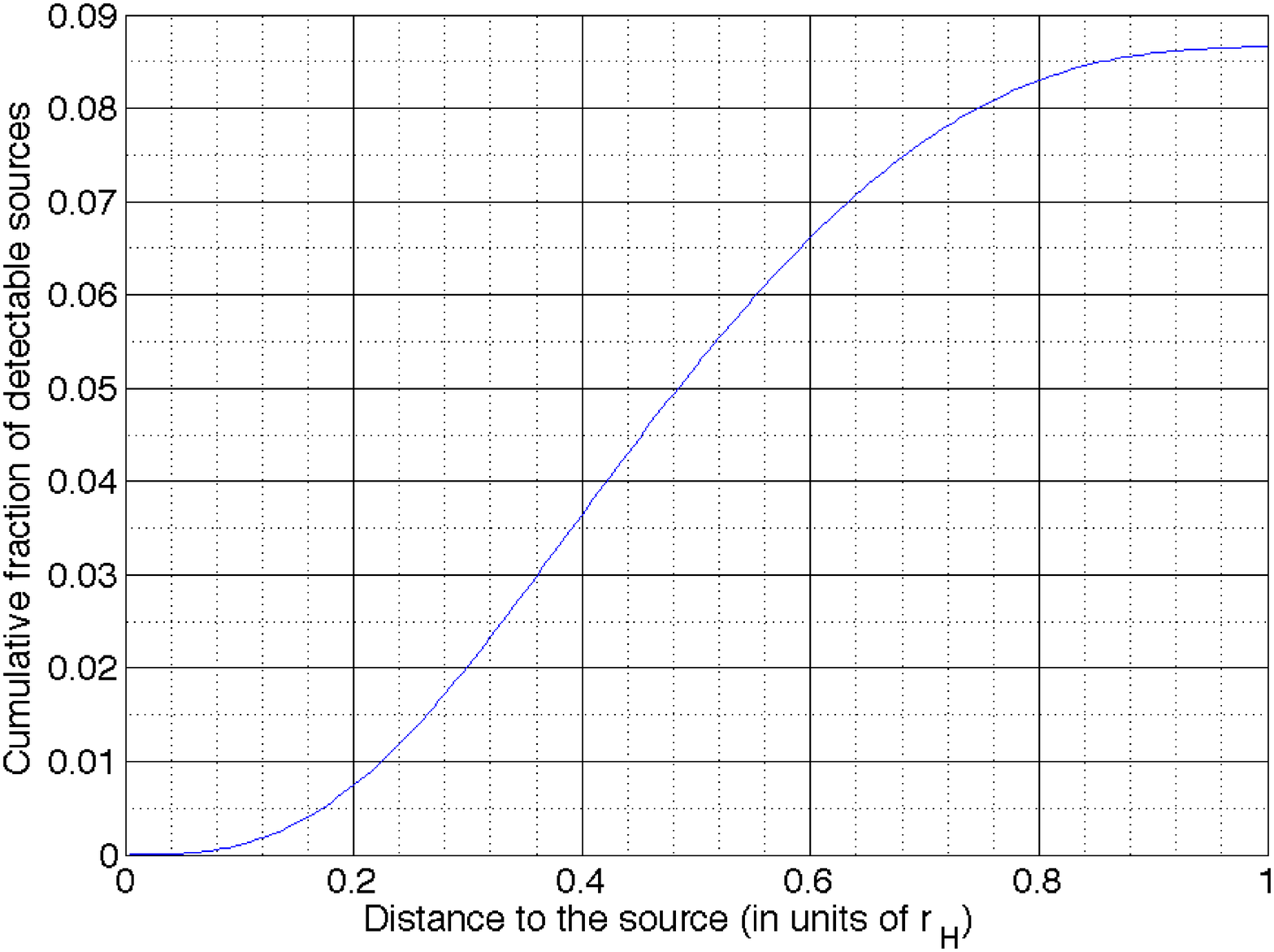}
\includegraphics[width=7.8cm]{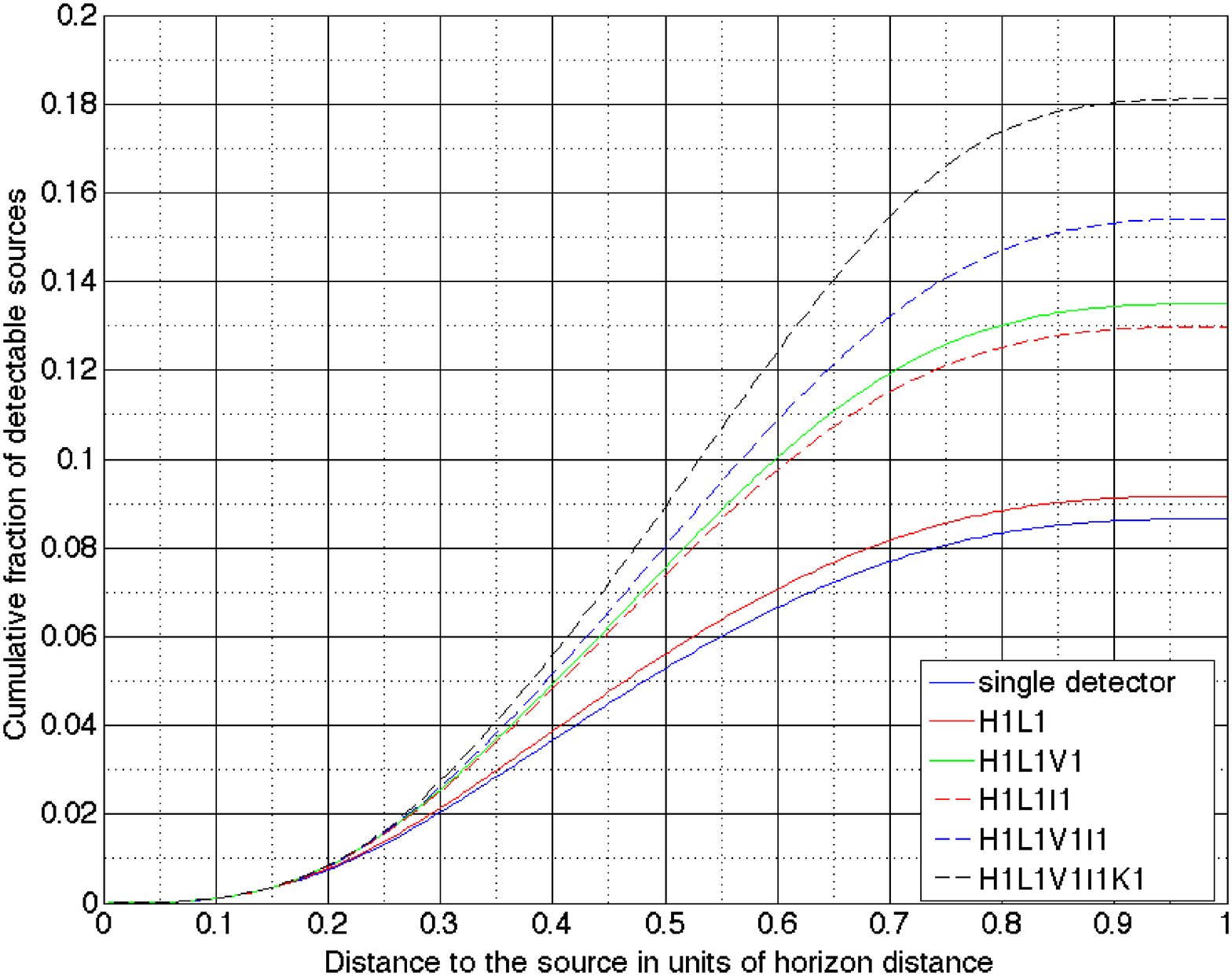}
\caption{ {\bf Top left:} Mean density of detectable sources on the two sphere at a distance $r$. This plot is obtained by solving the integral in Eq. (\ref{eq:meanDens}) numerically. {\bf Top right:} The maximum value that can be attained by the density of detectable sources on the two sphere at a distance $r$. Note that the similarity of this plot with the one in Fig. \ref{fig:probIotaMax} is due to the fact that the density of the detectable sources for a particular value of $\mathcal{P}(\theta, \phi)$ is proportional to the probability of detecting a CBC source at the direction defined by the same $\mathcal{P}(\theta, \phi)$, as can be seen from Eqs. (\ref{eq:density}) and (\ref{eq:probCBC}). {\bf Bottom left:} The cumulative fraction of detectable CBC sources in a spherical volume of radius $r$ plotted as a function of $r$. The sources are distributed uniformly in volume and are oriented randomly. It attains a maximum value of 8.65\% when $r$ reaches the horizon distance. This plot is obtained by numerically integrating Eq. (\ref{eq:CBCfraction}). {\bf Bottom right:} Comparison between the cumulative fraction of detectable CBC sources in a spherical volume of radius $r$ for a single detector and for various networks of detectors plotted as a function of $r$.
}
\label{f_gammaConstraintOne}
\end{figure}

Suppose all CBC sources are distributed uniformly throughout the universe. As discussed in Ref. \cite{Abadie:2010cf}, this is a good approximation to the true distribution for most of the volume that will be accessible to ADE detectors except for sources within 20Mpc.
If the total number of sources that exist in a spherical volume of radius $r_H$ is $N$, then the total number of sources detectable in a volume element $dV = r^2\sin\theta \,dr\, d\theta\, d\phi$ is
\bea\label{numdet}
dN_{\text{det}}(r, \theta, \phi) = \mathcal{D}(r, \theta, \phi) r^2\sin \theta\, dr\, d\theta\, d\phi\, ,
\eea
where
\bea
  \mathcal{D}(r, \theta, \phi) \equiv \begin{cases}
    \frac{3N}{4\pi r_H^3} \left[ 1 -  \left(\sqrt{8 + \frac{8r^2} {\mathcal{P}(\theta, \phi) r_H^2} } -3\right)^{1/2} \right], & \text{if $\frac{ r_H \sqrt{\mathcal{P}(\theta, \phi)} }{2\sqrt{2}} \leq r \leq r_H \sqrt{\mathcal{P}(\theta, \phi)}$},\\
    \frac{3N}{4\pi r_H^3}, & \text{if $r < \frac{ r_H \sqrt{\mathcal{P}(\theta, \phi)} }{2\sqrt{2}}$},
  \end{cases}
\label{eq:density}
\eea
is the density of sources detectable at a particular point $(r, \theta, \phi)$ in space. For nearby distances the density of detectable sources is higher than that at greater distances. Let us define the mean density of detectable sources at a distance $r$ as
\bea
\mathcal{D}_{\text{mean}}(r) = \frac{1}{4\pi} \int_0^{2\pi} d\phi \int_0^\pi d\theta\, \mathcal{D}(r, \theta, \phi) \sin \theta\,.
\label{eq:meanDens}
\eea
In the left plot of Fig. \ref{f_gammaConstraintOne} we show how $\mathcal{D}_{\text{mean}}(r)$ depends on the distance. Note that at small distances $\mathcal{D}_{\text{mean}}(r)$ approaches the density of all CBC sources. As we go away from the detector, we begin to lose sources that are sub-optimally oriented and located in the sky. Eventually $\mathcal{D}_{\text{mean}}(r)$  gets vanishingly small at the horizon distance where only those sources that are located overhead and have a face-on orientation are detectable.

The maximum density of detectable sources at a given distance is always overhead or underneath the detector. One can also see from the middle plot of Fig. \ref{f_gammaConstraintOne} that the maximum density of detectable sources up to the distance of $r_H/2\sqrt{2}$, is equal to the density of all CBC sources. Beyond $r = r_H/2\sqrt{2}$ the signal falls below threshold for sub-optimally oriented sources and hence the density of detectable sources decreases. Eventually, at $r = r_H$ the density of detectable sources overhead or underneath gets vanishingly small, with only the face-on sources being detectable. The fraction of CBC sources that are detectable in gravitational waves is
\bea \label{NS_fraction}
f_{\rm CBC} = \frac{1}{N} \int_{V_H} \mathcal{D}(r, \theta, \phi) \;dV\, ,
\label{eq:CBCfraction}
\eea
where ${V_H}$ is the volume of a sphere of radius $r_H$. Note that the effect of source redshift on this fraction is negligible for distances accessible to ADE detectors \cite{Abadie:2010cf}. 
Numerically integrating Eq. (\ref{NS_fraction}) we find $f_{\rm CBC} = 0.0865$, i.e, $8.65\%$ of CBC sources within the detection volume are detectable with an SNR louder than the threshold $\rho_{\rm{th}}$. In the right plot of Fig. \ref{f_gammaConstraintOne} we show how the ratio of the number of detectable sources up to a distance $r$ to the total number of CBC sources within the volume $V_H$ varies as a function of $r$.  Since this fraction scales as the cube of the average GW signal amplitude of a detectable source, the ratio of the signal amplitude of a CBC source at a distance $r<r_H$, when averaged over the location and orientation angles, is $(1/0.0865)^{1/3} = 2.26$ times smaller than that from the same source optimally located and oriented at the same distance. This is the same factor that is used to compute the rate of CBC sources \cite{Abadie:2010cf,Finn:1992xs}. We conducted similar studies for multiple detectors. We combined the antenna power pattern $\mathcal{P}(\theta, \phi)$ for the individual detectors numerically and solve for the integration in Eq. (\ref{NS_fraction}). We get $f_{\rm CBC}^{\rm HL} = 0.0917, f_{\rm CBC}^{\rm HLV} = 0.1350, f_{\rm CBC}^{\rm HLI} = 0.1298, f_{\rm CBC}^{\rm HLVI} = 0.1541$ and $f_{\rm CBC}^{\rm HLVIK} = 0.1811$, where `HLV' stands for the usual Hanford-Livingston-Virgo network. `I' stands for LIGO-India and `K' stands for KAGRA.

\subsection{Fraction of SGRBs with orphaned afterglows}
\label{probOrphAG}

In the remainder of this section we will assume that CBCNS sources are the only progenitors of SGRBs and their afterglows.
We first recognize the various factors that determine the number of SGRBs associated with CBCNS progenitors that are detectable by GW observations:
\bea
&&\text{Number of SGRBs associated with CBCNS GW detections} = \text{Number of CBCNS sources detectable in GWs by us} \noQ
&&~~~\times \text{fraction of CBCNS sources that emit gamma rays} \noQ
&&~~~\times \text{fraction of CBCNS gamma-ray emitters that are beamed toward us and detected by us as SGRBs}\,.
\eea
The above equation, factor for factor, can be symbolically expressed as
\be
N_{\gamma,\text{GW}} = N_{\text{GW}} \times P(\gamma|\text{CBCNS}) \times P(\text{SGRB}|\gamma,\text{CBCNS}) \,.
\ee
When the CBC population is limited to include only BNS and NSBH sources, the $N_{det}$ derived from Eq. (\ref{numdet}) reduces to $N_{\text{GW}}$.
Also, by a CBCNS gamma-ray emitter we mean a CBCNS source that would be detectable in gamma-rays by us provided the rays were beamed at us.
Let the gamma-ray emission of SGRBs be strongly beamed, with a beaming half-angle of $\beta_{\rm B}$. Then the fraction of CBCNS gamma-ray emitters that will be observable to us as SGRBs is
\bea
P(\text{SGRB}|\gamma,\text{CBCNS}) = \frac{ 2\pi \int_0^{\beta_{\rm B}}\sin \theta d\theta } {4\pi} = \sin^2\frac{\beta_{\rm B}}{2}\,.
\eea
If SGRBs were beamed isotropically, $\beta_{\rm B}=\pi$ and the above fraction is unity.
Therefore, the fraction of CBCNS gamma-ray emitters that will beam their gamma rays away from us is
\bea \label{orph_prob}
1 - \sin^2\frac{\beta_{\rm B}}{2}\, .
\eea
Let $f_{AG}$ be the fraction of CBCNS gamma-ray emitters that produce afterglows that are bright enough to be detectable by our observatories. Then the fraction of CBCNS gamma-ray emitters with an orphaned afterglow is
\bea \label{orph_prob_frac}
f_{OAG} \equiv f_{AG}\left(1 - \sin^2\frac{\beta_{\rm B}}{2}\right)\,.
\eea
The fraction $f_{AG}$ depends on the environment of SGRBs, which is not well understood \cite{Metzger:2011bv}. Note that some afterglows may be beamed (e.g., some X-ray afterglows), some others may be more isotropic (e.g., radio afterglows), and a fraction of them may be too weak to detect. 
On the other hand, the way $f_{OAG}$ is defined, it does not include CBCNS gamma-ray emitters that beam at us but we fail to detect them as SGRBs.
Moreover, a subset of orphaned afterglows may arise from CBCNS sources that do not produce a GRB, e.g., due to baryon loading \cite{Rhoads:2003ya}. Including these last two types of systems will only increase the number of orphaned afterglows associated with CBCNS sources.

\subsection{Probability of detecting an orphaned afterglow in conjunction with a gravitational wave signal} 
\label{fgammaCalc}


Let $P(\gamma|{\rm BNS})$ and $P(\gamma|{\rm NSBH})$ denote the fractions of binary neutron star and neutron star - black hole binary coalescences, respectively, that emit gamma rays.
Thus, the fraction of CBCNS sources that emit gamma rays can be written as 
\be\label{prob_cbcns}
P(\gamma|\text{CBCNS}) =  P(\gamma|{\rm NSBH}) P({\rm NSBH})+ P(\gamma|{\rm BNS}) P({\rm BNS}) \,,
\ee
where the values of $P({\rm NSBH})$ and $P({\rm BNS})$, the prior probabilities of a CBC source to be a NSBH source and a BNS source, respectively, are each assumed to be unity in the present discussion. Below, we first study the factors that influence $P(\gamma|{\rm NSBH})$ followed by those that affect $P(\gamma|{\rm BNS})$.

The central engine of a SGRB is the accretion disk that is created by the tidal disruption of an inspiraling neutron star. Some studies predict that an accretion disk of mass $m_{\rm{disk}} \gtrsim 0.01 M_{\odot}$ can provide sufficient energy to launch a jet for a duration of around $100$ msec by neutrino radiation \cite{PhysRevLett.104.141101}.  Thus, whether a system emits a gamma-ray burst or not will depend foremost on whether the neutron star can transfer $0.01 M_{\odot}$ in to the accretion disk. A massive NS with low compactness is more likely to transfer the required mass to an accretion disk massive enough to launch a jet than a low-mass neutron star with high compactness.
Let us assume that the smallest mass of an accretion disk that is required to fire the GRB engine is $m_*$. If we further assume that the neutron star mass is distributed normally from a lower limit $m_{\rm{min}}$ to an upper limit $m_{\rm{max}}$, and that the mean neutron star mass is $\bar{m}$, then we can write the probability density of neutron star mass as, \footnote{The formalism developed here can be applied to a different set of parameter values, as and when it is refined with new observations.}
\bea 
\label{nsMassDistribution}
  p_{\text{\tiny NS}}(m) =\begin{cases}
    \frac{1}{I}e^{-(m-\bar{m})^2/2\sigma^2}, & \text{if $m_{\rm{min}} \leq m\leq m_{\rm{max}}$},\\
    0, & \text{otherwise},
  \end{cases}
\eea
where, 
\bea
I = \int_{m_{\rm{min}}}^{m_{\rm{max}}} e^{-(m-\bar{m})^2/2\sigma^2} dm = \sigma \sqrt{ \frac{\pi}{2} } \left[ {\rm erfc} \left( \frac{m_{\rm{min}} - \bar{m}}{\sigma\sqrt{2}} \right ) - {\rm erfc} \left( \frac{m_{\rm{max}} - \bar{m}}{\sigma\sqrt{2}} \right ) \right]\, .
\eea
We follow Ref. \cite{minMassNS} in setting $m_{\rm{min}}=0.88M_{\odot}$. The values of all other parameters that define the above distribution are obtained from the emipiral study given in Ref. \cite{NSdistribution}, namely,
$\bar{m}=1.28M_{\odot}$, $m_{\rm{max}}=3.2M_{\odot}$, and the standard deviation $\sigma = 0.24M_{\odot}$.
The normalization factor is then deduced to be $I=0.5729M_{\odot}$.

Low-mass neutron stars are less likely to form a massive enough accretion disk that can launch an ultrarelativistic jet. If $m_*$ is the lower limit on the mass of a neutron stars forming SGRBs, then the fraction of all neutron stars with $m>m_*$ is:
\bea \label{massThreshold}
P(m_*) = \int_{m_*}^\infty p_{\text{\tiny NS}}(m)dm = \frac{1}{I} \int_{m_*}^{m_{\rm{max}}} e^{-(m-\bar{m})^2/2\sigma^2} dm\,.
\eea
If one assumes that the neutron stars in binaries are a good representation of all neutron stars from the same volume of the universe, then $P(m_*)$ is also the fraction of neutron stars in any CBCNS system in that volume that can form SGRBs. Below, we argue that $m_*$ must be related to other CBCNS parameters that are relevant to the formation of a SGRB. These are the compressibility $\kappa$ of the neutron star, the spin parameter $\chi$ of the companion black hole and the symmetrized mass-ratio $\eta = m_1m_2 / (m_1 + m_2)^2$ of the binary, with component masses $m_1$ and $m_2$. 

Let $P({\rm NS}|m_*,\kappa_{\rm min},\kappa_{\rm max})$ be the fraction of neutron stars that have mass greater than $m_*$ and compressibility in the range $(\kappa_{\rm min},\kappa_{\rm max})$.
Let $p(m,\kappa)$ be the joint probability density of neutron stars in $m$ and $\kappa$. Also, let the probability densities of stellar mass black holes in their spin parameter $\chi$ be $p_{\chi}(\chi)$ and 
of NSBH binary systems in their symmetrized mass-ratio $\eta$ be $p_{\eta} (\eta)$. Next let us assume that NSBH binaries are constituted of components drawn randomly from neutron star and stellar mass black hole populations. \footnote{This assumption may not be completely valid and, therefore, the resulting simplified probability expression is only approximate.} 
If so, the fraction of NSBH systems that can form an accretion disk capable of generating a GRB is
\bea
P({\rm NSBH}|m_*,\kappa_{\rm min},\kappa_{\rm max}) &=& P({\rm NS}|m_*,\kappa_{\rm min},\kappa_{\rm max}) \int_{0}^{0.25} d\eta \, p_{\eta} (\eta) \, \int_{0}^{1} d\chi \,p_{\chi} (\chi) \noQ
&=& \int_{0}^{0.25} d\eta \, p_{\eta} (\eta) \, \int_{0}^{1} d\chi \,p_{\chi} (\chi) \int_{\kappa_{\rm min}}^{\kappa_{\rm max}} d\kappa \int_{m_*}^{m_{\rm max}} dm\, p(\kappa,m) \,.
\eea
If there is no correlation between neutron star masses and compressibility, and
$p(m,\kappa) \equiv p_\kappa(\kappa)p_{\text{\tiny NS}}(m)$, then
\be\label{fracgammansbh}
P({\rm NSBH}|m_*,\kappa_{\rm min},\kappa_{\rm max}) = \int_{0}^{0.25} d\eta \, p_{\eta} (\eta) \, \int_{0}^{1} d\chi \,p_{\chi} (\chi) \int_{\kappa_{\rm min}}^{\kappa_{\rm max}} d\kappa \, p_\kappa(\kappa)\int_{m_*}^{m_{\rm max}} dm\, p_{\text{\tiny NS}}(m) \,.
\ee
A more general formula, which allows $m_*$ to vary with $\eta$, $\chi$ and $\kappa$ of a CBCNS system that is capable of emitting gamma rays, is:
\be\label{fracgammansbhcoupled}
P({\rm NSBH}|m_*,\kappa_{\rm min},\kappa_{\rm max}) = \int_{0}^{0.25} d\eta \, p_{\eta} (\eta) \, \int_{0}^{1} d\chi \,p_{\chi} (\chi) \int_{\kappa_{\rm min}}^{\kappa_{\rm max}} d\kappa \, p_\kappa(\kappa)\int_{m_*(\kappa, \chi, \eta)}^{m_{\rm max}} dm\, p_{\text{\tiny NS}}(m) \,,
\ee
where the integrals are now coupled due to the dependence of $m_*$ on $\eta$, $\chi$ and $\kappa$ in the lower limit of the NS mass integral. As argued below, we speculate that the above probability is more applicable in nature than the one given in Eq. (\ref{fracgammansbh}). 


The likelihood of forming a large accretion disk depends on the equation of state (EOS) of the neutron star, in addition to its mass. For a given mass, a neutron star with a stiffer EOS is more likely to give a large accretion disk: A neutron star that is more compact is less likely to get ripped apart by the tidal force of its binary companion before it reaches the last stable orbit (LSO). If the companion is a black hole, then such a neutron star will cross the LSO and plunge into the companion without any GRB arising from such a system. However, if the black hole has a large spin component along the orbital angular momentum of the binary, then the LSO is smaller. In the test particle limit, the LSO around a non-spinning black hole of mass $M_{\rm BH}$ is at a distance of $R=6GM_{\rm BH}/c^2$, whereas around a maximally spinning black hole of the same mass it is at the horizon, $R=GM_{\rm BH}/c^2$, for prograde orbits. For binary systems of interest, the location of such an orbit is less sharply defined. In lieu of it one uses the distance at which a slowly inspiraling system makes its transition into a rapid plunge. In any case, a neutron star is more likely to inspiral closer to the companion black hole if the latter has a large spin component along the direction of the orbital angular momentum. This makes it more likely for a neutron star to get tidally deformed and ripped apart into an accretion disk and, consequently, trigger the central GRB engine through accretion of neutron star matter into the companion. Furthermore, a smaller black hole has a larger tidal radius and is more likely to shred the neutron star before it reaches the LSO than a larger black hole. This implies that the GRB is more likely to be triggered from a binary source with a low mass-ratio. Thus, the threshold mass $m_*$ of the neutron star that is necessary, but not sufficient, to form a GRB triggering accretion disk is a function of the compactness of the neutron star, the companion spin and the mass-ratio of the binary. Hence, Eq. (\ref{massThreshold}) can be reexpressed as
\bea \label{probMassFunctional}
P\left(m_*(\kappa, \chi, \eta)\right) =  \frac{1}{I} \int_{m_*(\kappa, \chi, \eta)}^{m_{\rm{max}}} e^{-(m-\bar{m})^2/2\sigma^2} dm = \frac{\sigma}{I} \sqrt{ \frac{\pi}{2} } \left[ {\rm erfc} \left( \frac{m_*(\kappa, \chi, \eta) - \bar{m}}{\sigma\sqrt{2}} \right ) - {\rm erfc} \left( \frac{m_{\rm{max}} - \bar{m}}{\sigma\sqrt{2}} \right ) \right]\,.
\eea
Equation (\ref{fracgammansbhcoupled}) then gives the fraction of all NSBH sources that can form an accretion disk capable of generating a GRB to be
\bea \label{nsbh_fraction}
P(\gamma|{\rm NSBH}) = \epsilon_1 P({\rm NSBH}|m_*,\kappa_{\rm min},\kappa_{\rm max})
= \epsilon_1  \int_{0}^{0.25} d\eta \, p_{\eta} (\eta) \int_{0}^{1} d\chi \,p_{\chi} (\chi) \int_{\kappa_{\rm min}}^{\kappa_{\rm max}} d\kappa \, p_{\kappa} (\kappa) P\left(m_*(\kappa, \chi, \eta)\right) \,,
\eea 
where we included the factor $\epsilon_1$, which is the efficiency of the NSBH systems in Eq. (\ref{fracgammansbhcoupled}), with the values of the parameters $(m,\kappa, \chi, \eta)$ in their appropriate ranges, in producing a GRB. That efficiency may depend on the strength of the NS magnetic field,
precession of the system, and other factors that might be unraveled through future numerical simulations. It is important to clarify that the true dependence of $P(\gamma|{\rm NSBH})$ on NSBH parameters may be more complex than the one given above, the primary purpose of which is to recognize those parameters and to explain their physical relevance.

\begin{figure}[tbh]
\centering
\includegraphics[width=12.7cm]{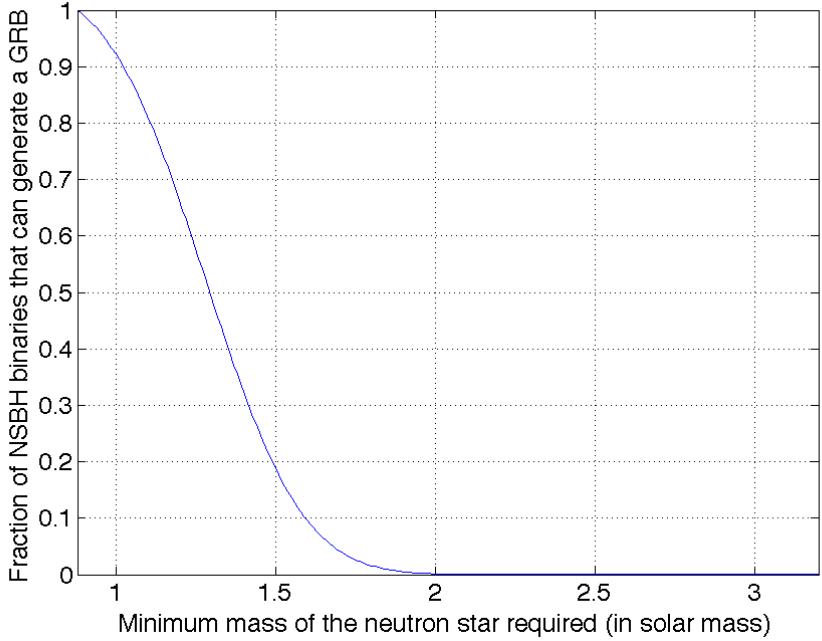}
\caption{Fraction of neutron star - black hole binary sources that give rise to a gamma-ray burst, which may or may not be beamed at us, for the special case of $p_{\eta}(\eta) = \delta(\eta - 0.25)$,  $p_{\chi}(\chi) = \delta(\chi - 1.0)$, and $p_{\kappa}(\kappa) = \delta(\kappa - 1.0)$. The minimum mass of the neutron star required for the emission, and for these choices of the probability densities, is plotted on the $x$-axis. }
\label{f_gammaConstraint}
\end{figure}

For the special case, where the aforementioned probability density functions are single valued, with
\bea\label{dens}
p_{\eta}(\eta) = \delta(\eta - 0.25), \;\;\;\; p_{\chi}(\chi) = \delta(\chi - 1.0),  \;\;\;\; p_{\kappa}(\kappa) = \delta(\kappa - 1.0)\,,
\eea
one finds that
\be
\label{fgammaspec}
\begin{aligned}
P(\gamma|{\rm NSBH}) = & \epsilon_1\frac{\sigma}{I} \sqrt{ \frac{\pi}{2} } \left[ {\rm erfc} \left( \frac{m_0 - \bar{m}}{\sigma\sqrt{2}} \right ) - {\rm erfc} \left( \frac{m_{\rm{max}} - \bar{m}}{\sigma\sqrt{2}} \right ) \right] \\
= &  0.53 \,\epsilon_1 \times \left[ {\rm erfc} \left( \frac{m_0 - 1.28}{0.34} \right ) - 1.23 \times 10^{-15} \right]\,,
\end{aligned}
\ee
where $m_0 = m_*(\kappa = 1, \chi = 1, \eta = 0.25)$. The value of the minimum mass $m_0$ is constrained to lie between $m_{\rm{min}}$ and $m_{\rm{max}}$; its distribution is shown in Fig. \ref{f_gammaConstraint} for the densities in Eq. (\ref{dens}) and for the fiducial value $\epsilon_1 = 1$.

Next consider the fraction $P(\gamma|{\rm BNS})$. For a coalescing binary neutron star system to be a progenitor of a gamma-ray burst, it first needs to result in the formation of a black hole with an accretion disk around it.
Some studies indicate that there can be two paths for that to happen (see, e.g., Ref. \cite{Bartos} and the references therein):
\begin{itemize}
\item The mass of the binary system
$(m_1 + m_2)$ is greater than $3M_{\odot}$ and the individual neutron star masses are unequal ($m_1 \neq m_2$). Let the probability of BNS systems that satisfy these criteria be $P_A$.
\item The condition $m_{\rm max} < (m_1 + m_2) < 3M_{\odot}$ is satisfied, but $(m_1 + m_2)$ is not so large that the hypermassive neutron star resulting from the merger promptly collapses to a black hole. Let the probability of these binaries be $P_B$.
\end{itemize}
Assuming this hypothesis to be true, the joint probability of a binary neutron star system coalescing to form an accretion disk around a central black hole can be written as,
\bea
\label{BNSprob}
P_{\rm BNS,~disk} = P_A + P_B
\eea
Note, however, that merely the creation of an accretion disk is not enough for triggering the GRB engine. For that to happen one needs an accretion disk that is massive enough to generate the accretion rate required for triggering a GRB. Whether such an accretion disk can be formed depends on multiple factors: Nuclear matter from both neutron stars can contribute to the accretion disk in varying amounts. Low compactness of a neutron star improves the chances of formation of a massive accretion disk. The component masses, spins and the orbital eccentricity also play a role. A precise determination of their influence is beyond the scope of this paper.
To make progress we introduce a second efficiency factor, $\epsilon_2$, that determines how likely it is for a BNS system to form an accretion disk massive enough to trigger a GRB. Then the fraction of BNS coalescences associated with GRBs is
\bea
\label{bns_fraction}
P(\gamma|{\rm BNS}) = \epsilon_2\,P_{\rm BNS, disk} \,.
\eea
Thus, the probability of observing an orphaned afterglow 
in conjunction with a gravitational wave signal from its CBCNS progenitor
can be obtained using Eqs. (\ref{NS_fraction}) and (\ref{orph_prob}) 
to be:
\bea
\label{eq:POAG}
P({\rm OAG},~{\rm GW}) &=& f_{\rm CBC}~f_{\rm OAG}~P(\gamma|\text{CBCNS}) \noQ\\
&=& 0.0865~\left(1 - \sin^2\frac{\beta_{\rm B}}{2}\right)~f_{\rm AG} ~P(\gamma|\text{CBCNS})\,,
\eea
where $P(\gamma|\text{CBCNS})$ is obtained by using $P(\gamma|{\rm NSBH})$ from Eq. (\ref{nsbh_fraction}) and $P(\gamma|{\rm BNS})$ from Eq. (\ref{bns_fraction}) in Eq. (\ref{prob_cbcns}).
The first two factors are purely geometric, with one arising from GW beaming and the other from SGRB beaming. The last two factors are based on the microphysics of the medium surrounding SGRBs and of neutron star matter. The last three factors will probably have to wait for joint EM-GW observations before their values are finally known. Note that through the presence of the fraction $f_{\rm CBC}$, Eq. (\ref{eq:POAG}) ensures that the probability $P({\rm OAG},~{\rm GW})$ applies for gravitational wave sources within the horizon distance.

If GRBs are used to trigger GW searches, then the detection threshold is lowered by at least 11.3/9.0 = 1.25, which increases the GW rate by $1.25^3 = 1.95$. So 50 CBCNS per year (which is the likely rate of BNS and NSBH detections in aLIGO \cite{Abadie:2010cf}) increases to 98. However, recall that $N_{\gamma,\text{GW}}/N_{\text{GW}}$ is the fraction of CBCNS GW detections that are expected to be associated with SGRBs. Thus, if that fraction is 3\%, one can expect 3 GW detections due to GRBs in one year. 

Afterglows at smaller energies or longer wavelengths than gamma rays will be observable more isotropically. All detected GWs will have observable afterglows provided they all are 100\% efficient in triggering bursts that encounter a dense enough surrounding medium to produce a strong afterglow. But the efficiency will most likely be far less than 100\%. The detected GWs will give us a sense of the likelihood of afterglows, orphaned or not, occurring with GWs. Recall that $f_{\text{OAG}}$ is the fraction of CBCNS with afterglows that are observed by us as orphans. An afterglow will improve GW detectability by reducing the detection threshold by 8\% of the all-sky, all-time GW search, as shown below Eq. (\ref{FAPOAG}). Thus, the realistic estimate of 50 detected CBCNS per year will increase to about $1.08^3\times 50 = 63$ per year, and $63f_{\text{OAG}}$ additional GW events can be expected due to follow-up of EM transients, some of which will be orphaned afterglows. If $f_{\text{OAG}}$ is 3\%, then a couple of additional GW events can be detected every year due to an orphaned afterglow. Note that our choice of a few percent as example values of $f_{\text{OAG}}$ and $N_{\gamma,\text{GW}}/N_{\text{GW}}$ is conservative and is, e.g., smaller or similar to values obtained in Ref. \cite{Kelley:2012tc}, which conducted a comprehensive study of expected rates of EM emissions observed in coincidence with GW events. While the absolute number of expected EM-GW associations may be small, the astrophysical insights these events will provide, as noted in Sec. \ref{sec:intro}, highlights the importance of pursuing their joint detection.

The above analysis makes the case for pursuing a two-pronged approach to finding CBCNS GW sources associated with orphaned afterglows. One approach is to use a GW detection to trigger an afterglow search in EM observatories. This requires the computationally expensive GW detections to be fast so that they can have a shot at detecting even prompt emissions. This is realizable (see, e.g., \cite{Cannon:2011vi} and the references therein). A more fundamental challenge in this approach arises from the fact that the sky localization error of GW searches can be several square-degrees, when at least 3 sites have tracked the GW event, or worse, with a smaller number of sites tracking the event. Therefore, sky-localization demands a high duty factor of the GW detectors. The second approach is to use EM observatories to find afterglow candidates and then search GW data, which may even be archived, for GW events. This is easier to implement but can be equally rewarding in terms of accessing new knowledge about these most energetic events in the universe.

\section{Externally triggered search for GW signals from poorly localized CBC sources}
\label{sec:poorlyLocSource}



A sizable fraction of GRBs do not have accurately known sky positions. A selected list of such GRBs that occurred during LIGO's S5 run is presented in tables \ref{tab:fermiErrorBoxes} and \ref{tab:IPNErrorBoxes}. 
The error in the sky position can arise from constraints inherent to the triangulation method, used by gamma-ray burst detectors. Typically, electromagnetic observatories have good sky localization. However, for a transient phenomenon like a GRB, the primary objective is detection rather than source localization, and a large fraction of GRBs are found with inaccurate sky positions. Detection of an afterglow in conjunction with a GRB can provide that information with a better accuracy since an afterglow can improve the localization of the associated GRB \cite{Rhoads:2000gh,VandenBerk:2001tx,Dalal:2001ym}.
The observation of a significant fraction of GRBs with error radii of tens of degrees, however, motivated us to study the effect of poor sky localization on the detectability 
of the CBCNS progenitor in GW searches externally triggered by GRBs.
For this analysis, any localization with an error radius of more than a few degrees is termed as poor.

Currently, externally triggered searches for gravitational waves are conducted exclusively in archived data, in a $6$ sec time-window around the epoch of a short duration GRB~\cite{Abadie:2010uf,Briggs:2012ce,Valeriu:2012}. When the sky position of the GRB is known accurately from EM observations, a GW search is launched for that sky position. If the CBCNS progenitor model of SGRBs is correct, and if the source is within a detector's range, one expects to observe a GW signal from it. The knowledge of the time of the event helps to reduce the false-alarm probability at a given SNR, as discussed in Sec. \ref{ObsOrphAG}.
There we also showed that a further reduction of false-alarm probability, by an order of magnitude, can be achieved by searching for GWs from a single direction in the sky. This allows for setting a lower detection threshold, which in turn helps target more distant SGRBs and increases their GW detection rate. While this process is optimal for a source with an accurately known sky position, e.g., from EM observations, it is not so 
when the sky position is unknown or has a large error radius.

In Fig. \ref{fig:systknown}, we show how the detection efficiency suffers when a SGRB's true location is $20^\circ$ away from the reported sky position. We selected the time and the sky-position of GRB090709B for this study. This GRB occurred on $09$ Jul $2009$ at 15:07:42 (UTC), i.e., at a GPS time of $931187277$ sec, and at a Right Ascension (RA) of $93.5^\circ$ and a Declination (dec) of $64.1^\circ$.
In a 2190 sec observation window centered at this GRB time, but after dropping 
246 sec immediately around that instant, we injected $3000$ CBCNS signals. The CBCNS masses ranged from $m_1= 1-3 M_{\odot}$ for the neutron star and $m_2 = 1-40 M_{\odot}$ for the black hole. 
For this study the binary components were taken to be non-spinning. 
The signals were injected in Gaussian noise with LIGO-I \cite{LAL} power spectral density (PSD) \cite{Helstrom} simulated for the H1L1V1 network.
The results are grouped into three categories based on the chirp-mass ${\cal M}_c = (m_1m_2)^{3/5}/(m_1+m_2)^{1/5}$ of the triggers arising from injections and background.
These are the low chirp-mass, with range $(0.0, 3.48]M_{\odot}$, the medium chirp-mass $(3.48, 6.0]M_{\odot}$ and the high chirp-mass $(6.0, 20.0]M_{\odot}$ systems. The reason for using the chirp mass for categorizing triggers is that the duration of the signal in the detector band is primarily determined by it. The characteristics of the noise artifacts that trigger the search pipeline also depend on the chirp masses of the search templates used.

\subsection{Effect of GRB sky-position error on the detection efficiency of targeted GW searches: The case of a single GRB sky position}
\label{varAcrossSky:single}

To quantitatively assess the effect of inaccurate GRB sky position on CBCNS searches, we study the detection efficiency of the search conducted as a function of the source distance for two cases, one with $20^{\circ}$ sky position error and another with no sky position error. The detection efficiency in a given distance bin and mass bin is the fraction of the injected triggers in that distance and mass bin found louder than the loudest background trigger in the same mass bin. Measurement of the distance to a GW source depends on a network's ability to resolve the signal's polarization \cite{Ajith:2009fz}. Such a resolution is not possible for every network or every sky position. For injections, the distances used to simulate the signals are used to bin them. The calculation of the detection efficiency in any distance bin uses the loudness of the loudest background trigger, in the mass bin of interest, regardless of whether it was possible to estimate its distance or not. 


\begin{figure*}[tbh]
\centering
\includegraphics[width=5.8cm]{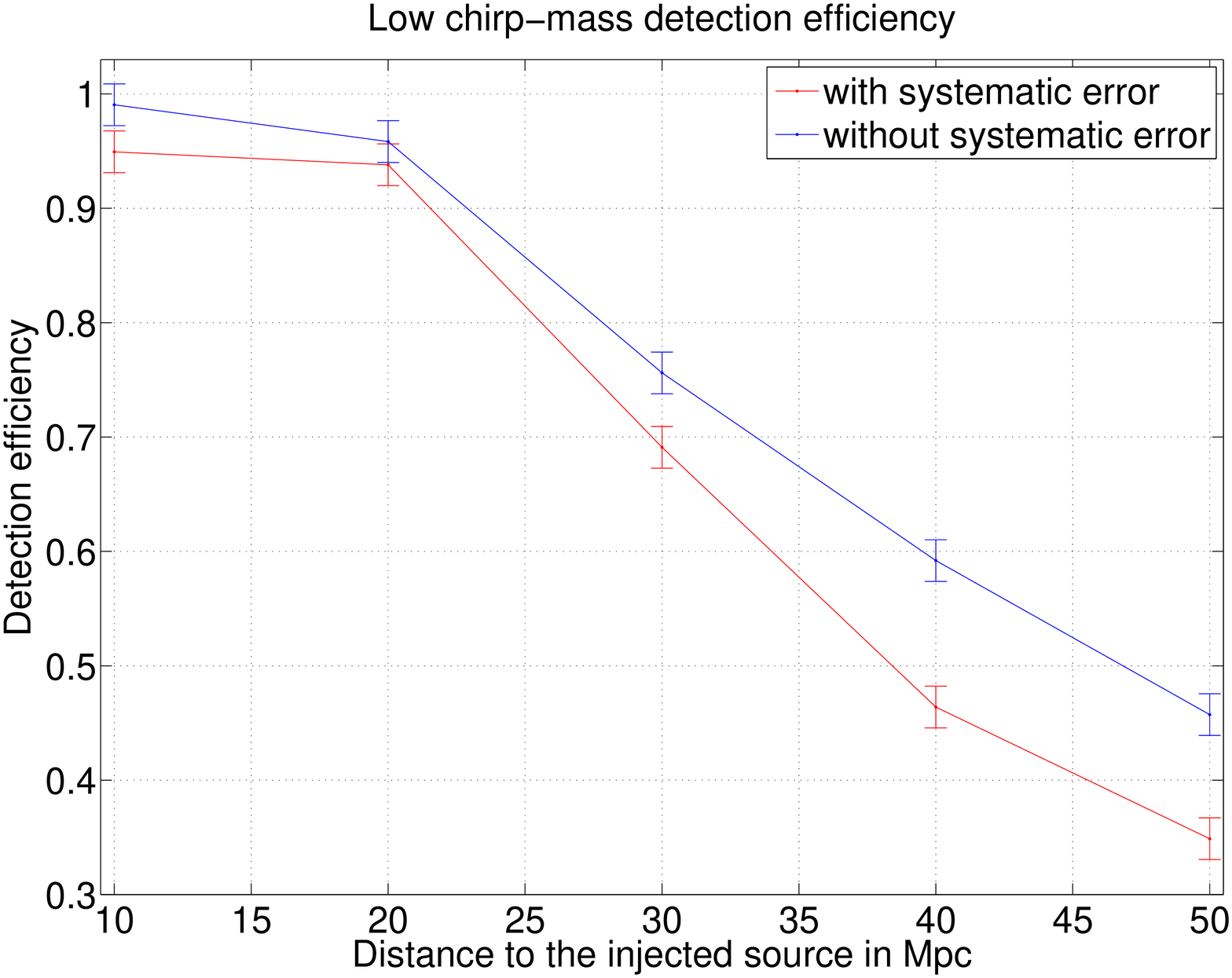}
\includegraphics[width=5.8cm]{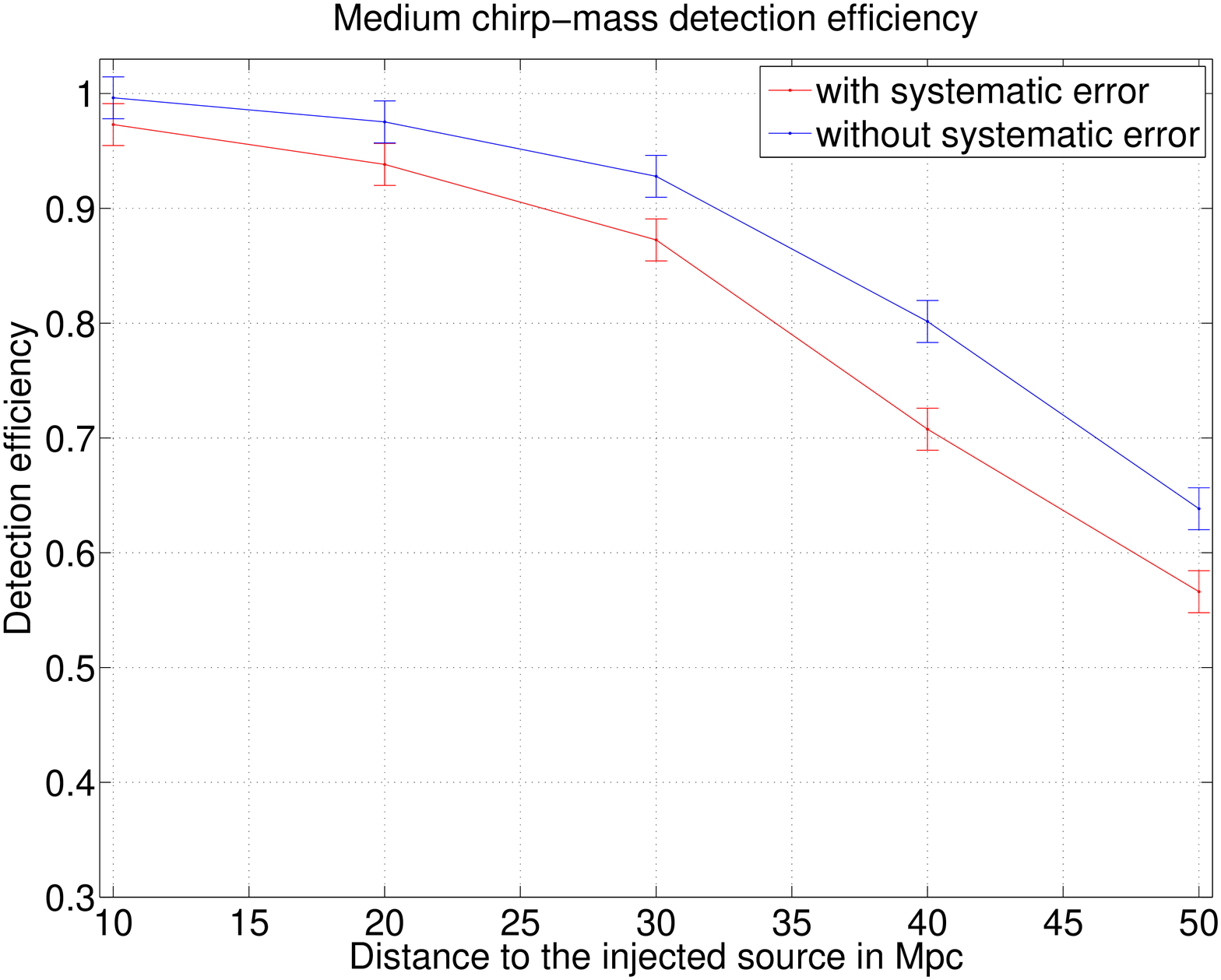}
\includegraphics[width=5.8cm]{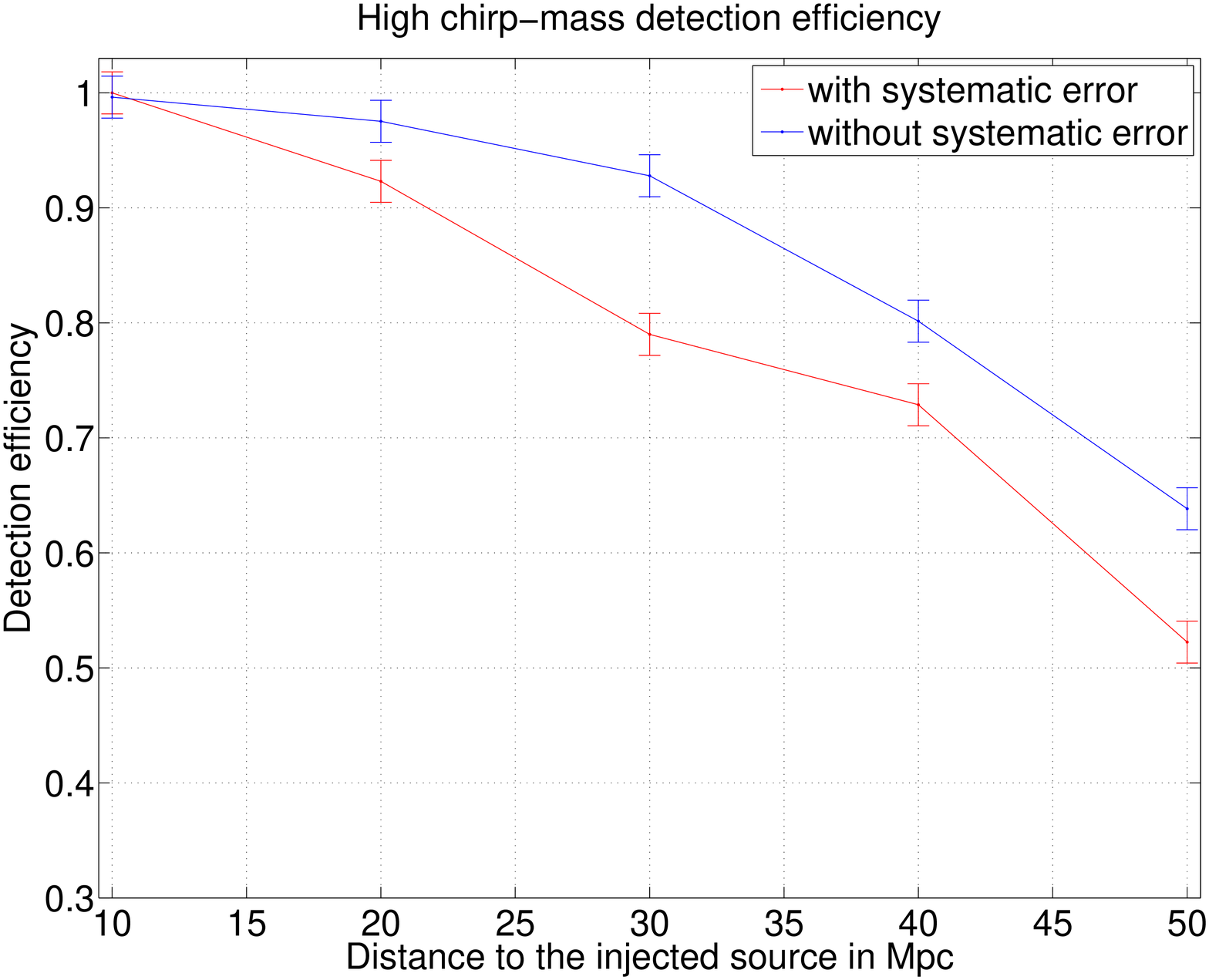}
\caption{\label{fig:systknown} Comparison of the detection efficiencies of (from left to right) low chirp-mass, medium chirp-mass and high chirp-mass CBCNS systems with accurately known sky positions (blue or upper curves) and those with a sky-position error of $20^\circ$ (red or lower curves). 
}
\end{figure*}

The results of this simulation study are shown in Fig. \ref{fig:systknown}. As expected, the detection efficiency deteriorates due to the systematic error in the sky position of the GRB used in the search for its GW signal. Compared to the case of no sky-position error, it drops in the low chirp-mass bin by 22\% and 24\% at injected distances of 40 Mpc and 50 Mpc, respectively, in the medium chirp-mass bin by 12.5\% at an injected distance of 50 Mpc, and in the high chirp-mass bin by 15.7\% and 20.5\% at injected distances of 40 Mpc and 50 Mpc, respectively. 

One way to tackle the problem of loss of GW detection efficiency due to an inaccurately known sky position from EM observations is to search in a wider region in the sky. This improves the chances of targeting the true sky position of the source. But it can also reduce the confidence in a detection candidate. 
This is because searching in multiple sky positions, or signal time-delays across the GW detector baselines, makes the noise background worse. Specifically, it increases the false-alarm probability at a given SNR, thereby reducing the significance of a signal. 
To assess what the trade off is between improvement in detectability and the deterioration in FAP, we performed a search of the same 
set of simulated injections as studied above except that (a) their sky position is now erroneous by $20^\circ$, in declination only, and (b) we use a larger set of sky-position templates. These templates were chosen to be distributed isotropically in a patch of sky around the reported GRB sky coordinates, with one template per pixel, which is four square-degrees in size.\footnote{The reason for this choice is explained below.}

\begin{figure*}[tbh]
\centering
\includegraphics[width=5.8cm]{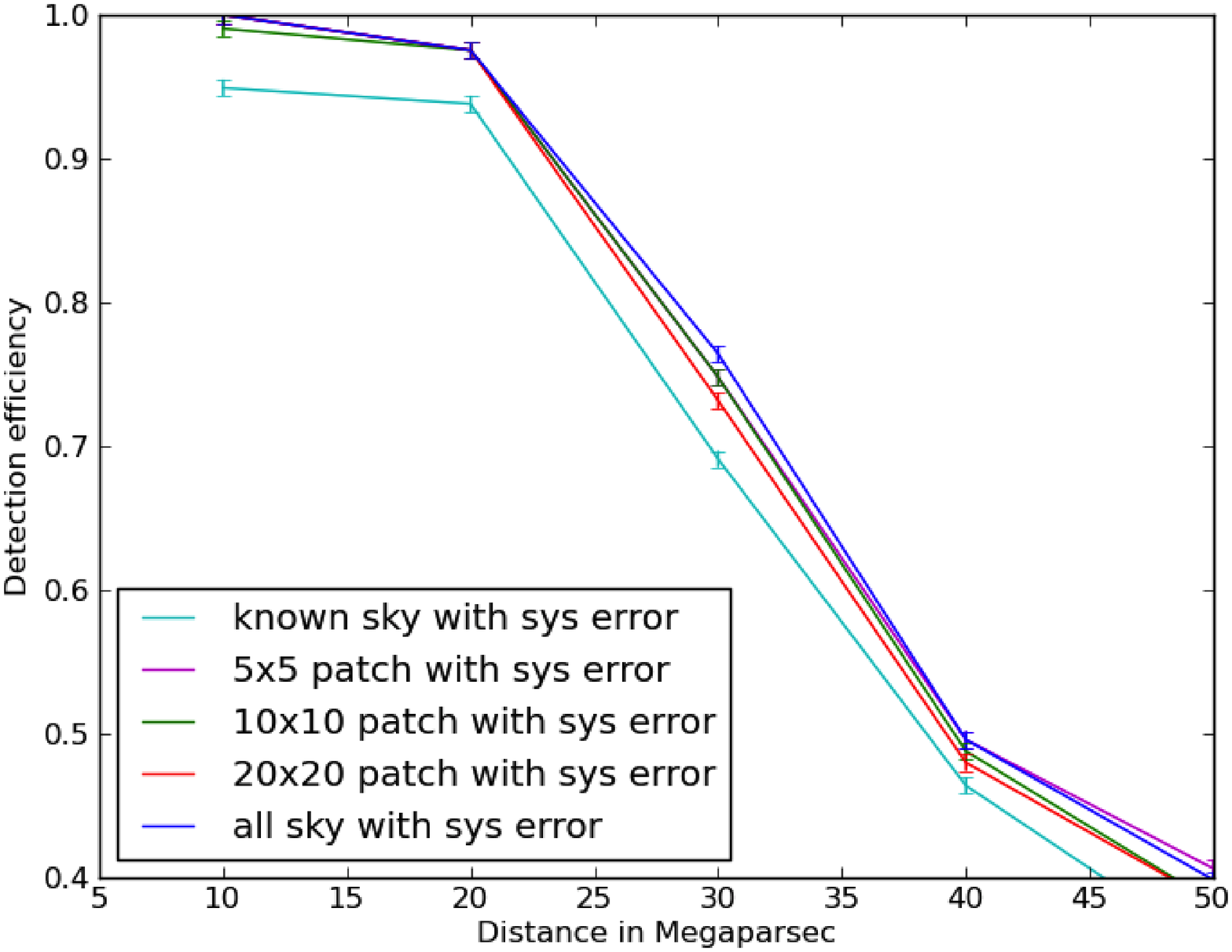}
\includegraphics[width=5.8cm]{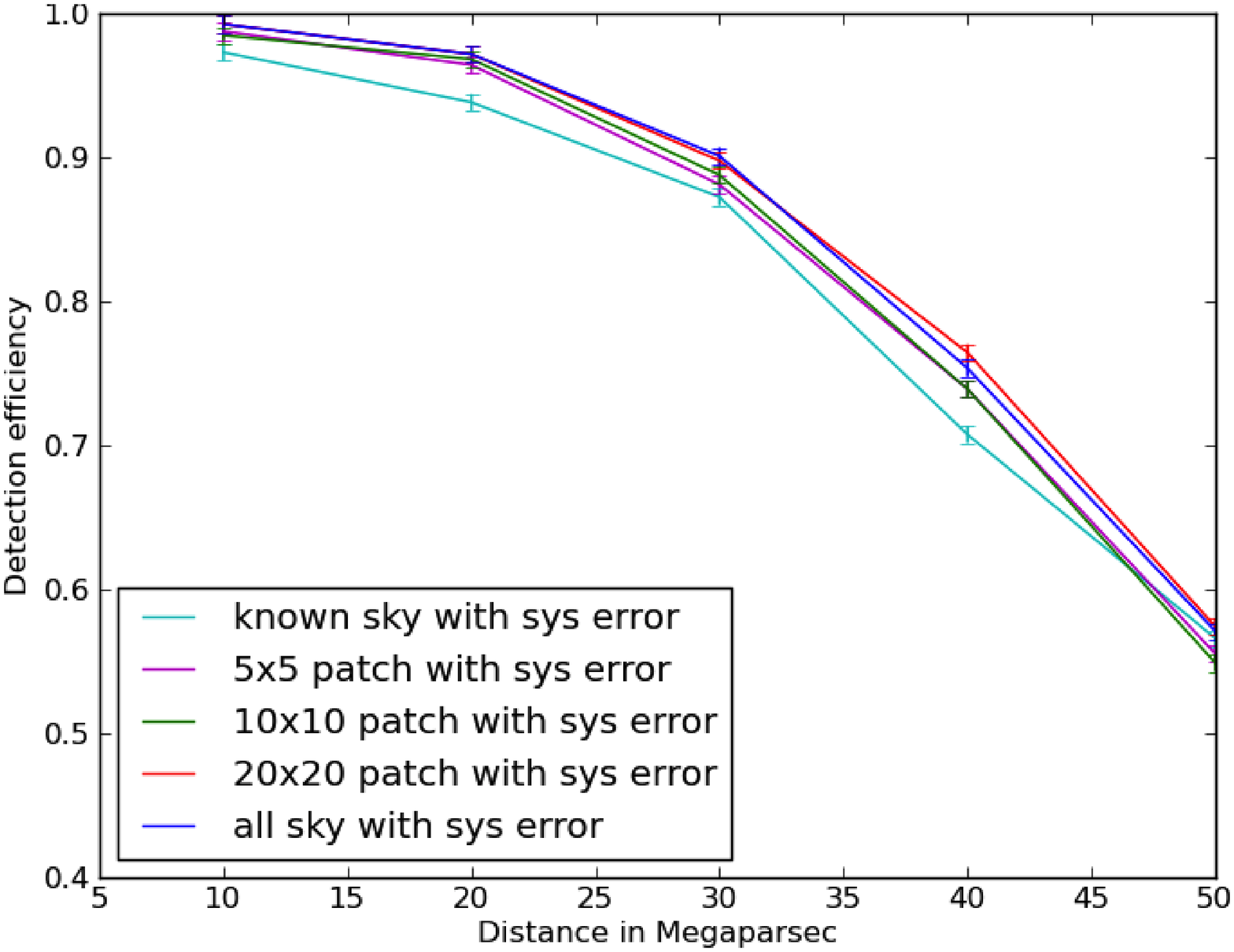}
\includegraphics[width=5.8cm]{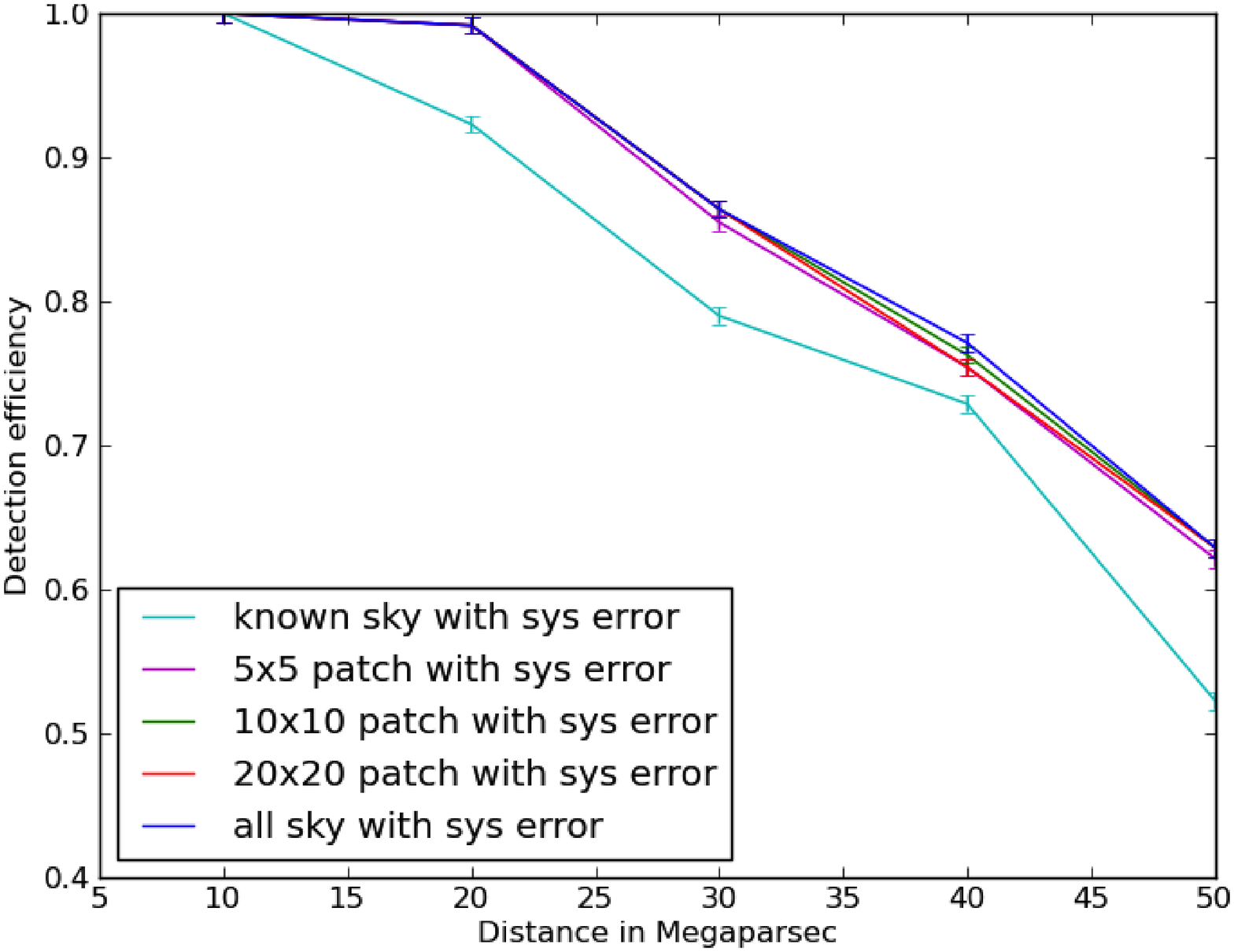}
\caption{Detection efficiency comparison for injections with a sky-position error of $20^{\circ}$ (from left to right) in low chirp-mass, medium chirp-mass and high chirp-mass bins. }
\label{fig:sysErrorComp}
\end{figure*}

The triggers output by the search were clustered in sky position by maximizing their SNR over the patch. We named this the {\em sky-patch} mode of search, as opposed to the {\em known-sky} mode, where the search was confined to a single point in the sky, as discussed above.
For the limiting case, we also conducted experiments where the sky-patch was extended to cover the whole sky. 
Results from the known-sky, sky-patch and all-sky searches are presented in Fig. \ref{fig:sysErrorComp}. Three sky-patches were used, with $5\times 5$, $10\times 10$, and $20\times 20$ pixels.
The improvement in the detection efficiency is up to about 20\% in any of the sky-patch modes or the all-sky mode compared to the known-sky mode. However, the detection efficiencies of the sky-patch and all-sky modes are smaller in many distance bins compared to that of the known-sky mode with no sky-position error (see the top curves in Fig. \ref{fig:systknown}). 
This proves that while a wide-area sky search performs worse, due to an increased FAP, than the known-sky search with no sky-position error, nevertheless it performs better than a known-sky search with a $20^\circ$ error in the GRB sky position.


\begin{figure}[tbh]
\centering
\includegraphics[width=8.8cm]{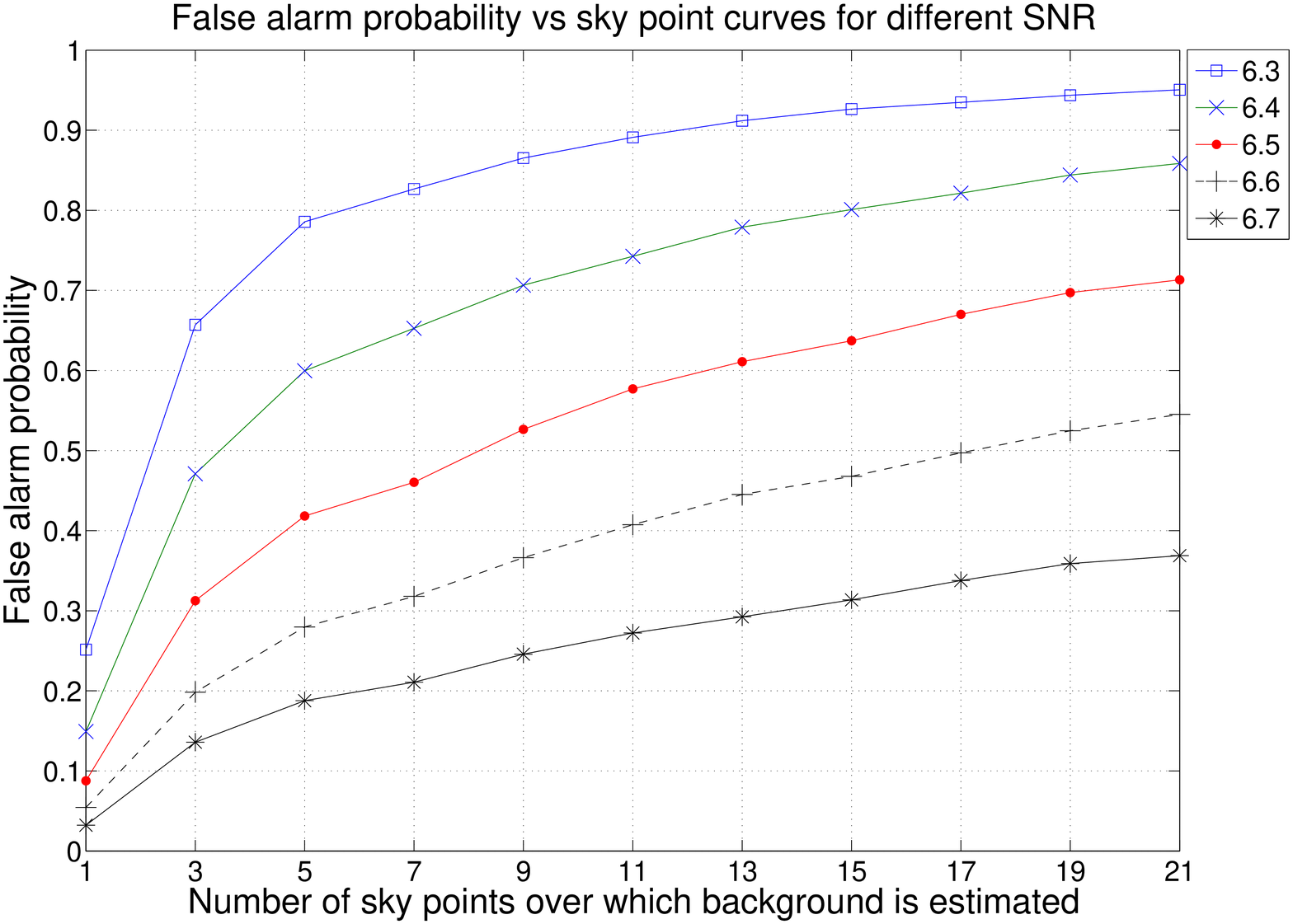}
\includegraphics[width=8.8cm]{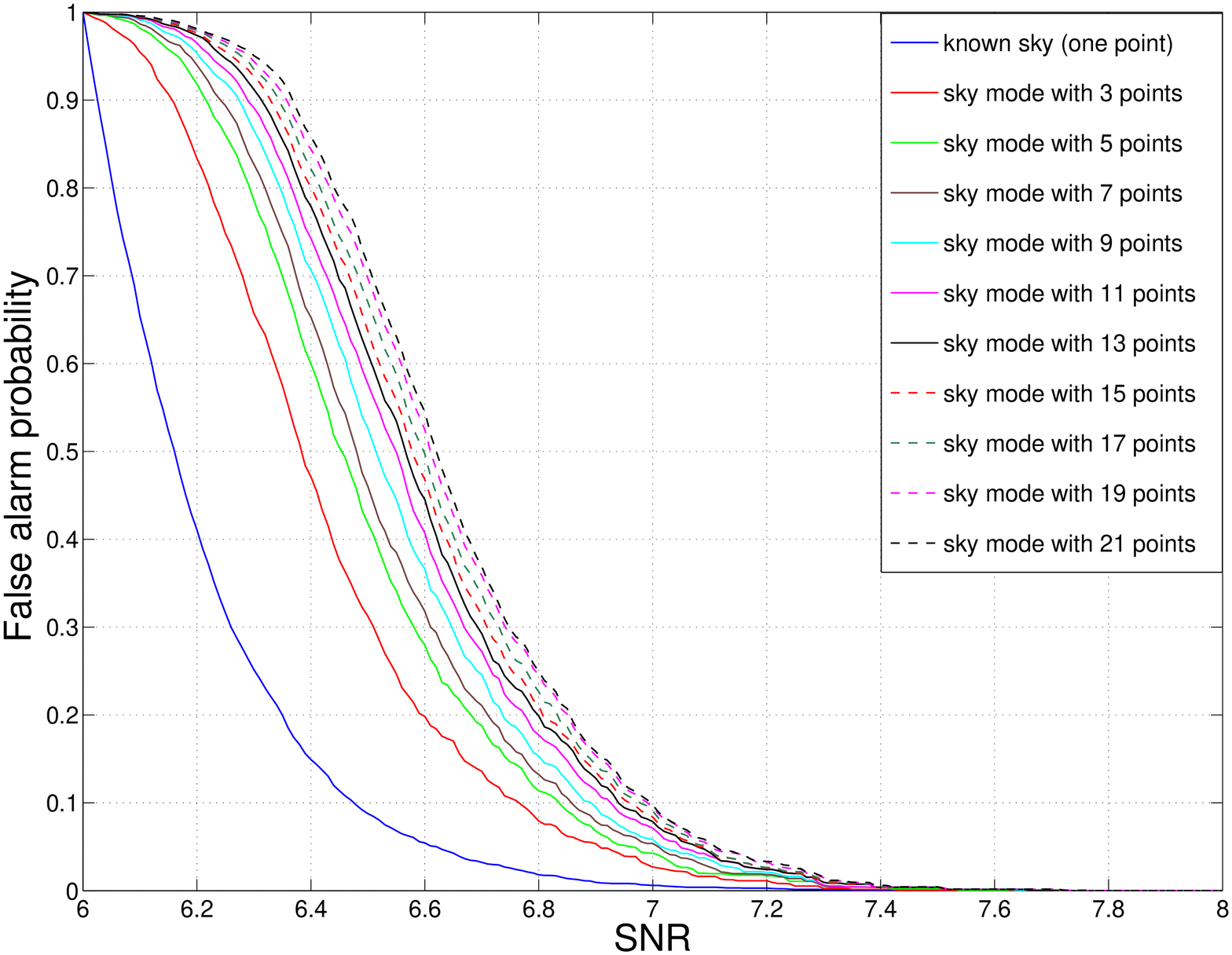}
\caption{{\bf Left}: Plots of the false-alarm probability of a sky-patch search as a function of the number of the sky position templates or points in the patch, for SNR values 6.0 - 7.5. The $x$-axis displays the number of sky points in the patch, which includes the true sky-position. An increase in the number of sky points in a search causes a monotonic increase in the false-alarm probability, for any given SNR, and it asymptotes to the all-sky FAP value at that SNR.
{\bf Right}: False-alarm probability plotted as a function of SNR for sky-patches with a varying number of sky points. Here the false-alarm probability at a given SNR, plotted on the $x$-axis, is defined as the number of background triggers that are found louder than that SNR divided by the total number of background triggers of any SNR found in the search with that sky-patch.}
\label{fig:FAP}
\end{figure}

Figure \ref{fig:FAP} shows how the false-alarm probability (FAP) increases as one widens the size of the sky patch. The first (left-hand side) plot depicts how the FAP at a given SNR increases with the number of sky points. Note how the FAP climbs sharply in increasing from one point in the sky to two points in the sky, before asymptotically approaching the all-sky FAP value at that SNR. 
On the other hand, as shown in the second (right-hand side) plot in that figure the FAP has the expected fall-off behavior with increasing SNR for any given sky-mode search. It also shows that at any given SNR the FAP increases as the sky-patch area is widened.

\subsection{Effect of GRB sky-position error on the detection efficiency of targeted GW searches: Multiple GRB sky positions}
\label{varAcrossSky}

\begin{figure*}[tbh]
\centering
\includegraphics[width=7cm]{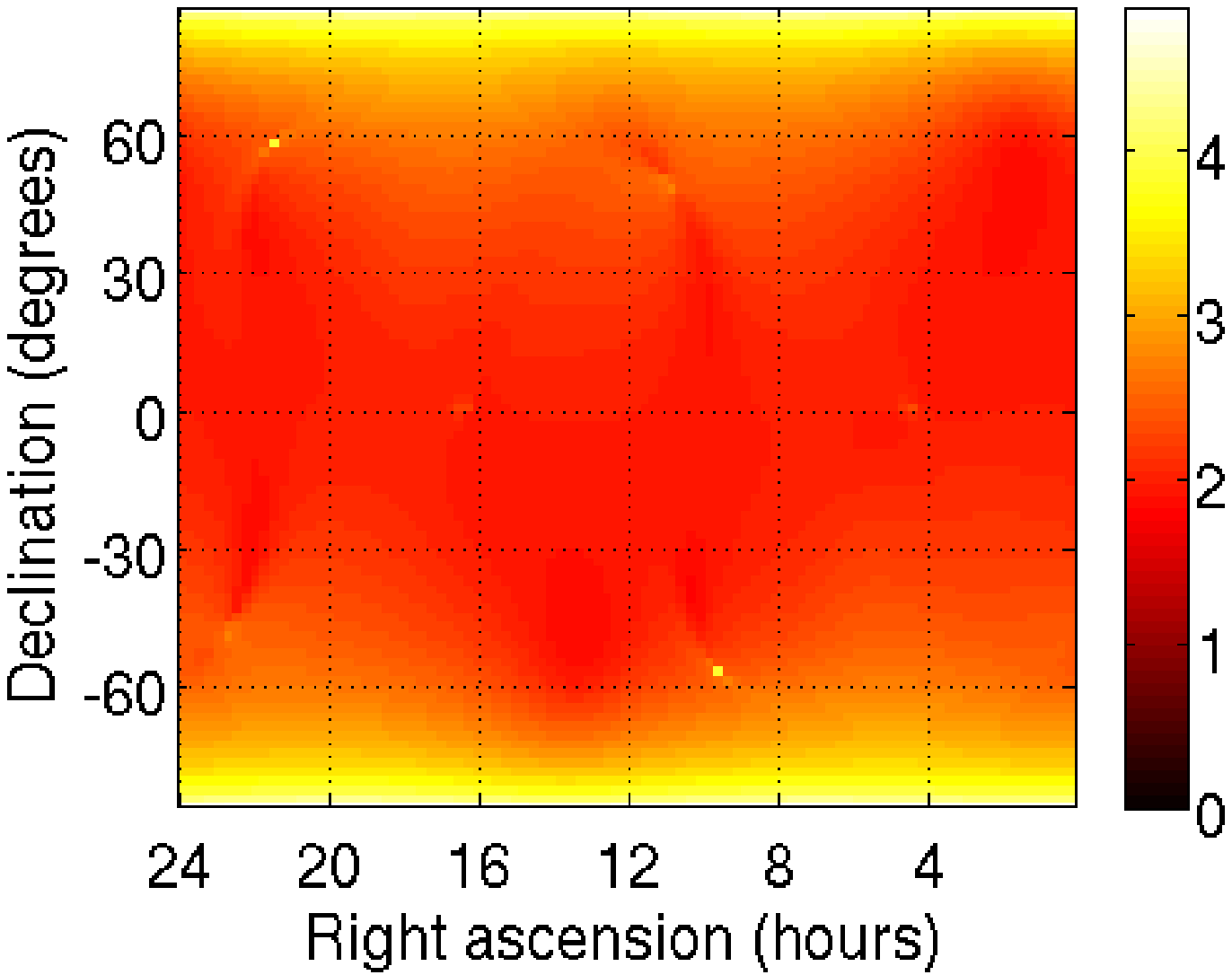}
\includegraphics[width=7cm]{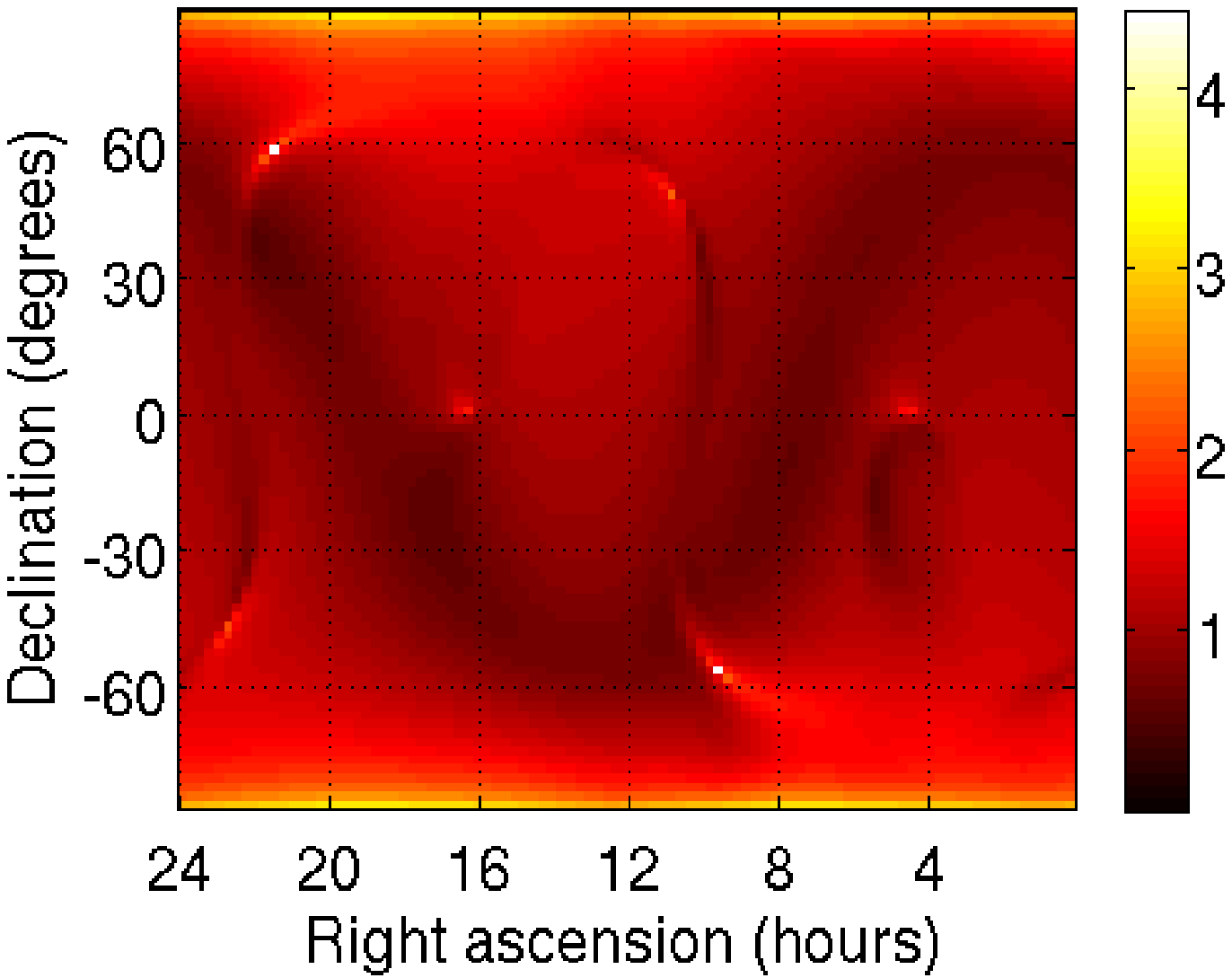}
\caption{{\bf Left plot:} The log$_{10}$ of the percentage change in the H1L1V1 network SNR owing to a $20^\circ$ error in the Declination of the source is mapped as a function of the true source sky position. {\bf Right plot:} Same as the left plot except that here the error in the sky position of the source is taken to be 4 square-degrees of solid angle. For both plots the detector noise PSD is taken to be LIGO-I \cite{LAL} for all three detectors. The source is a non-spinning CBCNS with component masses 2.5$M_\odot$ and 40$M_\odot$, optimally oriented and located at a distance of 17Mpc.}
\label{fig:fisher}
\end{figure*}

\begin{figure*}[tbh]
\centering
\includegraphics[width=8.8cm]{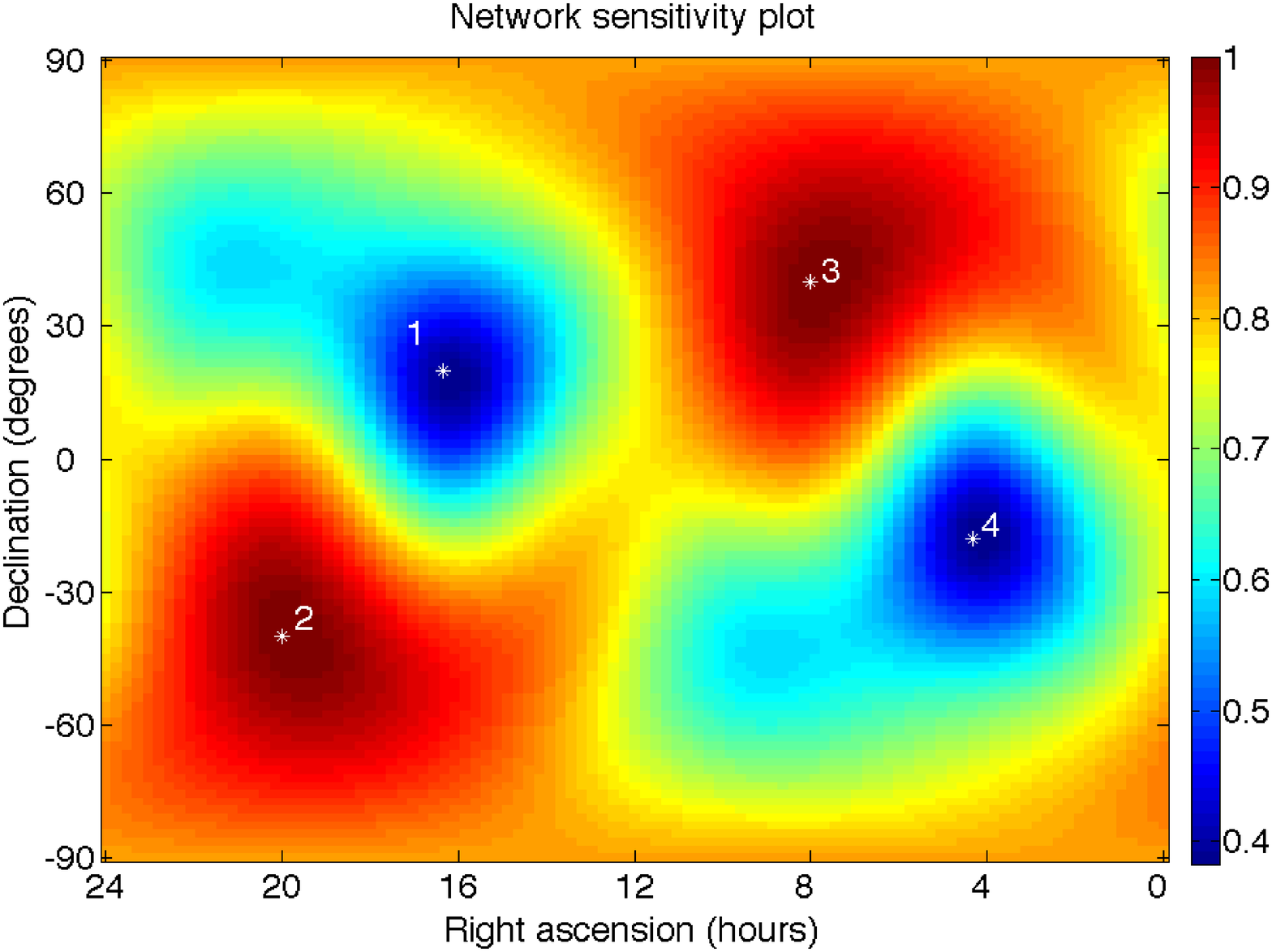}
\includegraphics[width=8.8cm]{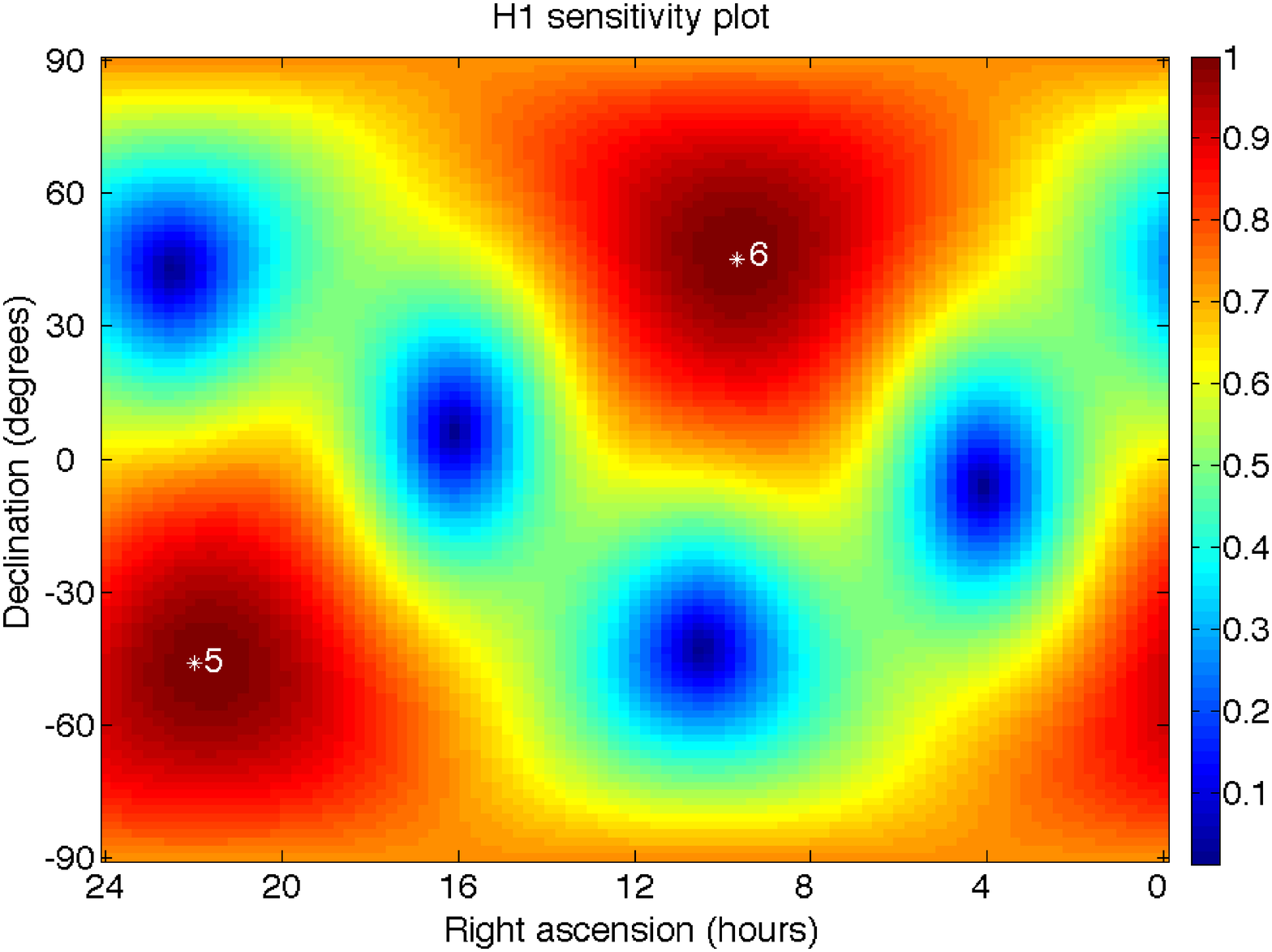}
\includegraphics[width=8.8cm]{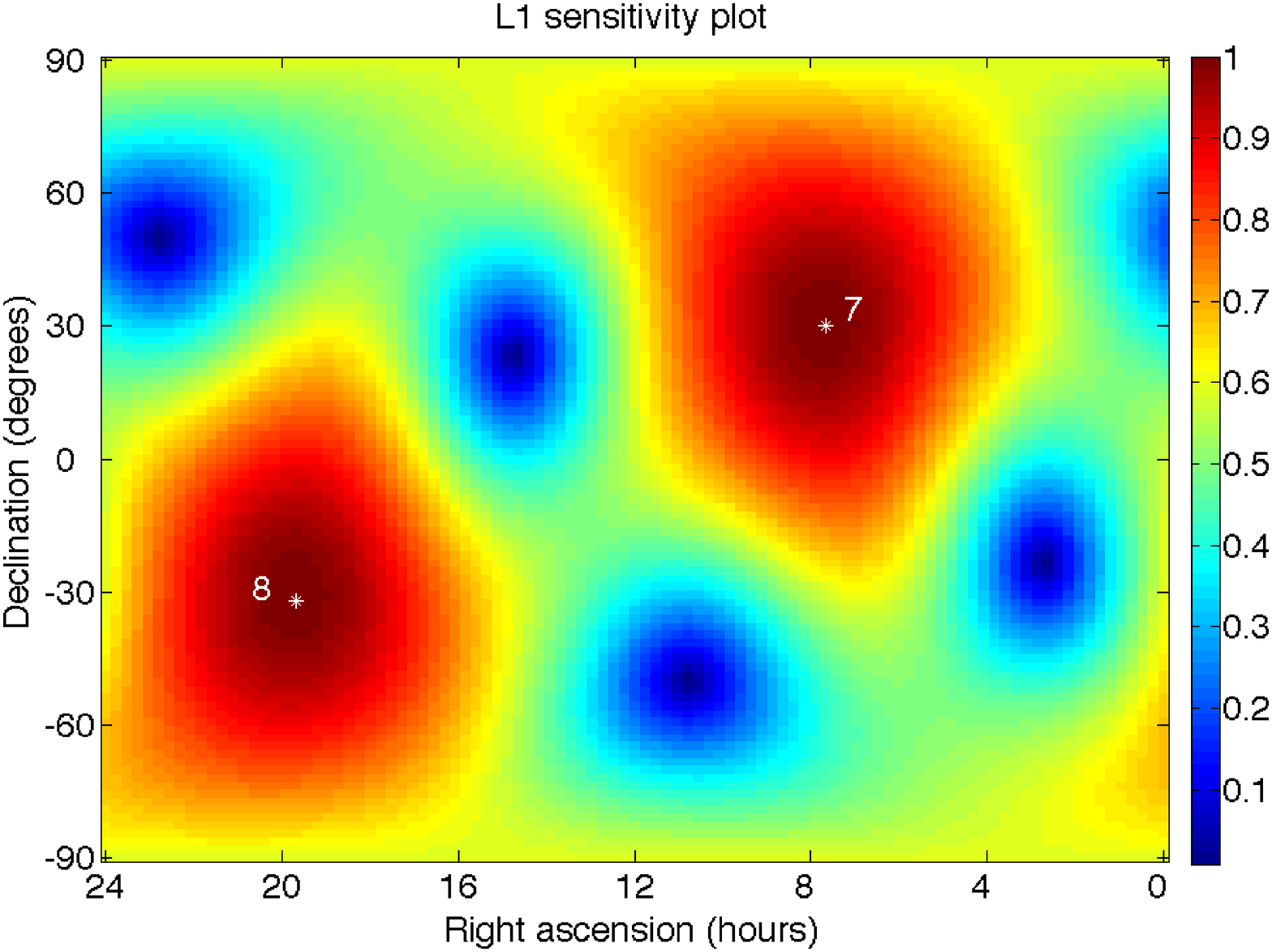}
\includegraphics[width=8.8cm]{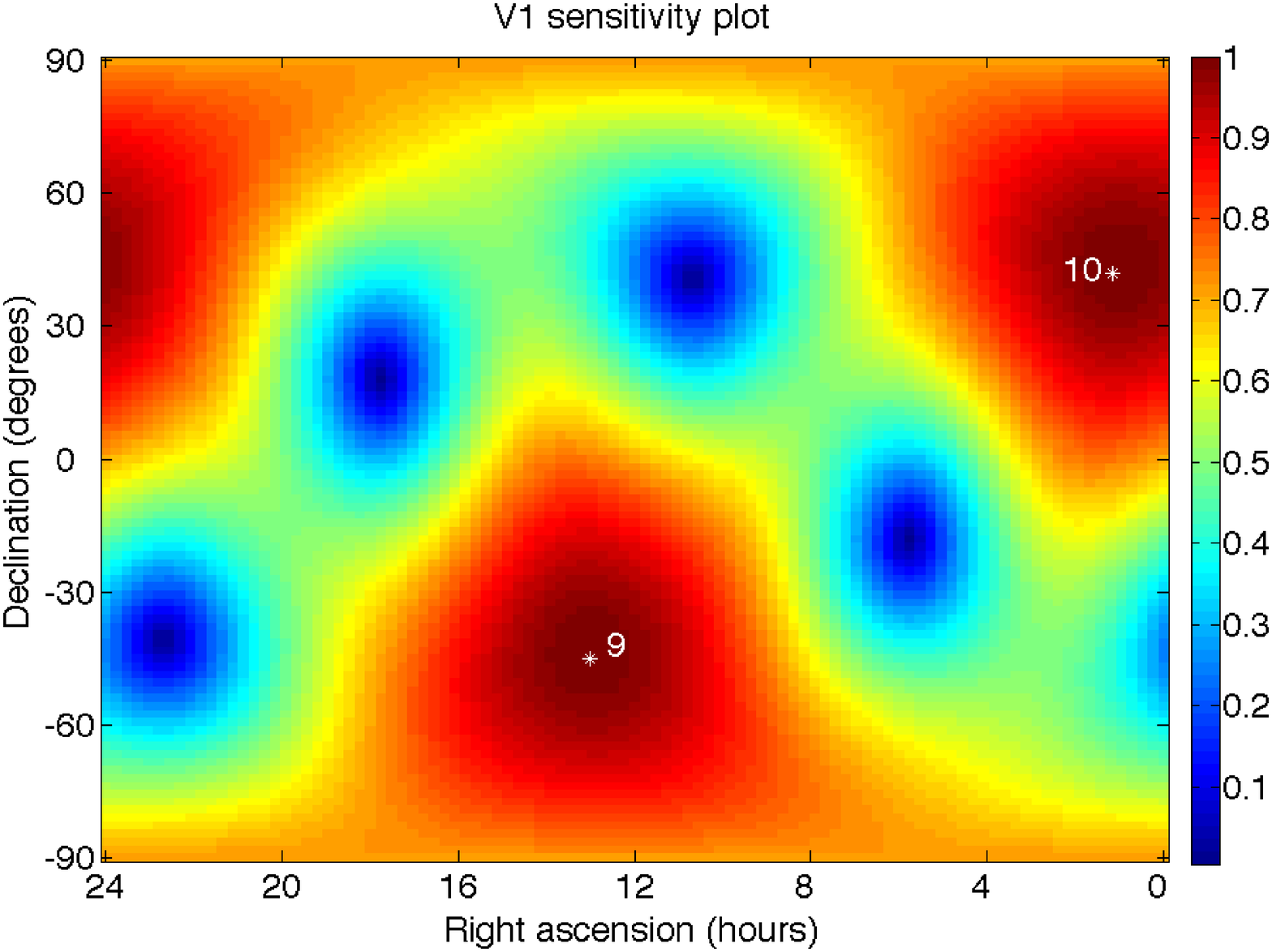}
\caption{{\bf Top left}: Network response across the sky; the quantity $\sqrt{\sum_{I}(F_+^{I~2} + F_\times^{I~2})}$, normalized to have a maximum value of unity, plotted as a function of the sky for the H1L1V1 network. The four white asterisks in this plot are located at the two most sensitive and two least sensitive sky positions. Their RA and dec, in degrees, are $(300.0^{\circ},-40.0^{\circ}), (120.0^{\circ},40.0^{\circ})$ and $(245.0^{\circ},20.0^{\circ}), (65.0^{\circ},-18.0^{\circ})$, respectively. These locations were chosen for our study of the variation of detection efficiency across the sky. {\bf Top right}: LIGO-Hanford sensitivity; the two white stars give the locations of greatest H1 response. At these two spots L1 and V1 have a relatively weak response. Thus, injections in these locations provide a good measure of how much the detection efficiency suffers when only one interferometer in the H1L1V1 network hass a good response. Their RA and dec (in degrees) are $(330.0^{\circ},-46.0^{\circ}), (145.0^{\circ},45.0^{\circ})$. {\bf Bottom row}: Same as the top right plot but for L1's best response (bottom left), at $(115.0^{\circ},30.0^{\circ})$ and $(295.0^{\circ},-32.0^{\circ})$, and V1's best response (bottom right), at $(195.0^{\circ},-45.0^{\circ})$ and $(17.0^{\circ},42.0^{\circ})$.
}
\label{fig:networkSens}
\end{figure*}

Here we enquire how the above results transform as one varies the sky-position of a GRB. Although a GW detector is much less directional, as shown in Fig. \ref{fig:networkSens}, than, say, an optical telescope \cite{Saulson,creightonAnderson}, the variation of its response across the sky cannot be neglected in determining the efficiency of a GW search. While the integrated response of a network of detectors is smoother across the sky than the response of any single detector, there are patches in the sky where its minima are nearly zero and maxima are nearly unity.\footnote{The response in Eq. (\ref{antpattern}) when summed over the number detectors can attain a maximum value equal to the number of detectors. However, the network response plotted in Fig. \ref{fig:networkSens} is normalized to have a maximum value of unity.} How sensitive a network is to a sky-position does not depend on the detector antenna-patterns alone but also on how their noise PSDs compare in the band where the signal is present. If the latter are identical, the network SNR has the same variation in the sky as shown in Fig. \ref{fig:networkSens}, which is computed for a face-on binary.
On the other hand, the fractional drop in SNR of a signal owing to a systematic error in the sky position used in the search template can have a different variation in the sky than that shown in Fig. \ref{fig:networkSens}. This is because that drop is determined by how fast the network response and the time-delays across the baselines in the network change with $(\theta,\phi)$ at a sky-position, which, in turn, can be estimated by the network Fisher matrix of the CBC signal as a function of the two sky-position angles \cite{Helstrom,Ajith:2009fz,Keppel:2012ye}. That fractional loss in SNR is plotted in Fig. \ref{fig:fisher} as a function of sky for a couple of cases, namely, when the error in the sky-position of the source is (a) $20^\circ$ in Declination and (b) 4 square-degrees of solid angle. 
Inferring the effect of that loss on the detection efficiency is non-trivial since the latter also depends on the change in FAR due to the error.
We next assess their combined effect through Monte Carlo simulations across a set of different sky positions.

\begin{figure*}[tbh]
\centering
\includegraphics[width=12.5cm]{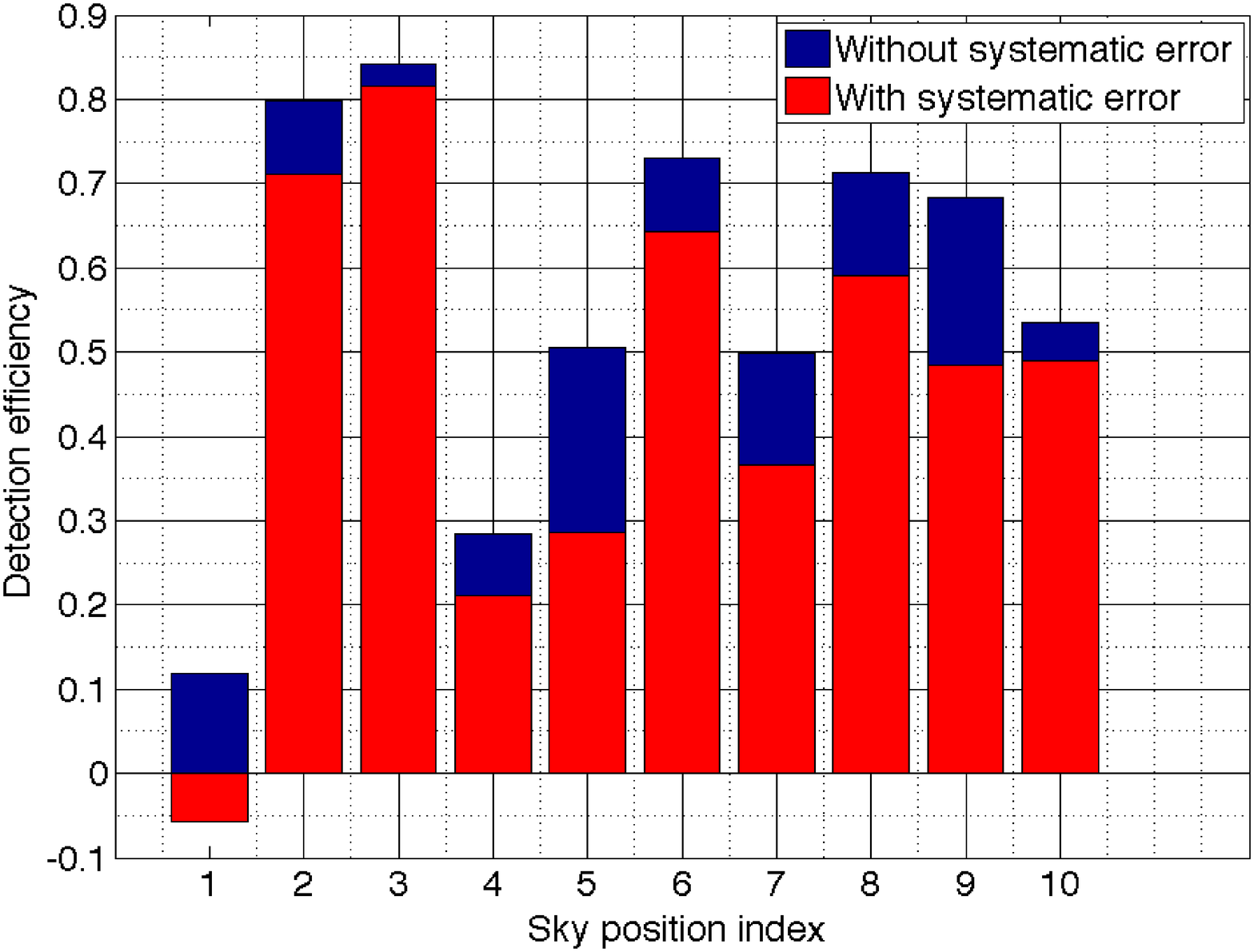}
\caption{The $y$-axis above denotes the detection efficiency of injectionsin the ten sky positions shown in Fig. \ref{fig:networkSens}. The source is the same as the one used in Fig. \ref{fig:fisher}. The ten different sky positions where injections were made to study the variation of detection efficiency as a function of the source sky position are listed on the $x$-axis. Their RA and dec (in degrees) are {\bf{\color{blue} 1}}: $(245.0^{\circ},20.0^{\circ})$, {\bf{\color{red} 2}}: $(300.0^{\circ},-40.0^{\circ})$, {\bf{\color{red} 3}}: $(120.0^{\circ},40.0^{\circ})$, {\bf{\color{blue} 4}}: $(65.0^{\circ},-18.0^{\circ})$, {\bf{\color{ForestGreen} 5}}: $(330.0^{\circ},-46.0^{\circ})$, {\bf{\color{ForestGreen} 6}}: $(145.0^{\circ},45.0^{\circ})$, {\bf{\color{ForestGreen} 7}}:  $(30.0^{\circ}, 115.0^{\circ})$, {\bf{\color{ForestGreen} 8}}: $(295.0^{\circ},-32^{\circ})$, {\bf{\color{ForestGreen} 9}}: $(195.0^{\circ},-45.0^{\circ})$, and {\bf{\color{ForestGreen} 10}}: $(17.0^{\circ},42.0^{\circ})$. The color code of the indices is as follows: Red-colored indices are points where the network response is the highest, as shown in Fig. \ref{fig:networkSens}. Blue indices are points of lowest network response, and green indices are points where one of the interferometers in the H1L1V1 network has a high response and the other two have a low response. Note that for the first sky position the fractional loss in SNR of the loudest injection is negative. This is because there the SNR of the injection trigger happens to be smaller than that of the loudest background trigger.
}
\label{fig:skymc}
\end{figure*}

To check the inferences drawn from the Fisher matrix calculations above on the variation of detection efficiency across the sky, we conducted a set of Monte Carlo simulations, the results of which are summarized in Fig. \ref{fig:skymc}. 
In this experiment, we ran a targeted coincident detection search \cite{Abbott:2008zzb} 
with the H1L1V1 network in simulated Gaussian data with LIGO-I noise PSD. The noise background for each of the ten different sky positions, depicted in Fig. \ref{fig:networkSens} as white stars, was obtained. We also injected in each of those ten positions a non-spinning CBCNS source, with component masses $(2.5M_\odot,40.0M_\odot)$ and optimal oriention at a distance of 17Mpc, and ran the same search pipeline with the correct sky template and the incorrect one (with $20^\circ$ error) to assess the effect of that systematic error on the recovered SNR. 
These experiments output the SNRs of the injections at the different sky positions, with and without the sky-template error. 

Ideally, the same number of injections should have been made at each of the ten sky positions as in the study in Sec. \ref{varAcrossSky:single}, namely, 3000, but that requires an order of magnitude more computational resources. Instead, we use the SNR of the single optimally oriented injection at each sky position to infer the detection efficiency. We do so by noting that the detection efficiencies presented in the previous section are for injections distributed uniformly in distance. 
Thus, the detection efficiency at any sky-position can be approximated by comparing the SNR of the loudest (optimally oriented) injection with that of the loudest background trigger at that sky position, as follows: Consider $N+1$ injections of GW signals from the same type of CBCNS source at a single sky position, but uniformly distributed in distance. Then, the SNRs of these injections will be uniformly distributed such that their sorted list forms an arithmetic progression $\{ \rho_0, \rho_1, \rho_2,..., \rho_N\}$, where $\rho_{i+1} - \rho_i \equiv \Delta\rho$ is a constant for $0 \leq i \leq N$. Let $\rho_B$ be the loudest background SNR. If $\rho_0 < \rho_B < \rho_N$ and $\Delta\rho \to 0$, one will always find a $\rho_i$ that is equal to $\rho_B$, say, for $i = k$. Therefore, all the triggers in the range $[\rho_0, \rho_{k-1}]$ have SNRs below that of the loudest background trigger. One can then define the ratio
\bea
D = \frac{\rho_N - \rho_B} {\rho_N} = \frac{N\Delta\rho  - k\Delta\rho}{N\Delta\rho} = \frac{N - k}{N} \, ,
\eea 
where we took $\rho_0 = 0$. The final expression shows that $D$ is the ratio of the total number of injections louder than the loudest background to the total number of injections, which is just the detection efficiency. Thus, $D$ can be approximated by the first expression, i.e., by comparing the SNRs of the optimally oriented injection and the loudest background trigger. This is what we have plotted on the vertical axis in Fig. \ref{fig:skymc}. There the sky positions are indexed from one to ten on the horizontal axis.

A comment is in order about the choice of an {\em isotropic} grid of sky-position templates in the sky-patch search modes: A better method of distributing templates in the sky is to require that the maximum fractional drop in SNR due to the mismatch in the sky position of the signal and the template be constant \cite{Ajith:2009fz,Keppel:2012ye}. However, as shown in Figs. \ref{fig:fisher} and \ref{fig:networkSens} in the neighborhoods of all ten injection sky positions a template spacing of 4 square-degrees will contribute less than 10\% to the overall SNR drop.  which will be dominated by the systematic error in the sky-position of the injections. The latter causes a drop of about 100\% or more across much of the sky. Reducing the pixel size increases the number of pixels needed to cover the same sky-patch, thereby, increasing the computational cost (which is the number of floating-point operations required per unit time) of the search.

To summarize, the above studies show that the effect of a systematic error in sky position is to reduce the detection efficiency of a known-sky search by a significant amount. The detection efficiency is less affected when a sky-patch is used to search for the GW counterpart.


\section{Searching all-sky, all-time CBC triggers in LIGO's S5 data for GW candidates concurrent with GRBs}
\label{sec:poorsky}


In this section we develop the motivation for conducting a search of coincident all-sky, all-time \cite{iHopePaper} CBC triggers in LIGO's S5 data 
 \cite{Abbott:2009tt,Abbott:2009qj,s5Lowmass,s5Highmass}
for GW candidates concurrent with GRBs.
No GW detections were reported in CBC searches of any kind in LIGO's S5 data because no triggers had a low enough FAP. However, as explained in Sec. \ref{sec:orphanagprospects} the FAP of a blind search is higher than that of a targeted search, at the same SNR. Thus, an all-sky, all-time CBC trigger that is concurrent with a GRB can have a lower FAP in a targeted GW search and can lead to a GW detection.~\footnote{The other shortcoming of the all-sky, all-time coincident GW triggers is that their FAP is computed over a duration of data that spans weeks as opposed to several minutes ($\sim$2000 sec) around the time of the GRB.} Indeed, a fully coherent targeted search has been performed to look for GW signals coincident with GRBs with accurate sky-position information, as external triggers \cite{Harry:2010fr}. However, when the sky position is not accurately known and one needs to search in a large error box a coherent search becomes computationally more expensive than the targeted search. 
Therefore, here we focus on GRBs with poor sky-position information that were not used to trigger GW searches. If a GRB were found to be concurrent with a GW trigger that was not significant enough in the all-sky, all-time coincident search, it might still be interesting to invest the computational resources to perform a targeted fully coherent or hierarchical coherent search \cite{Bose:2011km} around that GRB time. This is because the concurrent GW trigger can gain in significance in the latter types of search owing to their lower noise background.

We next show why the choice of SNR thresholds in the coincident search allows for the possibility of finding interesting GW triggers concurrent with GRBs. We argue that it is possible for some of these triggers to have a low enough FAP in a targeted search to constitute a detection.
As discussed earlier, the detection threshold in a targeted search in H1L1 is a network SNR of 9.0, for a FAP of $10^{-4}$. If all detectors are equally sensitive to a source, then the SNR in each detector of a signal at the threshold of detection is 6.4, which is 9.0 divided by the square root of the number of detectors \cite{Helstrom,Pai:2000zt}.
During the fifth science run in LIGO (S5) three detectors were taking data. These were H1, H2, and L1.
A fourth detector, Virgo (V1), joined them in the last several months of S5, where it shared its data from the first Virgo Science Run (VSR1) with LIGO. Although the sensitivities of H2 and V1 were about half or worse than those of H1 and L1, triggers from all five detectors were analyzed for this GRB coincidence study.

Specifically, an H1L1 coincident candidate with a SNR of 6.4 in each of H1 and L1 will have a FAP of $10^{-4}$ or less in a targeted search. It is therefore important to enquire if any coincident GW candidates were found in an all-sky, all-time (i.e., un-triggered) search that were concurrent or near concurrent with a GRB \cite{Valeriu:2012}. Note that even if an H1L1 candidate is found with a FAP higher than $10^{-4}$ in the all-sly, all-time search, it can have a FAP of $10^{-4}$ or less in the tergeted search, i.e., when found concurrent with a GRB. With this in mind we examined H1L1 candidates found by the un-triggered low- and high-mass searches to check for coincidences with GRBs.

\begin{figure*}[tbh]
\centering
\includegraphics[width=12.8cm]{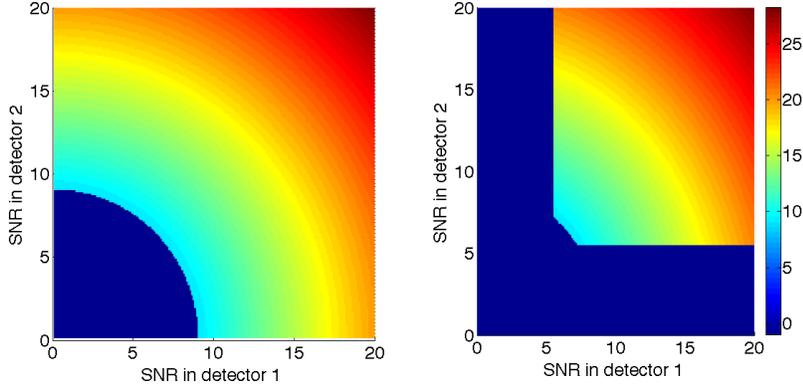}
\caption{The above plots of SNRs of triggers in two detectors show the region where a coincident search with individual detector SNR thresholds of $5.5$ will not find any triggers with a network combined SNR, shown in the colorbar, that is  greater than $9$ but a search with no individual detector thresholds will. Thresholding at an individual detector level rejects all coincidences that have a low SNR in one of the detectors.
The three-detector case is easy to understand from this study: If the SNRs of a signal are less than the threshold in two of the three detectors but greater than that in the third, the combined SNR can still have an appreciable value (with a low FAP). 
}
\label{fig:coincSNR_range}
\end{figure*}

Next consider what the inclusion of H2 does to a search. The weakest signal that can produce a triggered H1L1 candidate with a FAP of $10^{-4}$ (or H1L1 SNR of 9.0) will have a SNR of 6.4 in each of the two detectors, as mentioned above. 
As one increases the number of detectors, $N$, a network SNR that is lower by a factor of $N^{1/2}$ would have the same FAP, assuming that the detectors are similar in their noise and antenna profiles. Thus, a network SNR of $9.0/\sqrt{(3/2)} = 7.3$ in H1H2L1 would have a FAP of $10^{-4}$ if H2 had the same sensitivity as H1 and L1. Since H2 actually has a lower sensitivity, the H1H2L1 SNR at the same FAP is closer to 8.0. So the H2 SNR at which a signal can contribute to a H1H2L1 un-triggered coincident candidate at the same FAP is 2.7. This signal will have a SNR of 5.4 in H1 and L1, which is close to the SNR threshold placed on H1 and L1 in the un-triggered searches, namely, 5.5. 
(Note that the orientations of H1 and L1, while not identical, are very similar, as borne out by their responses displayed in Fig. \ref{fig:networkSens}.
Higher values of the SNR in H2 reduce the H1H2L1 FAP and lower values increase the FAP. So, between an H1 (and L1) SNR of 5.5 and 6.4,
contribution of H2's SNR helps keep the network FAP at or below $10^{-4}$. Therefore, it makes sense to examine lists of H1H2L1 triggers from the coincident low-mass and high-mass searches that have SNRs in H1, L1, and H2, greater than or equal to 5.5, 5.5, and 2.7, respectively. This was the main motivation behind the experiment reported in this section. As an aside, note that since the all-sky, all-time searches had a SNR threshold of 5.5 in all detectors, it makes sense to do targeted searches to detect, especially, those signals in H1H2L1 that would have an H2 SNR in the range $(2.7, 5.5)$. This point is illustrated in the plots in Fig. \ref{fig:coincSNR_range}. In the left-hand side plot we show the case where no SNR threshold is set in any detector. Thus, every trigger from a detector is retained and the thresholding (which is at a network SNR of $9.0$ in this example) is done on the combined SNR, which is just the square-root of the sum of squares of the individual detector SNRs. The right-hand side plot in the same figure depicts the case where triggers from a detector are retained in the coincident network analysis only when their SNR is above 5.5.

This also provides an important motivation for the hierarchical coherent search \cite{Bose:2011km}. The idea behind the hierarchical coherent search is that for all coincident candidates in H1 or L1, which have a SNR above 5.5 in each detector, include the data from H2 in the H1H2L1 analysis, as long as H2 was active, even if H2 did not contribute a threshold-crossing trigger to a candidate found by the coincident pipeline. In other words, the hierarchical coherent pipeline is expected to have the most impact on the detectability of those signals that have a SNR of $(5.5, 6.4)$. While this is a narrow range, note that the first detections will likely happen at SNRs close to the upper limit of that range. (As shown in Ref. \cite{Schutz:2011tw}, the most likely detected SNR is 1.26 times greater than the SNR threshold. For a threshold of 5.5, it is 6.9.)
This keeps the computational costs at a minimum, while taking full advantage of the signal's phase coherence across the network detectors in improving its detectability. Also when H1 (L1) is not active, H2L1 (H1H2) triggers can be detections. In practice, the hierarchical coherent pipeline searches with SNR thresholds of $\{\rho_{\text{th}}^{\text{H1}}=4.0,\rho_{\text{th}}^{\text{L1}}=4.0,\rho_{\text{th}}^{\text{H2}}=3.0\}$. Ideally, the H2 threshold should be lower, at 2.7, as discussed above, but it is still in the range of interesting SNR values.
Note, however, that we do not analyze H1H2 triggers because our background estimation of those triggers is not robust owing to cross-correlated noise at the same site.
%

GRB alerts issued by gamma-ray observatories can have large sky-position errors.  It is counterproductive to search using the known-sky mode for GWs from such GRBs, as discussed in Sec. \ref{sec:poorlyLocSource}.
Indeed, fully coherent pipelines are in use \cite{Valeriu:2012} to search over sky-patches with multiple sky positions for GWs from these objects.
Typically, a GRB alert provides the source sky-position along with an error box or error radius associated with it. Those errors for some GRBs are tabulated in Tables \ref{tab:fermiErrorBoxes} and \ref{tab:IPNErrorBoxes}.

\subsubsection{List of SGRBs detected with large sky-position errors by the Fermi gamma-ray satellite}
\label{subsubsec:fermigrbs}

While the gamma-ray burst monitor (GBM) on-board the Fermi satellite has an excellent sky coverage, its sky resolution can be poor on occasion. In the first two years of Fermi's observation time, $491$ GRBs were detected by GBM \cite{Fermi:2012}. Of these, we found that $40$ GRBs with $T_{90} \leq 2.0$ sec have a large sky-position uncertainty (i.e., the error radius is greater than $10^\circ$). (Here, $T_{90}$ is the duration over which a burst emits between $5\%$ to $95\%$ of its total measured photon counts after the background has been subtracted.) Given the large error radii of these GRBs in Table \ref{tab:fermiErrorBoxes}, we expect a significant drop in detection efficiency 
of these GRBs if one employs a known-sky targeted search. Therefore, GRBs like the ones listed in Table \ref{tab:fermiErrorBoxes} make good candidates for GW searches of the sky-patch or all-sky type. 


\begin{table}[tbh]
\begin{center}
  \begin{tabular}{ |c | c |c | c | c | c |}
    \hline
    GRB name & GRB time (UTC) & RA ($^{\circ}$) & Dec ($^{\circ}$) & Error radius($^{\circ}$) & $T_{90}$ sec \\ \hline
    \hline

	GRB080806A   &   14:01:11.2038 &  94.6  &  57.8  &  13.6   &   2.304 $\pm$ 0.453	\\ \hline
	GRB080828B    &  04:32:11.2646 &  221.3 &  -12.3  & 16.9   &   3.008 $\pm$ 3.329\\ \hline
	GRB080831A    &  01:16:14.7521 &  211.2 &  -51.7  & 11.5 &   0.576 $\pm$ 1.168\\ \hline
	GRB080919B    &  18:57:35.1052 &  219.5  &  44.4  &  18.1  &   0.512 $\pm$ 0.405\\ \hline
	GRB081105B   &   14:43:51.2874 &  215.8  &  38.7  &  11.4  &   1.280 $\pm$ 1.368\\ \hline
	GRB081113A   &   05:31:32.8973  & 170.3  &  56.3  &  12.4  &   0.576 $\pm$ 1.350\\ \hline
	GRB081115A   &   21:22:28.1472  & 190.6  &  63.3   & 15.1  &   0.320 $\pm$ 0.653\\ \hline
	GRB081119A   &   04:25:27.0590  & 346.5  & 30.0   &   15.2  &   0.320 $\pm$ 0.680\\ \hline
	GRB081122B   &   14:43:26.2316  & 151.4  & -2.1  &  11.2   &   0.192 $\pm$ 0.091\\ \hline
	GRB081204B   &   12:24:25.7930  & 150.8  & 30.5  &  10.2  &   0.192 $\pm$ 0.286\\ \hline
	GRB081206B   &   14:29:30.6929  & 353.3 &  -31.9 &  12.6  &  7.936 $\pm$ 4.382\\ \hline
	GRB081213A   &   04:09:41.6360 &  12.9  &  -33.9  & 13.2  &   0.256 $\pm$ 0.286\\ \hline
	GRB081229B    &  16:12:17.3756 &  310  &   22.8  &  20.7  &   Data problems precluded duration analysis\\ \hline
	GRB090120A   &   15:02:22.7594 &  38.1 &   -72.2  & 11.2 &   1.856 $\pm$ 0.181\\ \hline
	GRB090126C    &  05:52:33.7347 &  224.9 &  41.2  &  11.1  &   0.960 $\pm$ 0.231\\ \hline
	GRB090304A    &  05:10:48.1569 &  195.9 &  -73.4  & 12.3   &   2.816 $\pm$ 0.923\\ \hline
	GRB090320C    &  01:05:10.5272  & 108.3  & -43.3  & 17.9  &  2.368 $\pm$ 0.272\\ \hline
	GRB090405A    &  15:54:41.3408  & 221.9 &  -9.2  &  10.4  &   0.448 $\pm$ 1.498\\ \hline
	GRB090412A   &   01:28:05.2531 &  1.3  &   -51.9  &  10.6  &   0.896 $\pm$ 0.264\\ \hline
	GRB090418C   &   19:35:24.9183 &  262.8  & -28.2  & 14.4  &   0.320 $\pm$ 0.405\\ \hline
	GRB090427B   &  15:27:00.8558 &  210  &   -45.7 &  11.8   &  1.024 $\pm$ 0.362\\ \hline
	GRB090616A   &   03:45:42.5323 &  103.1 &  -3.7  &  10.3  &   1.152 $\pm$ 1.168\\ \hline
	GRB091006A   &  08:38:46.9285  & 243.1 &  -31  &   12.8  &   0.192 $\pm$ 0.091\\ \hline
	GRB091015B   &   03:05:42.9372  & 316.1 &  -49.5  & 12.7  &  3.840 $\pm$ 0.590\\ \hline
	GRB091018B    &  22:58:20.6027 &  321.8 &  -23.1 &  13.1  &   0.192 $\pm$ 0.286\\ \hline
	GRB091019A    &  18:00:40.8812 &  226  &   80.3  &  12.8  &    0.208 $\pm$ 0.172\\ \hline
	GRB091126B    &  09:19:48.5326 &  47.4  &  31.5  &  14.3   &    Too weak to measure duration; visual duration is 0.025 sec\\ \hline
	GRB091224A   &   08:57:36.5574  & 331.2 &  18.3  &  15.6  &   0.768 $\pm$ 0.231\\ \hline
	GRB100101A    &  00:39:49.3358 &  307.3 &  -27  &   17.4  &   2.816 $\pm$ 0.320\\ \hline
	GRB100126A    &  11:03:05.1248  & 338.4  & -18.7  & 18.3  &  10.624 $\pm$ 12.673\\ \hline
	GRB100204C    &  20:36:03.7668  & 91.3  &  -20.9  & 16.6   &   1.920 $\pm$ 2.375\\ \hline
	GRB100208A   &   09:15:33.9419 &  260.2 &  27.5  &  29.3  &   0.192 $\pm$ 0.264\\ \hline
	GRB100326A   &   07:03:05.5029 &  131.2 &  -28.2  & 12.6  &  5.632 $\pm$ 2.064\\ \hline
	GRB100411A    &  12:22:57.3442  & 210.6  & 47.9  &  31.6  &   0.512 $\pm$ 0.231\\ \hline
	GRB100516A   &   08:50:41.0629 &  274.4  & -8.2  &  18.4  &   2.112 $\pm$ 1.134\\ \hline
	GRB100516B   &   09:30:38.3170  & 297.7  & 18.7  &  13.7  &   0.640 $\pm$ 0.487\\ \hline
	GRB100530A   &   17:41:51.2263 &  289.7  & 31    &  11.6  &  3.328 $\pm$ 0.810\\ \hline
	GRB100616A   &   18:32:32.8957 &  342.9  & 3.1  &   45.7  &   0.192 $\pm$ 0.143\\ \hline
	GRB100621C   &  12:42:16.4305  & 160.9  & 14.7  &  11.4  &   1.024 $\pm$ 0.202\\ \hline
	GRB100706A   &  16:38:18.9243 &  255.2  & 46.9   & 12.2   &   0.128 $\pm$ 0.143\\ \hline

    \hline
  \end{tabular}
\caption{Fermi SGRBs with large sky-position uncertainties that were concurrent with LIGO's S5 run.}
\label{tab:fermiErrorBoxes}
\end{center}
\end{table}

\subsubsection{List of short GRBs, detected with large sky position error, by IPN satellites}
\label{subsubsec:ipngrbs}

IPN is a network of spatially separated gamma-ray burst detectors on several satellites. It uses delays in the time of arrival of gamma-ray signals at these detectors to triangulate the GRB sky positions.
IPN has been detecting GRBs since the $1970$s. At its peak it involved $10$ satellites located at various distances from the Sun, between the orbits of Venus and Mars. Currently, IPN uses four spacecraft, namely, the NASA/ESA Ulysses mission, WIND, HETE-II and $2001$ Mars Odyssey.
The sky position determination by triangulation across the network can be erroneous when a smaller number of spacecraft detect the signals. The following factors contribute to IPN sky position errors:

\begin{itemize}

\item{ 
Inaccuracy in the synchronization of clocks in the gamma-ray detectors and in the calibration of those clocks contribute to errors in timing the arrival of signals and, hence, the GRB sky position.}
\item{ Uncertainty in the spatial location of the spacecraft leads to errors in the lengths of the baselines and, therefore, errors in estimating the signal time-delays across the baselines.}
\item{ The number of satellites detecting a particular GRB can be less than $3$ in some cases. Two detectors form a single baseline, which can do no better than localize the GRB on a ring in the sky. The two types of errors discussed above broaden that ring to an annulus.}
\end{itemize}

Some IPN GRBs can have very large error boxes, often larger than $100$ square degrees. A list of a subset of such short GRBs that occurred during LIGO's S5 run is given in Table \ref{tab:IPNErrorBoxes}. These GRBs were selected for archival look-up only \cite{Valeriu:2012}, where triggers from the coincident all-sky, all-time CBC search pipeline, as studied in Refs. \cite{iHopePaper,s5Highmass,Abbott:2009tt}, were checked for GW candidates concurrent with the GRBs listed in that table. After all, the computational requirement posed by a fully coherent GW search in a wide sky region are very high \cite{Pai:2000zt,Valeriu:2012}. 
One can however, analyze these GRBs using the coherent hierarchical search pipeline, accommodating for the sky position uncertainties, as proposed above. This can provide us a better chance of detecting GW from these sources. 

\begin{table}[tbh]
\begin{center}
  \begin{tabular}{ |c | c |c |c| }
    \hline
    GRB name & GPS time (sec) & Error box (square degrees) & Duration of GRB (sec) \\ \hline
    \hline
    GRB061001 & 843772482 & $\sim$2000 & 1.00 \\ \hline
    GRB060601B & 833183754 & $\sim$600 & 0.50 \\ \hline
    GRB070910 & 873480823 & $>$200 & 0.38 \\ \hline
    GRB070413 & 860531889 & 350 & 0.19 \\ \hline
    GRB070203 & 854579218 & $>$2000 & 0.69 \\ \hline
    GRB061014 & 844841836 & $>$3000 & 1.50 \\ \hline
    GRB060916 & 842452428 & $>$3000 & 0.13 \\ \hline

    \hline
  \end{tabular}
\caption{Short IPN GRBs with large sky position errors. 
}
\label{tab:IPNErrorBoxes}
\end{center}
\end{table}


\section{Improving the performance of targeted GW searches}
\label{sec:noskyerror}

If the sky position of a GW source is accurately known from EM observations, such as of an associated SGRB or afterglow, then one might naively suspect that the search for its GW signal should employ only that single position. A search that uses multiple sky-position ``templates'' is computationally more expensive. 
More importantly it will also incur a higher false-alarm probability, as we estimated below Eq. (\ref{eq:skyfar}).
In spite of these drawbacks, a case can be made to also search away 
from the true sky-position provided it increases the detection efficiency, which, at a given FAP, is the number of signals detected louder than a background trigger at that FAP.
Such an anomaly can occur, e.g., when there is a mismatch between the signal and the template owing to (a) inaccurate modeling of the signal \cite{Bose:2012vb} or (b) detector calibration errors \cite{allencalib,bosecalib}.

Detectors have calibration errors of about 5-10\% in the strain amplitude and up to several degrees in the strain phase.
The error can vary from one detector to another and in time. While the temporal variation is expected to be slow and minimally affect targeted GW detectability of transients, the detector dependence can affect the signal's amplitude and time-delay differently in the network detectors, thereby, partially mimicking an error in the sky-position. 
Note that calibration errors have been shown to affect estimation of signal parameters, such as CBC masses, strongly (in fact, linearly) \cite{allencalib,bosecalib}. 
As we show below, the covariance of the error in sky-position angles with that in other source parameters suggests that searching in a wider patch can sometimes mitigate the adverse effect on signal SNR and detectability.

Furthermore, the BH spin can be very high in SGRB sources that include a black hole. However, GW template banks for NSBH systems with high BH spin have not been applied in GW searches yet. Searching those systems with inaccurate templates, either with non-spinning ones, as is done in Refs. \cite{Abadie:2010uf,s6grbpaper}, or with effective template families
can severely diminish their detectability.
These searches can also benefit from a somewhat expanded search in the sky, again owing to the covariance of the sky position angles with other source parameters, even if the sky-position is known accurately.

\subsection{Targeted GW search over a finite sky patch even for accurately known GRB sky position}
\label{subsec:paramErrCov}

 A mismatch between a signal and a template can lead to a drop in the SNR and, therefore, affect the detection probability. However, for some types of errors that probability can be partially salvaged by allowing the template sky-position to be different from the true one. 
Essentially, if there exists a non-vanishing covariance between the errors in sky-position and other CBC parameters, then it can be exploited to mitigate the SNR loss. The nine parameters that characterize the non-spinning CBC signals are the total mass $M$, the symmetrized mass-ratio $\eta$, the sky-position angles $(\theta,\phi)$, the polarization angle $\psi$, the orbital inclination angle $\iota$, the luminosity distance $d_L$, the initial (or some reference) phase $\varphi_0$, the time of arrival (or some reference time) $t_0$, and the overall signal amplitude $\cA$. We group these as components of the parameter vector, 
$\mbox{\boldmath $\vt$} \equiv \{\cA, t_0, \varphi_0, M, \eta, \theta,\phi,\psi,\iota\}$.
Owing to noise, their estimates, $\hat{\mbox{\boldmath $\vt$}}$, may differ from the true values, i.e., $\bvthat = \bvt + \Delta  \bvt$, 
where $\Delta \mbox{$\vt^a$}$ is the random error in estimating the parameter 
$\mbox{$\vt^a$}$. The magnitude of these errors can be estimated from
the elements of the variance-covariance matrix,
$g^{\mu\nu}=\ \overline{\Delta \vt^\mu\,\Delta \vt^\nu}$ \cite{Helstrom}.
The mismatch between a template and a signal is
\be
{\cal M}\left(\bvt^\sigma\right) = g_{\mu\nu} \Delta \bvt^\mu \Delta \bvt^\nu \,,
\ee
where we have used the Einstein summation convention for the repeated indices $\mu$ and $\nu$. For simplicity, consider a template that matches a signal in all its parameters except $M$ and the two sky-position angles. Assume that an observer has no control on changing $\Delta M$ but can vary the template sky-position angles. Then the above mismatch ${\cal M}$ is minimum when 
\bea
\Delta {\theta} &=& \Delta \tilde{\theta} \equiv \frac{C_{M\theta}}{C_{MM}} \Delta M \,,\no\\
\Delta {\phi} &=& \Delta \tilde{\phi} \equiv \frac{C_{M\phi}}{C_{MM}} \Delta M \,,
\eea
where a tilde denotes the parameter value that minimizes the mismatch and $C_{\mu\nu}$ is the cofactor of the metric $g_{\mu\nu}$.
The minimized mismatch is
\be\label{eq:minmis}
{\cal M}\left(\bvt^\sigma\right) \Big|_{\Delta \tilde{\theta},\Delta \tilde{\phi}} = \frac{g({\bvt}^\sigma)}{C_{MM}({\bvt}^\sigma)} \Delta M^2 \,,
\ee
where $g$ is the determinant of $g_{\mu\nu}$. The above expression can be vanishingly small, even for a finite $\Delta M$, if $g({\bvt}^\sigma)$ is so. The same idea can work over a larger parameter space that includes $\eta$, spin parameters (for spinning waveforms), and calibration errors. Indeed, in Ref. \cite{Bose:2008ix} it was demonstrated that allowing $\eta$ to exceed somewhat beyond its physically permitted upper limit of 0.25 (which is its value for binaries with mass-ratio of unity) in a mass template bank can improve the detection efficiency of a CBC search.


In CBC searches, one typically uses a bank of templates, derived from a waveform model, that are discretely spaced in the template parameters. Searches in LIGO and Virgo data have used template banks in the component masses that keep the separation of templates close enough to suffer at most a 3\% loss in the SNR in any detector. The simulations studied in this paper have used the same type of mass template banks. As suggested by Eq. (\ref{eq:minmis}) above and proved in signal injection studies below, that SNR drop can be mitigated by using a sky-patch instead of a single sky position in targeted GW searches.

\subsection{Targeted search for GW sources using inaccurate templates: Masses}
\label{subsec:inaccurateMasses}

\begin{figure}[tbh]
\centering
\includegraphics[width=5.8cm]{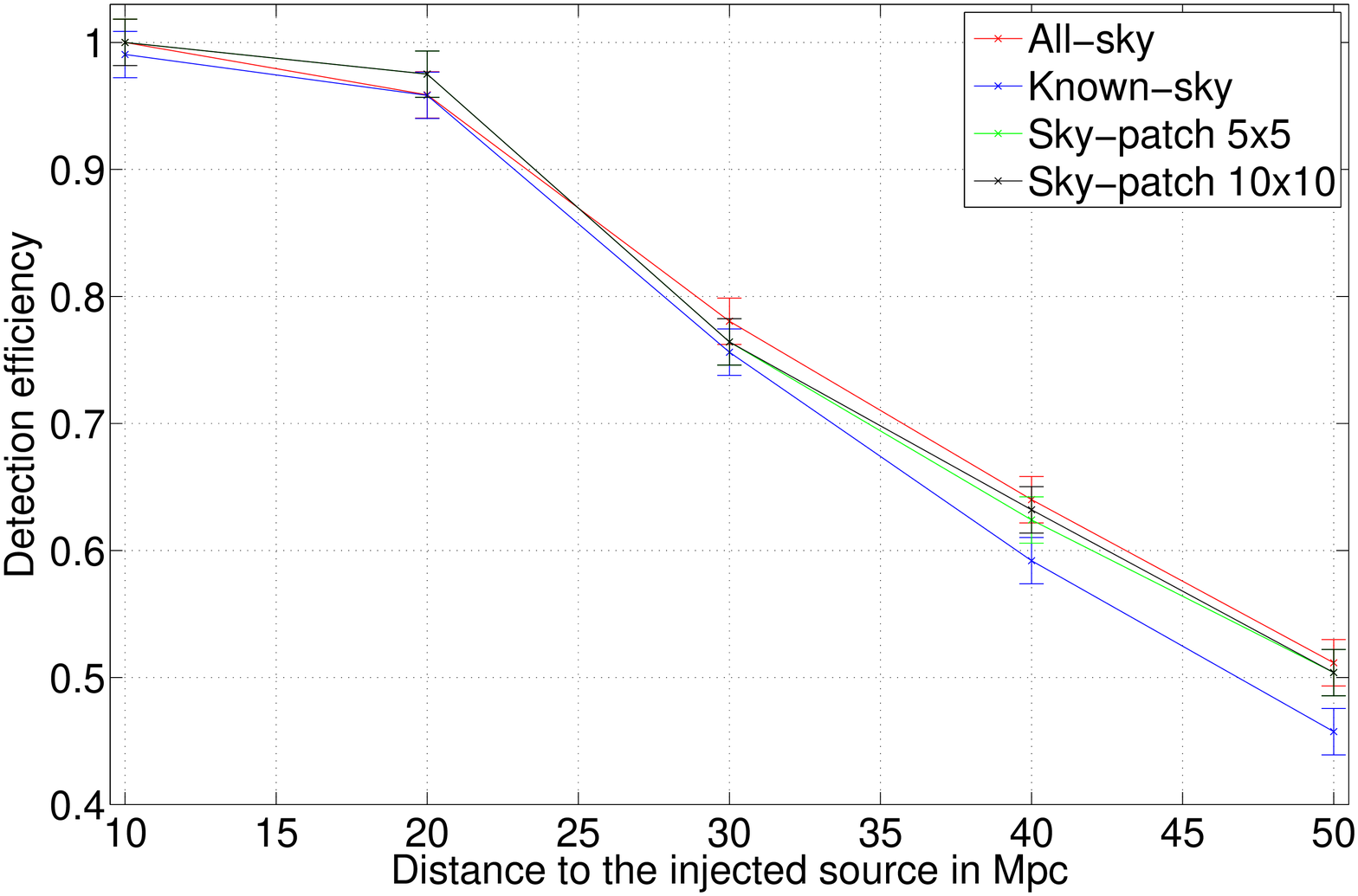}
\includegraphics[width=5.8cm]{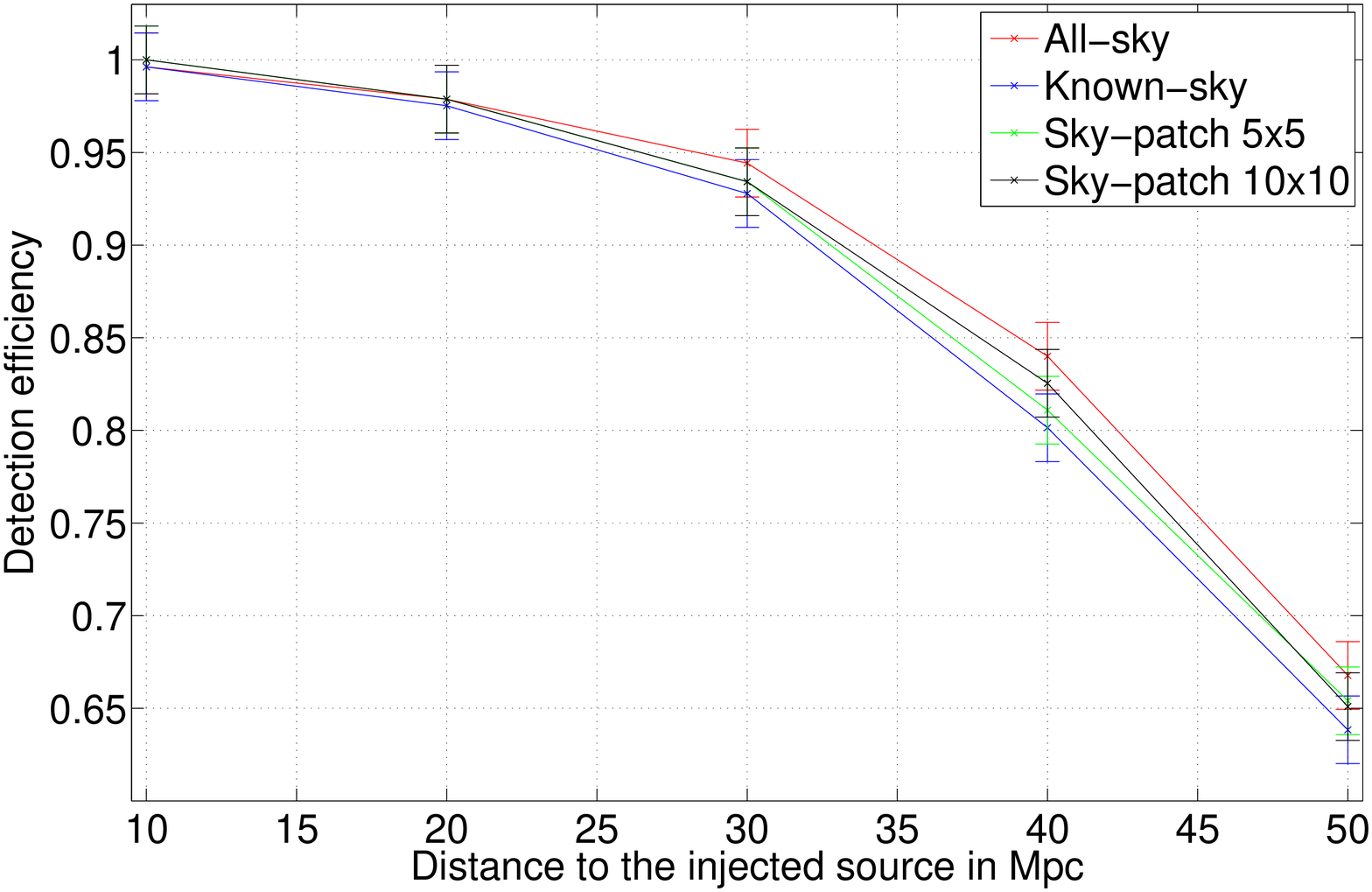}
\includegraphics[width=5.8cm]{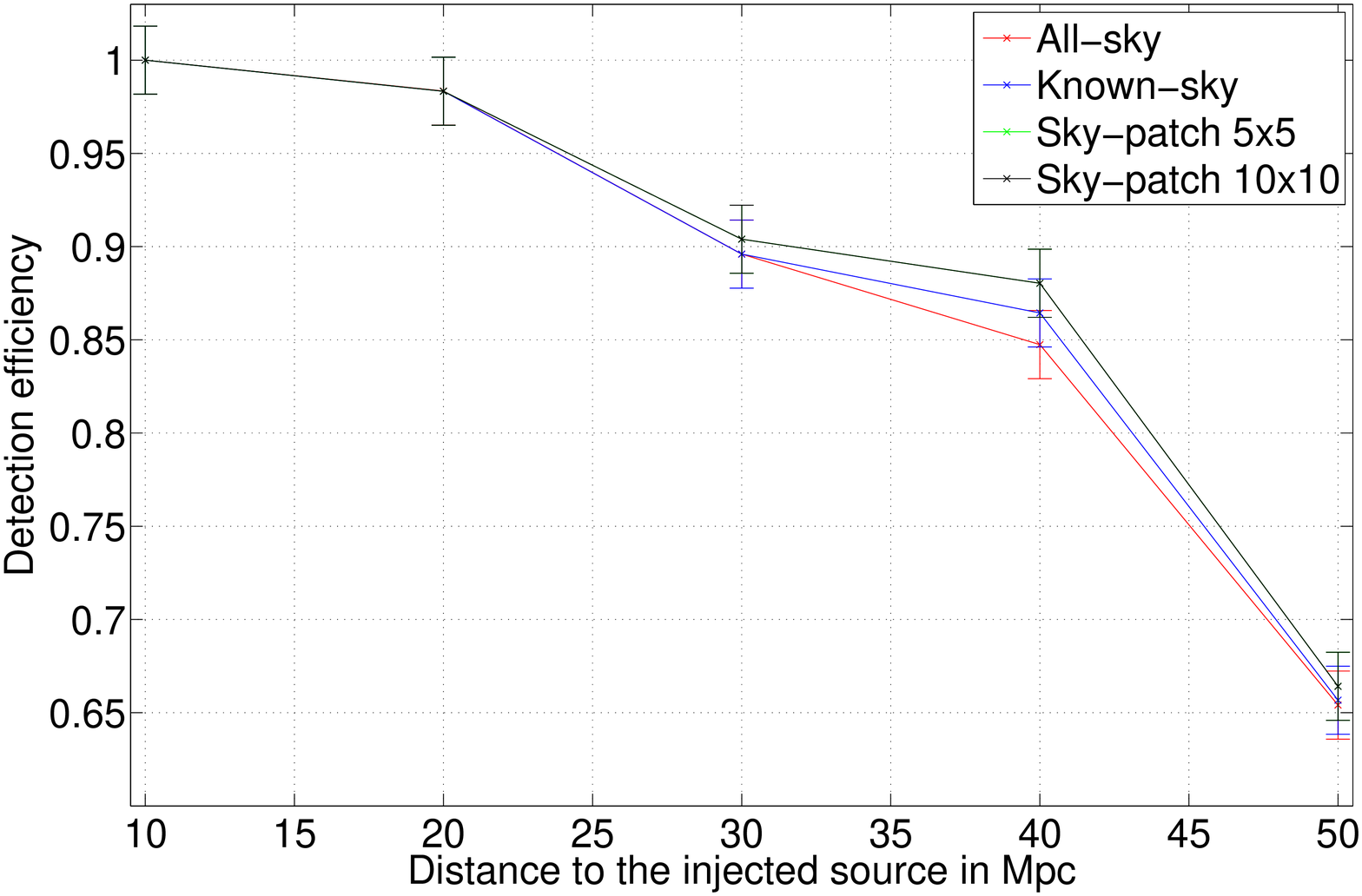}
\caption{Detection efficiency comparison (from left to right) for the low chirp-mass, medium chirp-mass and high chirp-mass bins for four sky modes. Unlike in Fig. \ref{fig:sysErrorComp} there is no sky-position error here. }
\label{fig:paramErrCov}
\end{figure}

To test the implications of the analytic calculations of Sec. \ref{subsec:paramErrCov}, we carried out a Monte Carlo simulation 
with $3000$ non-spinning injections, identical to the ones used for the systematic sky-position error study in Sec. \ref{sec:poorlyLocSource}. 
We conducted four types of sky-position template searches, namely, one in the known-sky search mode, two with different sky-patch sizes, and one in the all-sky search mode. 
The detection efficiencies of these searches are presented in Fig. \ref{fig:paramErrCov}. Note that in the low and medium chirp-mass bins, the known-sky detection efficiency performs worse than that of other sky modes despite the advantage it has of having the lowest FAP of all modes. Parameter error covariance helps the detection efficiency of the other search modes.
On the other hand, in the high chirp-mass bin the detection efficiency of the known-sky is better than that of the all-sky mode. Here, the higher FAP of the all-sky mode dominates over any gains arising from Eq. (\ref{eq:minmis}).
Nevertheless, the two sky-patch modes still outperform the known-sky mode in this mass bin as well, by as much as $5\%$. Therefore, we conclude that the best performance can be expected for a search that employs a sky-patch of the optimal size. More Monte Carlo simulation studies are needed to determine what that size is in different sections of the space of component masses \cite{ShaonOptimalPatch}.







\begin{figure}[tbh]
\centering
\includegraphics[width=5.8cm]{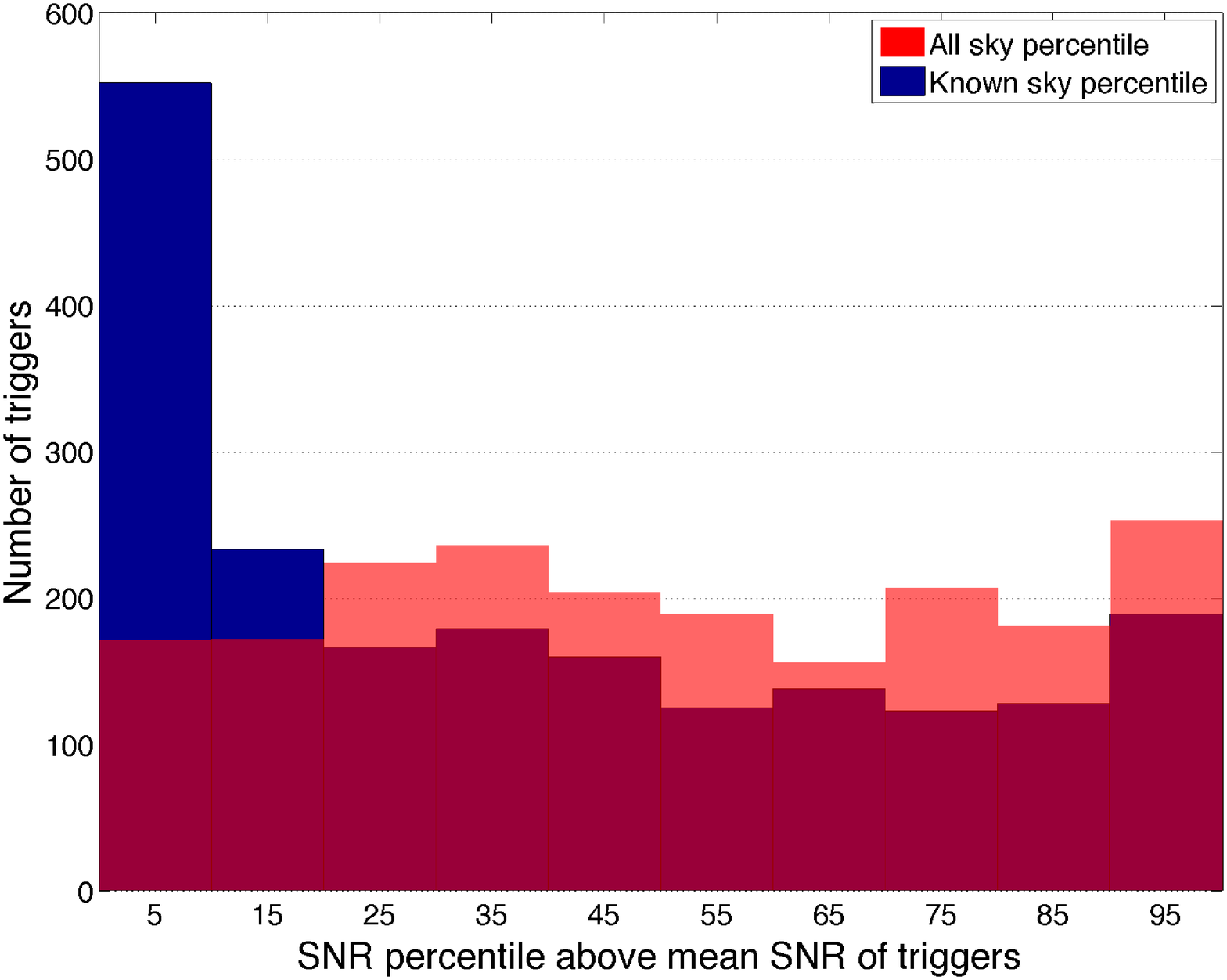}
\includegraphics[width=5.8cm]{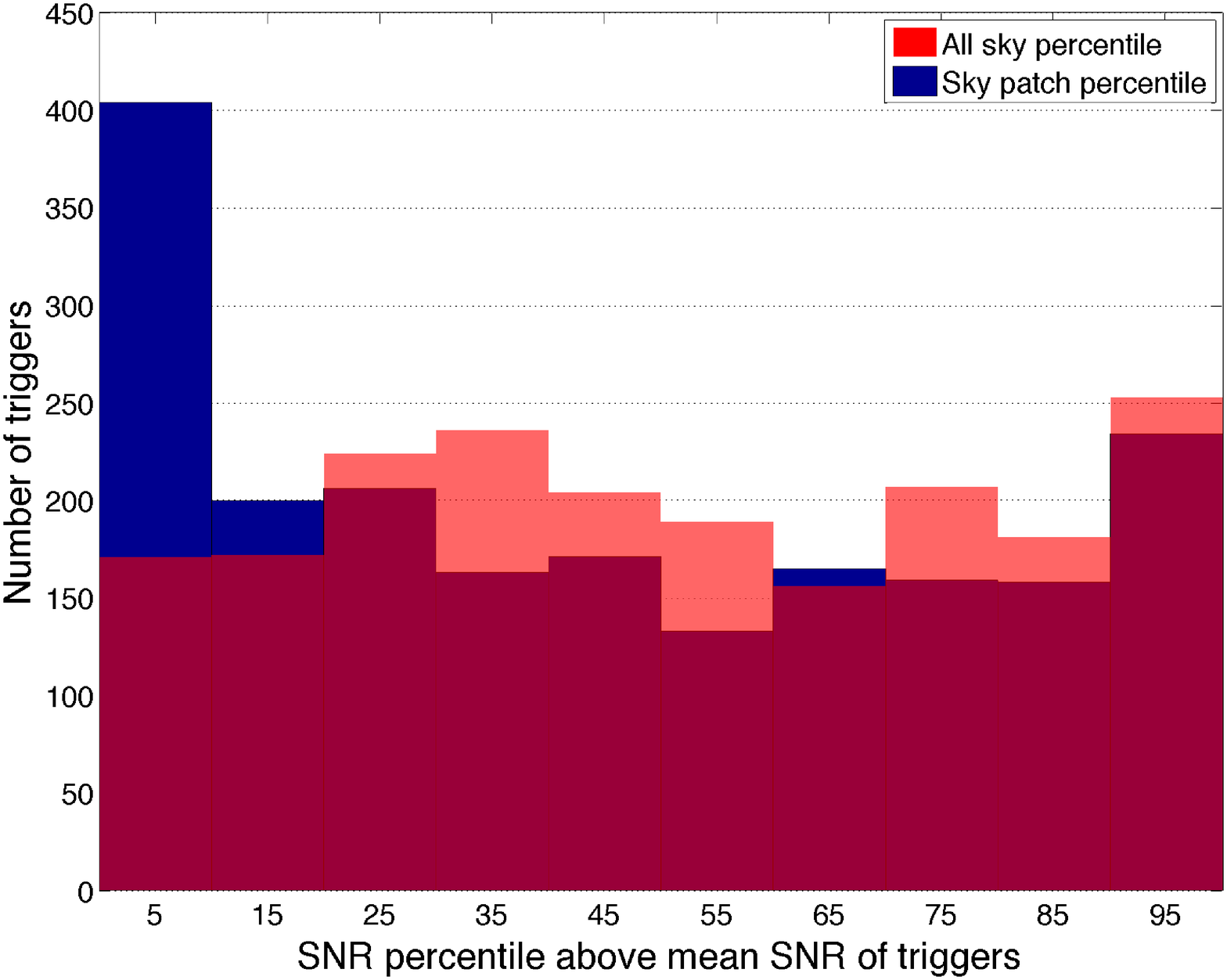}
\includegraphics[width=5.8cm]{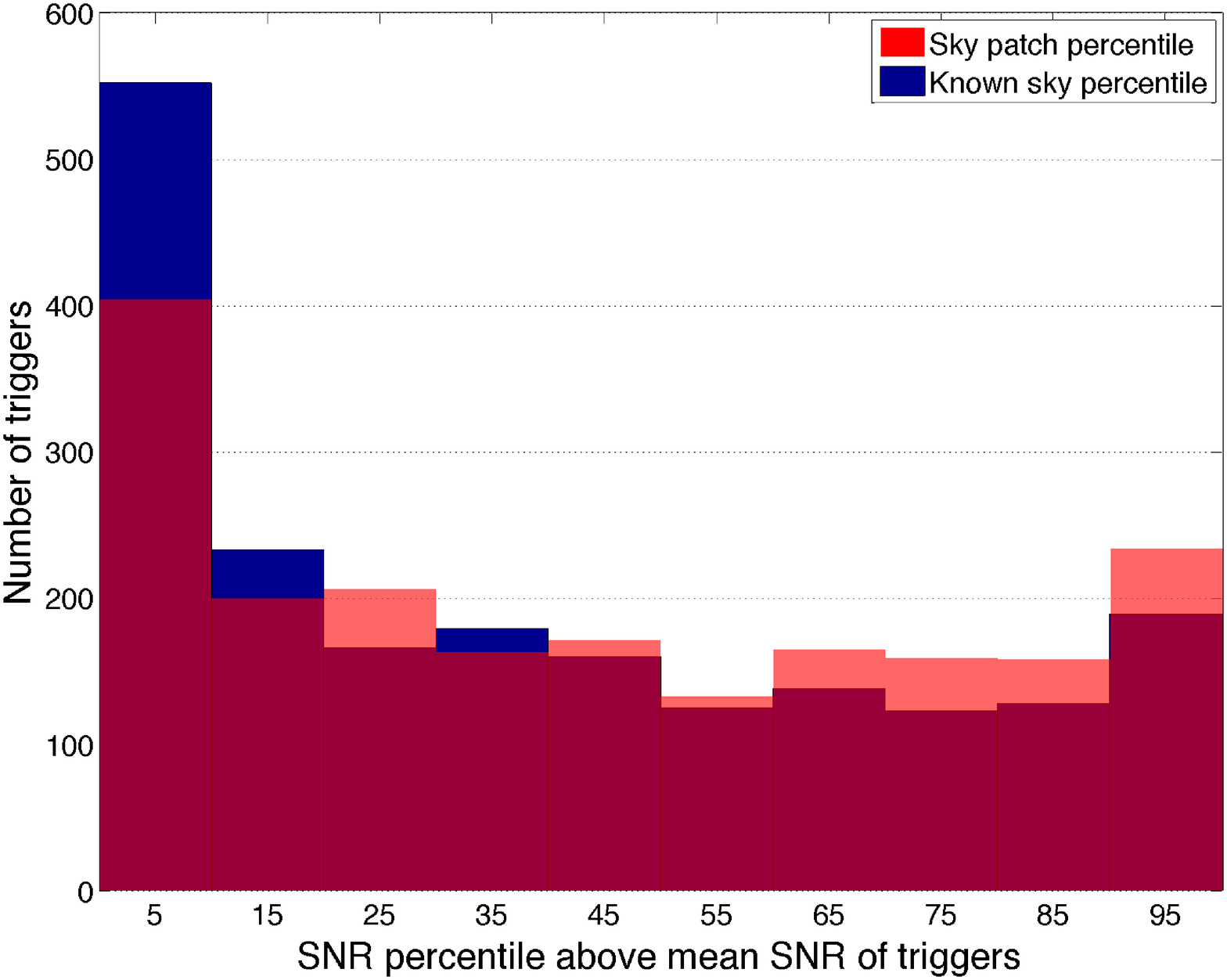}
\includegraphics[width=5.8cm]{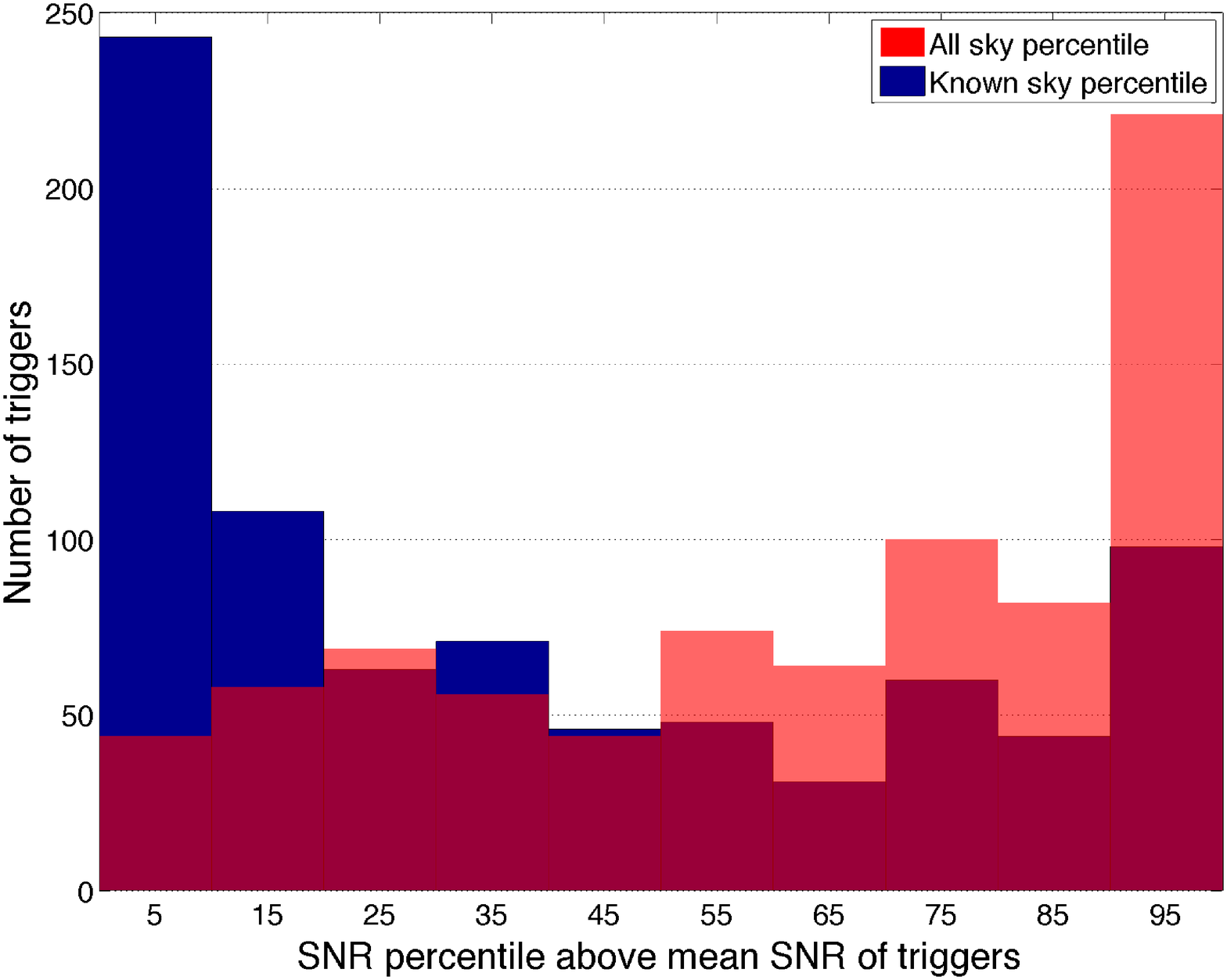}
\includegraphics[width=5.8cm]{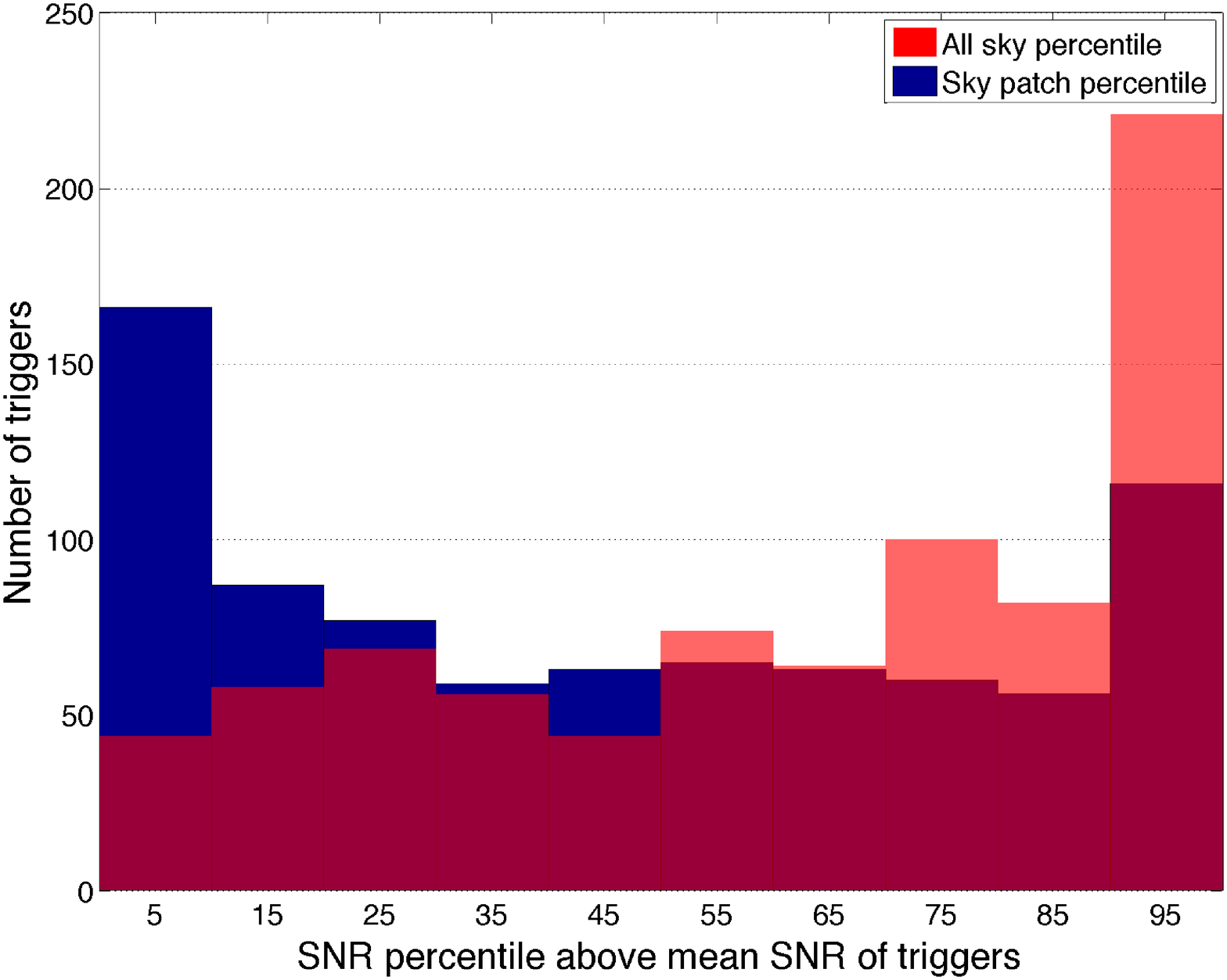}
\includegraphics[width=5.8cm]{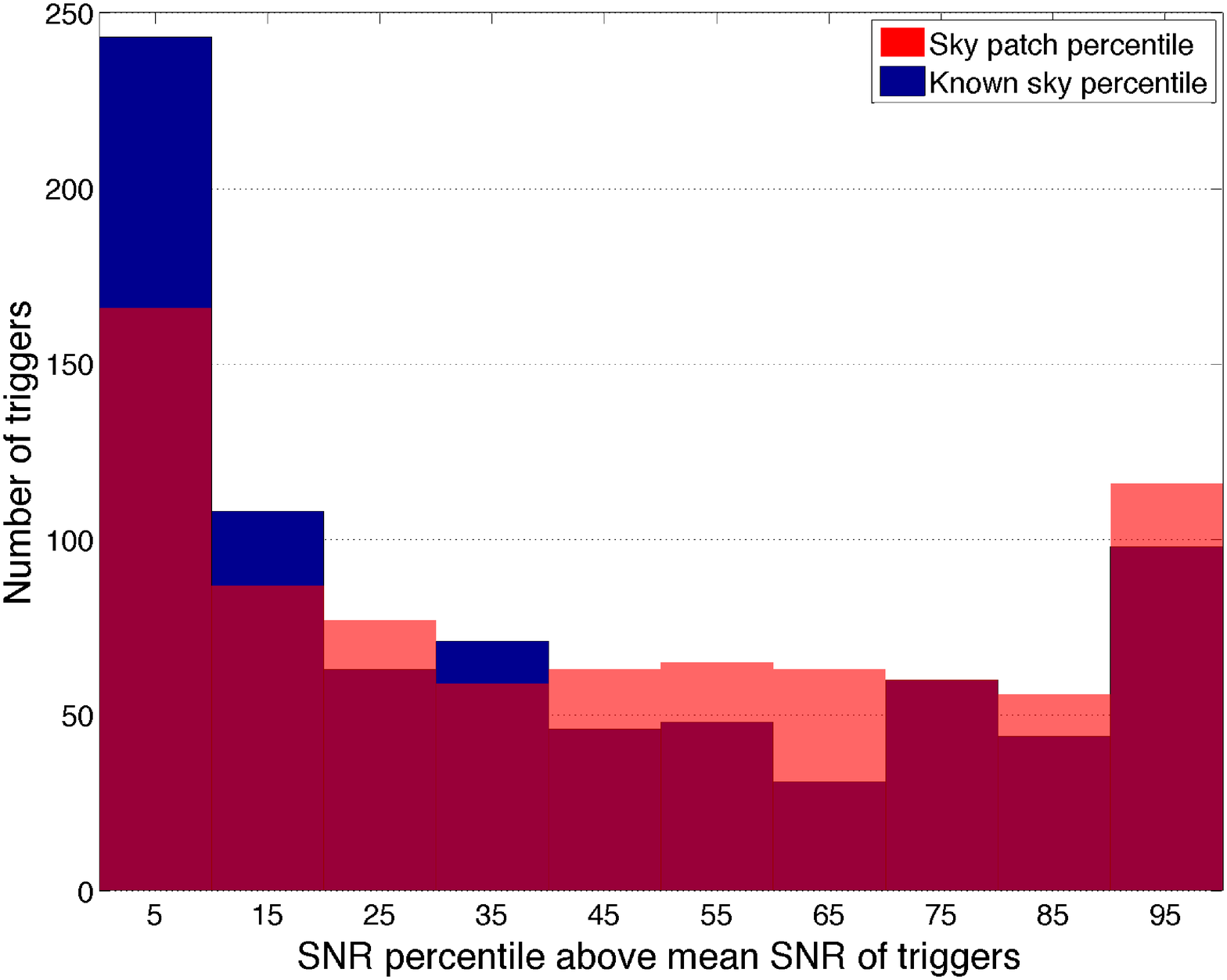}
\caption{
In these plots we study the combined distribution of the SNRs of the 3000 injections from all three sky-mode searches. For every injection we compute the mean of its network SNRs found by the known-sky, sky-patch, and all-sky modes of the search. We next compile the list of the 3000 mean SNRs. This is the reference list with which we compare the SNRs of injections from any of the search modes. Pairwise comparisons of distributions of injection SNRs from the different sky-mode searches are shown in these plots. In the first plot we find that 253 of the injections had all-sky SNRs greater than 90\% of the mean SNRs. The corresponding number of injections (189) was smaller for known-sky SNRs. At the other end, more known-sky SNRs are less loud than all-sky SNRs: While 552 injections had known-sky SNRs weaker than 90\% of the mean SNRs, the corresponding number of injections was 171 for all-sky SNRs. One finds qualitatively similar improvement in the SNR values of the same injections in the sky-patch mode compared to the known-sky mode of the search.
}
\label{fig:performance_mean_snr}
\end{figure}

To isolate the effect of parameter error covariance on improving signal-template match, we study the combined distribution of the SNRs of the injection triggers in Fig. \ref{fig:performance_mean_snr} from all three sky-mode searches, without any reference to the distribution of the SNRs of background triggers. There, for every injection the mean of its network SNRs found by the known-sky, sky-patch, and all-sky modes of the search was computed. 
Next a list of these 3000 mean SNRs was complied. This is the reference list we compare the SNR distribution of any of the search modes with in Fig. \ref{fig:performance_mean_snr}.
In the first plot we find that 253 of the injections had all-sky SNRs greater than 90\% of the mean SNRs. The corresponding number of injections (189) was smaller for known-sky SNRs. At the other end of the same plot more known-sky SNRs are less loud than all-sky SNRs: While 552 injections had known-sky SNRs weaker than 90\% of the mean SNRs, the corresponding number of injections was 171 for all-sky SNRs. One finds qualitatively similar improvement in the SNR values of the same injections in the sky-patch mode compared to the known-sky mode of the search.

\subsection{Targeted search for GW sources using inaccurate templates: Spin}
\label{subsec:spin}

The CBCNS progenitor of a SGRB can have rapidly spinning components. If it has a highly spinning black hole, the companion neutron star can inspiral closer to it before plunging into it because its LSO is closer than that of a slowly spinning black hole of the same mass. This increases the probability of the neutron star to be ripped apart by the black hole's tidal forces as the two spiral closer to each other. That, in turn, improves the chances of creating an accretion disk with sufficient mass for the GRB engine to fire (see Sec. \ref{fgammaCalc}). This suggests that among CBCNS systems associated with SGRBs there will exist an astrophysical bias toward those with a highly spinning black hole. Current targeted CBC searches do not use spinning gravitational-wave templates, although significant effort is being invested to include them. Matched filtering with non-spinning templates on data that have spinning signals introduces systematic errors due to waveform mismatch. Such a systematic error leads to reduced SNRs
just like the error in component masses does, 
as discussed in Sec. \ref{subsec:inaccurateMasses}. 
If we enable searching in multiple sky positions around the external trigger's sky position even when its position is known accurately, its GW signal can often be found with a larger SNR from a different sky position. (Note that signal detection rather than accurate parameter estimation is the primary goal here.)

\begin{figure*}[tbh]
\centering
\includegraphics[width=16.5cm]{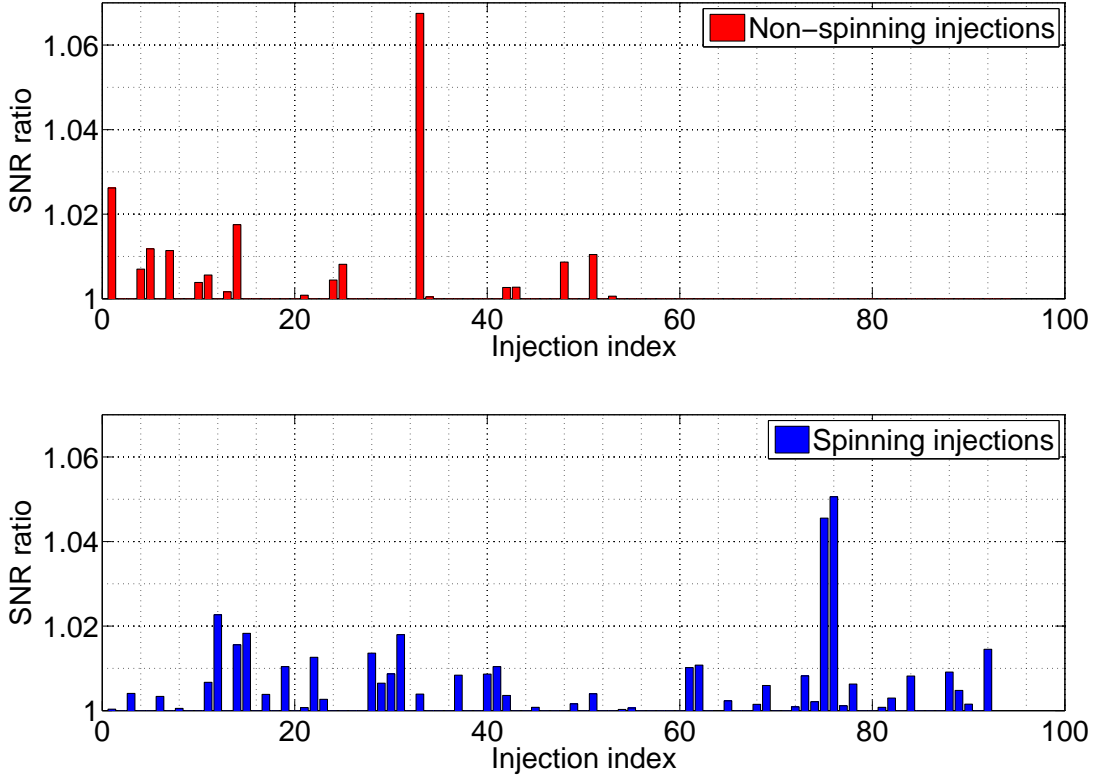}
\caption{Plot of ratio of SNR of the injection trigger when the search was conducted over a patch of $21$ points to that of the SNR of the same injection found in a search conducted in one point in the sky. The top plot with red bars is for the case of non-spinning injections, and the bottom plot is the same for the spinning injections. 
This study was conducted on $100$ injections in the low-mass region of the template bank. For the spinning case, the spin was chosen to be very high, i.e., with the spin parameter greater than $0.75$, so that the effect of spin is prominent.}
\label{fig:spinEffect}
\end{figure*}

To verify the above claims, we injected 100 non-spinning BNS and NSBH signals to simulated LIGO-Virgo data and ran targeted known-sky and sky-patch searches on them. In these studies 
the sky-patch used was a set of 21 sky-positions, all with the same RA, and included the true sky position. The remaining twenty sky positions were distributed such that their dec changed in steps of $2^\circ$ in both sides of the true dec value. The known-sky search used only the true sky-position. 
We repeated this study with a second set of injections that had the same parameters as the first set but now with a non-zero spin parameter for the binary components. The black hole and neutron star spin parameters of the injected CBC signals were in the ranges $(0.70, 0.98)$ and $(0.30, 0.75)$, respectively.
Only non-spinning templates were used in searching both types of injections, which were made quite strong in order to ensure that most of them are found.
However, four of the spinning injections were missed in the known-sky search. Three of these were recovered when the search was performed with a sky-patch.
Among the remaining injections, which were found in both types of searches (i.e., known-sky and sky-patch), $46$ were found with a SNR louder than the SNR of the same injection in the known-sky mode. A similar result is also observed for the same number of non-spinning injections. There, we found that $18$ injections in the sky-patch study were found with a SNR louder than the SNR of the same injection in the known-sky study. 
In Fig. \ref{fig:spinEffect}, the red bars show the gain in SNR of the triggers for non-spinning injections when using the sky-patch instead of the known-sky search.
The blue bars show the same for the spinning injections. 
More spinning injections than non-spinning injections are found with a louder SNR in the sky-patch mode than in the known sky mode. This observation confirms that in the presence of a spin parameter mismatch the parameter error covariance contributes significantly enough to reduce the SNR loss (and, concomitantly, alter the measured sky-position of the source). 


It is important to note that since the same mass template-bank is used in studying the spinning and non-spinning injections, the effect of its discreteness on SNR loss is present in both. Similarly, the mitigation of that effect in a sky-patch search due to sky-position error covariance is also present in both studies. However, since more spinning injections are recovered with a higher SNR in the sky-patch study than the known-sky study, the impact of the sky-position error covariance is more pronounced when there is mismatch in an additional parameter, namely, the spin values of the injections and the templates (which were always non-spinning).


\section{Discussion}
\label{sec:discussion}

Multiple future electromagnetic observatories are being planned that will target transient EM phenomena some of which may potentially be orphaned afterglows. A fraction of these phenomena may be due to compact binary coalescences, involving at least one neutron star. Some studies have argued that if SGRBs
are beamed and if some CBCNS systems that generate afterglows do not emit gamma-rays, then a lot many more orphaned afterglows, associated with CBCNS sources, may occur than SGRBs. Coupled GW and EM observations of SGRBs and afterglows, orphaned or not, will unequivocally confirm if the progenitors are indeed CBCNS sources or if there is actually a variety of progenitors. They will also teach us about their galactic or intergalactic environments, the nature of their host galaxies, stellar population synthesis, etc. This exploration critically depends on the EM observatories taking data concurrently with the GW observatories, such as aLIGO, AdV, and KAGRA. Some of the planned EM observatories will have a wide sky coverage and a good cadence to increase their chances of finding orphaned afterglows, which can be ephemeral, dropping in their apparent magnitudes quite rapidly.
Owing to their transient nature, it will be helpful to have GW detectors at multiple sites operate with large duty factors, and GW search codes running with low latency so that they can find CBCNS 
sources as they coalesce and alert the EM observatories in advance of a prompt EM (and even neutrino) emission or afterglow.
However, since GW detectors may not be able to localize every CBCNS or even unmodelled GW burst signal \cite{Abbott:2009kk} they detect, e.g., because not enough of them are operating to successfully triangulate the sky position, it will be important to follow-up orphaned afterglows in GW data. We highlight that an orphaned afterglow has not been found yet, and a search pipeline does not exist yet
that can use the sky position of such an EM source and search at that specific location but in a time window that can stretch for hours to weeks in the past to detect a GW signal associated with it in archived data. Such a pipeline needs to be developed.

Contrastingly, coherent CBCNS and burst search pipelines for detecting GW counterparts to SGRBs exist and have been run on archived GW data \cite{Briggs:2012ce}. Also, fast GW burst searches exist that have been used to alert EM observatories to look for EM counterparts and afterglows, but without any positive identification so far \cite{Virgo:2011aa}. This may change in the advanced detector era. Fast CBC searches are under development that will target detecting their GW signals in advance for the compact binary merger so that they can alert EM observatories to look for afterglows \cite{Cannon:2011vi}. This development notwithstanding, hunting for GW counterparts of SGRBs in archived GW data will always remain an important exercise. In this regard, in this paper we make the case that searching with a sky-patch with multiple sky positions can 
improve the GW detection efficiency even when the sky position of the SGRB, or an orphaned afterglow, is known accurately through EM observations. 



\acknowledgments

Special thanks are due to Alexander Dietz, Nickolas Fotopoulos, and Ian Harry for extensive discussions and collaboration on GRB triggered searches for GWs in LIGO-Virgo data. We would also like to thank Matthew Duez, Steve Fairhurst, Atish Kamble, Greg Mendell, Fred Raab and Sanjay Reddy for helpful discussions. One of us (SG) would like to thank the Inter-University Centre for Astronomy and Astrophysics, where a part of this work was carried out, for its support and hospitality. This work was supported in part by NSF grants PHY-0855679 and PHY-1206108.

\bibliographystyle{plainnat}
\bibliography{References}

\end{document}